\documentclass[useAMS,usenatbib,usegraphicx]{mn2e}

\usepackage{subfigure}
\usepackage{url}
\usepackage{times}
\usepackage{comment}
\usepackage{color}
\usepackage{enumitem}

\makeatletter

\renewcommand{\@thesubfigure}{(\alph{subfigure})\hskip\subfiglabelskip}
\renewcommand{\@@thesubfigure}{(\alph{subfigure})}

\newcommand\ion[2]{#1$\;${\small\rmfamily\@Roman{#2}}\relax}%

\newcommand\aap{{A\&A}}%
\newcommand\araa{{ARA\&A}}%
\newcommand\aaps{{A\&AS}}%
\newcommand\mnras{{MNRAS}}%
\newcommand\apj{{ApJ}}%
\newcommand\apjs{{ApJS}}%
\newcommand\apjl{{ApJ}}%
\newcommand\aj{{AJ}}%
\newcommand\pasp{{PASP}}%
\newcommand\memsai{{Mem.~Soc.~Astron.~Italiana}}%
\newcommand\procspie{{Proc.~SPIE}}%

\setlist[enumerate]{noitemsep}
\setlist[enumerate,1]{leftmargin=*}
\setlist[description]{noitemsep}
\setlist[description,1]{leftmargin=*}
\setenumerate[0]{label=(\arabic*)}

\makeatother

\title[Integrated Abundances for Chemical Tagging]{Optimal Integrated
Abundances for Chemical Tagging of Extragalactic Globular
Clusters\footnote{Based on observations obtained with the Hobby-Eberly
Telescope, which is a joint project of the University of Texas at
Austin, the Pennsylvania State University, Stanford University,
Ludwig-Maximilians-Universit\"{a}t M\"{u}nchen, and
Georg-August-Universit¡ät G\"{o}ttingen.}}

\author[C.M. Sakari et al.]{Charli M. Sakari$^{1,4}$\thanks{E-mail:
sakaricm@uvic.ca}, Kim Venn$^{1}$, Matthew Shetrone$^{2}$,
Aaron Dotter$^{3}$, and
\newauthor Dougal Mackey$ ^{3}$\\
$^{1}$Department of Physics and Astronomy, University of Victoria,
Victoria, BC V8W 3P2, Canada\\
$^{2}$McDonald Observatory, University of Texas at Austin, HC75 Box
1337-MCD, Fort Davis, TX 79734, USA\\
$^{3}$Research School of Astronomy and Astrophysics, The Australian
National University, Weston, ACT 2611, Australia\\
$^{4}$Vanier Canada Graduate Scholar}

\begin{document}

\maketitle

\label{firstpage}

\begin{abstract}
High resolution integrated light (IL) spectroscopy provides detailed
abundances of distant globular clusters whose stars cannot be
resolved.  Abundance comparisons with other systems (e.g. for chemical
tagging) require understanding the systematic offsets that can occur
between clusters, such as those due to uncertainties in the underlying
stellar population.  This paper analyses high resolution IL spectra of
the Galactic globular clusters 47~Tuc, M3, M13, NGC~7006, and M15 to
1) quantify potential systematic uncertainties in Fe, Ca, Ti, Ni, Ba,
and Eu and 2) identify the most stable abundance ratios that will be
useful in future analyses of unresolved targets.  When stellar
populations are well-modelled, uncertainties are $\sim0.1-0.2$ dex
based on sensitivities to the atmospheric parameters alone; in the
worst case scenarios, uncertainties can rise to $0.2~-~0.4$~dex.  The
[\ion{Ca}{1}/\ion{Fe}{1}] ratio is identified as the optimal
integrated [$\alpha$/Fe] indicator (with offsets $\la 0.1$ dex), while
[\ion{Ni}{1}/\ion{Fe}{1}] is also extremely stable to within $\la 0.1$
dex.  The [\ion{Ba}{2}/\ion{Eu}{2}] ratios are also stable when the
underlying populations are well modelled and may also be useful for
chemical tagging.
\end{abstract}


\section{Introduction}\label{sec:Intro}
Chemical tagging has been very successful in the Milky Way, enabling
the identification of stellar streams and globular clusters (GCs) that
were likely accreted from dwarf galaxies and demonstrating that
accretion has played some role in the formation of the Milky Way
(e.g., \citealt{Freeman,Cohen2004,Sbordone2005,Sakari2011}).  In order
to form a more  general picture of galaxy formation, additional
systems must be studied; however, similar studies of other massive
galaxies are much more difficult to perform because individual stars
cannot be resolved for high resolution spectroscopic
observations. These distant systems must therefore be studied through
their \textit{integrated light} (IL).  GCs are particularly useful for
distant IL work, since they appear as bright point-sources, and can be
observed at greater distances than their individual stars.

The main difficulty in IL studies is interpreting the observations in
terms of the physical properties of the underlying stellar
population.  These difficulties occur largely because of degeneracies
in the IL spectra.   One way to counter these effects is
to calibrate IL analysis techniques to nearby, well-studied systems or
theoretical models.  For example, IL photometry has been calibrated
to GC metallicities (see \citealt{BrodieStrader2006} for a review),
while low to medium resolution ($R < 5000$) IL spectroscopy has been
semi-empirically calibrated for determinations of age, metallicity,
and some element abundance ratios (C, N, O, Mg, Na, Ca, e.g., see
\citealt{Schiavon2002,LeeWorthey2005}).

Recently, high resolution ($R > 20000$) spectroscopy has been applied
to IL studies of Galactic GCs (e.g. MB08; \citealt{Colucci2009}; and
\citealt{Sakari2013}, hereafter Paper 1).   These high
resolution studies can be used to improve the precision of the
chemical abundance results by examining spectral lines that are less
blended and have a range of strengths.   Chemical abundances of more
elements can be determined as well, e.g. from neutron capture elements
such as Ba and Eu.  The program ILABUNDS (presented in MB08) has
been tested on Galactic GCs (MB08, \citealt{Cameron2009}, Paper I) and
has been applied to GCs in M31, the Large Magellanic Cloud, Local
Group dwarf galaxies, and the early-type galaxy NGC~5128
\citep{Colucci2009,Colucci2011a,Colucci2011b,Colucci2012,Colucci2013}.
Studies of nearby systems have shown that IL abundances can reproduce
the abundances of individual stars and will trace the abundance
patterns of field stars that formed in the same environment (for
elements that do not vary within clusters).  As IL analyses are
pushed to more distant systems, they are providing the first detailed
studies of chemical enrichment in systems outside the Local
Group---for example, IL observations of GCs show that the elliptical
galaxy NGC~5128 seems to have undergone rapid chemical enrichment
compared to the Milky Way \citep{Colucci2013}.

Abundance results for GCs can be as precise as those from individual
stars when determined from high resolution IL spectra ($\sim 0.1$ dex;
see Paper 1).    However, both kinds of spectroscopic analyses suffer
from systematic errors.   For individual stars, systematic errors are
usually due to the uncertainties in the temperature, gravity,
metallicity, and microturbulence of the model atmosphere, and are
often added in quadrature (other sources of systematic error exist but
are not usually folded into the total error).  These systematic errors
can occasionally exceed the random abundance errors
\citep{McWilliam1995a}, and complicate comparisons of abundance
results from various studies.  For IL spectra of GCs, a determination
of the total systematic errors is more complicated.   \textit{Without
a general understanding of the systematic errors that occur during an
IL analysis, it is extremely difficult to compare IL abundances to
those from individual stars, or to those from other GCs, regardless of
the precision in the abundances or the quality of the spectra}.

The goal of this paper is to understand and quantify the systematic
errors that are present in a typical high resolution IL spectral
analysis.  In general, abundance analyses suffer from two different
types of systematic uncertainties:
\begin{enumerate}
\item Uncertainties that arise from models, techniques, corrections,
or assumptions that apply to all targets in a given study (such as the
choice of model atmospheres, the methods for measuring spectral lines,
NLTE corrections, atomic data, etc.).  Such uncertainties can be
reduced or eliminated through \textit{differential}
analyses.\footnote{However, some of these uncertainties may depend on
GC properties such as metallicity.}
\item Uncertainties that arise from discrepancies between reality and
input simplifications or assumptions (e.g. models of the evolved stars
or the inclusion of interloping field stars).  Such uncertainties will
vary between targets, and cannot be removed through differential
analyses.
\end{enumerate}
\noindent IL spectral analyses suffer from both types of errors.
Under the assumptions that each GC's underlying stellar population is
known and that model atmospheres can be correctly assigned to the
stars in the cluster, most uncertainties should fall under the first
type.  In reality, however, GC stellar populations cannot be
perfectly modelled---moreover, each GC is unique, and the specific
deviations from the models will vary from cluster to cluster.
Previous authors have investigated the systematic effects on both low
and high resolution IL spectra as a result of various assumptions
about the underlying population.  \citet{PercivalSalaris2009}
investigated the effects on low resolution spectral indices as a
result of temperature and metallicity scale offsets between the
stellar evolution models and the spectral libraries.  The effects of
improperly modelling particular stellar subpopulations, e.g. the
horizontal branch (HB), have also been tested extensively
(\citealt{Schiavon2004,Colucci2009}, Paper I).  This paper goes
further, by isolating and investigating the systematic errors that
occur when atmospheric parameters are assigned to the stars in the
modelled population.

High resolution IL spectra of the well-studied, resolved Galactic GCs
47~Tuc, M3, M13, NGC~7006, and M15 are used to perform these
tests---these GCs span a large range in metallicity and HB morphology
and therefore form an ideal test sample.  The observations, data
reduction, and abundance analysis methods are discussed in Section
\ref{sec:Data}.  The original, baseline abundances for comparisons are
presented in Section \ref{sec:ILABUNDS}.  The specific systematic
uncertainty tests are described in Appendices \ref{sec:CMD},
\ref{sec:HRD}, \ref{sec:Both}; errors that are expected to occur in a
Colour-Magnitude Diagram (CMD) based analysis (for  resolved systems)
are presented in Appendix \ref{sec:CMD}, offsets that may occur in
theoretical Hertzsprung-Russell Diagram (HRD) analyses (for unresolved
systems) are presented in Appendix \ref{sec:HRD}, and uncertainties
that are expected to occur in both types of analyses (such as
potential foreground stars) are described in Appendix \ref{sec:Both}.
The results are summarized in Section \ref{sec:Discussion}, where the
implications of these tests for applications to distant systems are
discussed.  The best element ratios for IL chemical tagging are also
identified.

\section{Data and Analysis Methods}\label{sec:Data}

\subsection{Target Selection}\label{subsec:Targets}
The target GCs 47~Tuc, M3, M13, NGC~7006, and M15 were selected to
cover a wide range of metallicities (from $[\rm{Fe/H}] \sim -0.7$ to
$-2.4$; \citealt{Harris}) and horizontal branch (HB) morphologies
(from very red to very blue; see Table \ref{table:Observations}).  In
particular, M3, M13, and NGC~7006 form a ``second parameter'' triad,
i.e. the three clusters have approximately the same metallicity, yet
have very different HB morphologies.  These clusters therefore provide
an excellent test set, since the systematic effects of metallicity
and HB morphology can be investigated.

\subsection{Observations and Data Reduction}\label{subsec:Observations}
With the exception of 47~Tuc (which was obtained by R. Bernstein and
A. McWilliam at the Las Campanas Observatory, or LCO; see MB08), the
GC IL spectra were obtained with the High-Resolution Spectrograph (HRS;
\citealt{HRSref}) on the Hobby-Eberly Telescope (HET;
\citealt{HETref,HETQueueref}) at McDonald Observatory in Fort Davis,
TX.  The 1$\arcsec$ slit was used, providing an instrumental
resolution of $R=30,000$.  The 600 gr/mm cross disperser provides
wavelength coverage from $\sim 5320-6290$ \AA \hspace{0.025in} on the
blue chip and $\sim 6360-7340$ \AA \hspace{0.025in} on the red chip.
The large 3$\arcsec$ fibres were scanned across the cluster cores to
obtain IL spectra of the central regions (see Table \ref{table:Observations}
to see how far the coverage extended). More details on the
observations can be found in Paper I; the S/N ratios of the final
spectra are summarized in Table \ref{table:Observations}.

As described in Paper I, the data reduction was performed in the Image
Reduction and Analysis Facility program (IRAF).\footnote{IRAF is
distributed by the National Optical Astronomy Observatory, which is
operated by the Association of Universities for Research in Astronomy,
Inc., under cooperative agreement with the National Science
Foundation.}  Standard HRS data reduction methods were used, except
that bias frame removal was not performed and optimal variance
weighting was used during aperture extraction.  To subtract the sky,
separate sky exposures were taken after each observation; these sky
spectra were replaced with continuum fits with the emission lines
added back in.  Telluric standards were also observed in order to
remove telluric absorption lines.  The IL spectra were normalized using the
continuum fits to an Extremely Metal-Poor (EMP) star, as described
in Paper I.  The individual spectra were combined using average
sigma-clipping routines to mitigate the effects of cosmic rays.
Velocity information was determined through cross-correlations with an
Arcturus template spectrum (from the Arcturus
Atlas;\footnote{\url{ftp://ftp.noao.edu/catalogs/arcturusatlas/}}
\citealt{Hinkle2003}), and is also listed in Table \ref{table:Observations}.  More information on the data reduction
procedure can be found in Paper I.

\begin{table*}
\centering
\begin{minipage}{165mm}
\caption{Basic information about the target GCs.\label{table:Observations}}
  \begin{tabular}{@{}llccccccccccc@{}}
  \hline
Cluster & [Fe/H]$_{\rm{lit}}$ & HB index & & S/N$^{a}$ & S/N$^{a}$ & & $v_{\rm{helio, obs}}$ &
$v_{\rm{helio, lit}}$ & $\sigma_{\rm{obs}}$ & $\sigma_{\rm{lit}} $ & &
$r_{\rm{IL spectra}}^{b}$ \\
 &  &  & & (5500 \AA) & (7000 \AA) & & (km s$^{-1}$) & (km s$^{-1}$) &
(km s$^{-1}$) & (km s$^{-1}$) & & (r$_{\rm{c}}$)\\
  \hline
47~Tuc$^{c}$& -0.70 & -0.99 & & 120 & 180 & & -             & -      & $11.50\pm 0.30^{d}$ & 11.0 & & 1.1 \\
 & & & & & & & & & & &&\\	     
M3          & -1.60 & 0.08  & & 180 & 230 & & $-146.0 \pm 1.1$ & -147.6 & $5.66\pm 0.15$   & 5.5  & & 1.8 \\
 & & & & & & & & & & &&\\	     
M13         & -1.60 & 0.97  & & 130 & 250 & & $-247.5 \pm 1.3$ & -244.2 & $7.23\pm 0.33$   & 7.1  & & 1.7 \\
 & & & & & & & & & & &&\\	     
NGC 7006    & -1.50 & -0.28 & & 65  & 130 & & $-380.4 \pm 0.7$ & -384.1 & $4.49\pm 0.60$   & -    & & 2.4 \\
 & & & & & & & & & & &&\\	     
M15         & -2.40 & 0.67  & & 95  & 220 & & $-106.6 \pm 0.2$ & -107.0 & $12.54\pm0.60$   & 13.5 & & 7.1 \\
\hline
\end{tabular}
\end{minipage}\\
\medskip
\raggedright {\bf References: } More information can be found in Paper
I.  The [Fe/H] estimates are from isochrone fitting
\citep{Dotter2010,Dotter2011}.    The HB index,
$(B~-~R)~/~(B~+~V~+~R)$, comes from
\citet{MackeyVanDenBergh}. Literature values for $v_{\rm{rad}}$ and
$\sigma$ are from the Harris Catalog \citep{Harris}.\\
$^{a}$S/N ratios (per pixel) are measured in IRAF.\\
$^{b}$The maximum coverage of the IL spectra, expressed in units of the core
radius (which was obtained from \citealt{Harris}).\\
$^{c}$47~Tuc was observed with the Las Campanas 2.5 m Du Pont
Telescope by R. Bernstein \& A. McWilliam; see MB08 for more details.\\
$^{d}$This velocity dispersion has been determined in the same way as
the other GCs, for consistency.\\
\end{table*}

\subsection{Line List and EW Measurements}\label{subsec:DAOSPEC}
The spectral lines in this analysis were selected from the IL
spectral line lists from MB08 and \citet{Colucci2009} and the RGB line
lists from \citet{Sakari2011} and \citet{Venn2012}.  Fe, Ca, Ti, Ni,
Ba, and Eu lines were selected for this analysis because these
elements are useful for chemical tagging purposes.  The specific Ca,
Ti, Ni, and Ba lines are listed in Table \ref{table:LineList}, along
with the adopted atomic data.

\begin{table*}
\centering
\begin{minipage}{165mm}
\begin{center}
\caption{The Line List.$^{a}$\label{table:LineList}}
  \begin{tabular}{@{}lccccccccc@{}}
  \hline
Wavelength & Element & E.P. & log gf & \multicolumn{6}{l}{Equivalent width (m\AA)}\\
(\AA) & & (eV) & & Sun & 47~Tuc & M3 & M13 & NGC~7006 & M15\\
\hline
5581.979 & \ion{Ca}{1} & 2.523 & -0.555 & 97.0   & 113.0  & 50.3 & -$^{b}$& 63.3 & -$^{b}$\\
5588.764 & \ion{Ca}{1} & 2.526 &  0.358 & -$^{c}$& -$^{b}$& 103.7& 97.2   & 124.0& 47.0\\
5590.126 & \ion{Ca}{1} & 2.521 & -0.571 & -$^{c}$& 112.1  & 58.1 & 53.9   & 67.4 & -$^{b}$\\
5601.286 & \ion{Ca}{1} & 2.526 & -0.69  & -$^{c}$& 113.9  & 56.0 & 51.7   & 57.0 & -$^{b}$\\
5857.459 & \ion{Ca}{1} & 2.933 &  0.24  & -$^{c}$& 131.0  & 86.5 & 74.0   & 95.0 & -$^{b}$\\
\hline
\end{tabular}
\end{center}
\end{minipage}\\
\medskip
\raggedright {\bf Notes: } Equivalent widths were measured in DAOSPEC;
all strong lines were checked and refined in \textit{splot}.  Lines
stronger than 150 m\AA \hspace{0.025in} were not included in the
analysis.  Note that this limit may be too high, since some CMD boxes
have EWs $> 150$ m\AA \hspace{0.025in}.  However, with the exception
of 47~Tuc, only a handful of IL spectral lines are stronger than 110
m\AA. Furthermore, none of the clusters show any noticeable trends in
\ion{Fe}{1} abundance with EW in the CMD-based analyses.\\
$^{a}$Table \ref{table:LineList} is published in its
entirety in the electronic edition of \textit{Monthly Notices of the
Royal Astronomical Society}. A portion is shown here for guidance
regarding its form and content.\\
$^{b}$ The lines not measured in the target GCs were too weak or were
obscured by noises, cosmic rays, etc.\\
$^{c}$ The lines that were not measured in the Solar spectrum were
those stronger than EW$ = 150$ m\AA; for those lines the Solar values
of \citet{Asplund2009} were used.\\
\end{table*}

Paper I presented the integrated Fe, Na, Mg, and Eu abundances, and
showed that for the target GCs Na, Mg, and Eu are affected by
star-to-star variations within the clusters.  For that reason, these
abundances may not be useful for chemical tagging purposes, and Na and
Mg are not considered in this systematic error analysis.  Eu is
retained, however, as the star-to-star Eu variations within a cluster
are not always significant.  \citet{Roederer2011} demonstrated that
Eu variations are only large for the most massive GCs. Furthermore,
these dispersions are not seen in all massive, metal-poor GCs---for
instance, \citet{Cohen2011} detect no heavy element dispersion in M92.
Ba also varies within some GCs (e.g. M15; \citealt{Worley2013});
however, again this is likely only the case for the most massive GCs.
Furthermore, evidence suggests that Ba and Eu may vary together, such
that the ratio of [Ba/Eu] may still be useful for chemical tagging
purposes \citep{Worley2013}.

EWs were measured with the automated program DAOSPEC\footnote{DAOSPEC
has been written by P.B. Stetson for the Dominion Astrophysical
Observatory of the Herzberg Institute of Astrophysics, National
Research Council, Canada.} \citep{DAOSPECref}. Paper I showed that
DAOSPEC is capable of reproducing EWs measured in IRAF's
\textit{splot} and those measured with the program GETJOB
\citep{McWilliam1995a}.  Lines stronger than 150 m\AA \hspace{0.025in}
were removed from the abundance analysis (see the discussions in Paper
I and \citealt{McWilliam1995b}). The Fe EWs are tabulated in Paper I;
the EWs for the other lines are shown in Table \ref{table:LineList}.
\ion{Eu}{2} has only a single, weak line, and its abundance must be
determined via spectrum syntheses (see Paper I).  In this case, the
EWs that matched the synthesis-based abundances were used for all
differential errors analyses in Appendices \ref{sec:CMD},
\ref{sec:HRD}, and \ref{sec:Both}; these EWs are also listed in Table
\ref{table:LineList}.

\subsection{Atmospheric Parameters and Models}\label{subsec:ModelAtms}
In an IL spectral analysis stellar model atmospheres can either be
generated with observed photometry or theoretical isochrones,
depending on the target. Nearby clusters have high-quality CMDs, and
each star's colour and magnitude can be used to infer its temperature
and other atmospheric parameters.  This is the method that is used to
derive the baseline abundances of the target Galactic GCs (47~Tuc, M3,
M13, NGC~7006, and M15), which are presented in Section
\ref{sec:ILABUNDS} (see also Paper I and MB08).  The stars in very
distant clusters, however, cannot be resolved, and isochrones must be
used to model the underlying population's HRD.

\subsubsection{Input Photometry}\label{subsubsec:Photometry}
The \textit{Hubble Space Telescope} ({\it HST}) photometry used in the
CMD~-~based analyses comes from two sources.  The 47~Tuc $B$, $V$ data
is from \citet{Guhathakurta1992} and \citet{Howell2000}, and was
provided by R. Schiavon---this is the same photometry presented in
MB08. The $V$, $I$ data for all clusters are from the ACS Survey of
Galactic Globular Clusters
(\citealt{Sarajedini2007,Anderson2008,Dotter2011}).  The {\it HST}
magnitudes were converted to Johnson's $V$, $I$ magnitudes via the
transformations in \citet{Sirianni2005}.   Stars within the maximum
radii observed in the IL spectra were selected for input to ILABUNDS,
using the cluster centres from \citet{Goldsbury2010}; this circular
selection leads to slight discrepancies with the irregular coverage
patterns.  The CMDs were then binned into boxes, as described in
Paper~I.

\subsubsection{Input Isochrones}\label{subsubsec:Isochrones}
When isolated effects are investigated (e.g. HB morphology; see
Appendix \ref{subsec:HBMorphology}), the boxes from the input
\textit{photometry} are used to simplify the comparisons with the
original, CMD-based abundances.  For tests that require models of the
underlying stellar population, the isochrones from the following
sources are considered:
\begin{enumerate}
\item BaSTI/Teramo models \citep{BaSTIref,Cordier2007} with
\citet{ODFNEWref} opacities---the default isochrones utilize extended,
$\eta = 0.2$ AGB models, though other treatments are investigated.
\item Dartmouth Stellar Evolution Database models (DSED;
\citealt{DSED}) \item Victoria-Regina Stellar models \citep{VRref}.
\end{enumerate}

\subsubsection{Model Atmospheres}\label{subsubsec:ModelAtms}
Once the atmospheric parameters of a box are known, a corresponding
Kurucz model
atmosphere\footnote{\url{http://kurucz.harvard.edu/grids.html}}
\citep{KuruczModelAtmRef} is then assigned.  The grid values are
interpolated to each box's specific $T_{\rm{eff}}$ and $\log g$.

\subsection{Isotopic \& Hyperfine Structure Corrections}\label{subsec:HFS}
The isotopic and hyperfine structure (HFS) components for the
\ion{Ba}{2} lines are from \citet{McW1998} while the \ion{Eu}{2}
components are from \citet{LawlerA} and \citet{LawlerB}.  All HFS
corrections were found to be negligible ($\la 0.05$ dex) and were not
applied to any of the \ion{Ba}{2} or \ion{Eu}{2} abundances presented
in this paper.

\subsection{Solar Abundances}\label{subsec:Sun}
All of the [Fe/H] and [X/Fe] ratios presented in this paper are
calculated {\it line by line} relative to the Solar abundances derived
with the equivalent widths in Table \ref{table:LineList} and Paper I.
These equivalent widths were measured in the Solar spectrum ($R =
300,000$; \citealt{Kurucz2005}) from the Kurucz 2005 solar flux
atlas.\footnote{\url{http://kurucz.harvard.edu/sun.html}}  Solar
atmospheric parameters of $T_{\rm{eff}} = 5777$ K, $\log g = 4.44$
dex, $\chi = 0.85$ km s$^{-1}$, and $[\rm{M/H}] = 0.0$ were adopted
\citep{Yong}.  When the Solar lines were stronger than 150 m\AA,
\citet{Asplund2009} Solar abundances were adopted for those lines.

\section{Initial Abundances}\label{sec:ILABUNDS}
The integrated abundances are determined with the equivalent width
version of the program ILABUNDS (described in detail in MB08).  The
initial CMD-based abundances with the standard ILABUNDS input are
presented in Table \ref{table:OriginalAbunds}.  The [Fe/H] and [X/Fe]
ratios were calculated differentially \textit{for each line}, using
the Solar abundances derived from the EWs in Table
\ref{table:LineList} (see Section \ref{subsec:Sun}).  As in individual
stellar analyses, the [X/Fe] ratios are calculated by comparing
elements of similar ionization states.  Thus, the [X/Fe] ratios of
neutral species are relative to \ion{Fe}{1} and those of singly
ionized species are relative to \ion{Fe}{2}.\footnote{In RGB stars,
comparing singly ionized species to \ion{Fe}{2} reduces systematic
uncertainties, as this compares the dominant ionization stages.  IL is
dominated by RGB stars, and this methodology is therefore adopted.}
Unless otherwise noted, abundance uncertainties in Appendices
\ref{sec:CMD}-\ref{sec:Both} are calculated relative to these baseline
abundances.

\begin{table*}
\centering
\begin{minipage}{165mm}
\caption{Initial GC Abundances.\label{table:OriginalAbunds}}
  \begin{tabular}{@{}lcccccccc@{}}
  \hline
& $[$\ion{Fe}{1}/H$]$ & $[$\ion{Fe}{2}/H$] $ & $[$\ion{Ca}{1}/\ion{Fe}{1}$]$ & $[$\ion{Ti}{1}/\ion{Fe}{1}$]$ & $[$\ion{Ti}{2}/\ion{Fe}{2}$]$ & $[$\ion{Ni}{1}/\ion{Fe}{1}$]$ & $[$\ion{Ba}{2}/\ion{Fe}{2}$]$$^{\rm{a}}$ & $[$\ion{Eu}{2}/\ion{Fe}{2}$]^{\rm{a,b}}$  \\
\hline
47~Tuc     &  $-0.81\pm0.02^{c}$ & $-0.69\pm0.07$ & $0.28\pm0.05$ & $0.27\pm0.09$ & $0.29\pm0.07$ & $-0.04\pm0.07$ & $-0.01\pm0.08$ & $0.27\pm0.14$ \\
N          &  \phantom{-}68  & 4              & 9              &  6             & 2              & 7              & 2              & $1$ \\
{\it Lit.} &  $-0.72$        & $-0.72$        & $ 0.19$        & $ 0.24$        & $ 0.36 $       & $ $0.0         & $ 0.31$        & $0.14\pm0.03$\\
{\it MB08} &  $-0.75\pm0.03$ & $-0.72\pm0.06$ & $0.31\pm0.08$  & $0.41\pm0.07$  & $0.54\pm0.09$  & $ 0.0 \pm0.06$ & $0.02\pm0.02$  & $0.04$  \\
& & & & & & & & \\
M3         &  $-1.51\pm0.02$ & $-1.58\pm0.05$ & $ 0.37\pm0.06$ & $ 0.30\pm0.09$ & $ 0.36\pm0.06$ & $-0.03\pm0.08$ & $-0.06\pm0.09$ & $0.75\pm0.11$ \\
$N$        &  \phantom{-}95  & 5              & 17             & 6              & 2              & 7              & 3              & $1$ \\
{\it Lit.} &  $-1.50$        & $-1.50$        & $ 0.27$        & $ 0.32$        & $ 0.32$        & $-0.02$        & $ 0.17$        & $0.51\pm0.02$\\
& & & & & & & & \\
M13        &  $-1.57\pm0.02$ & $-1.55\pm0.07$ & $ 0.33\pm0.06$ & $ 0.29\pm0.13$ & $ 0.42\pm0.06$ & $-0.02\pm0.08$ & $ 0.06\pm0.08$ & $0.76\pm0.10$ \\
$N$        &  \phantom{-}71  & 3              & 13             & 6              & 2              & 7              & 3              & $1$ \\
{\it Lit.} &  $-1.53$        & $-1.53$        & $ 0.26$        & $ 0.39$        & $ 0.39$        & $0.02$         & $ 0.24$        & $0.49\pm0.03$\\
& & & & & & & & \\
NGC~7006   &  $-1.52\pm0.03$ & $-1.56\pm0.07$ & $ 0.46\pm0.12$ & $ 0.29\pm0.17$ & $ 0.33\pm0.04$ & $-0.04\pm0.09$ & $ 0.19\pm0.08$ & $0.72\pm0.15$ \\
$N$        &  \phantom{-}73  & 5              & 14             & 7              & 1              & 6              & 3              & $1$ \\
{\it Lit.} &  $-1.52$        & $-1.52$        & $ 0.30$        & $0.32$         & $0.32$         & $0.02$         & $ 0.33$        & $0.36\pm0.02$\\
& & & & & & & & \\
M15        &  $-2.30\pm0.03$ & $-2.38\pm0.10$ & $ 0.31\pm0.09$ & $-^{d}$        & $ 0.33\pm0.12$ & $+0.01\pm0.08$ & $-0.21\pm0.06$ & $1.31\pm0.20$ \\
$N$        &  \phantom{-}31  & 1              & 6              & $-\phantom{^{d}}$            & 2              & 1              & 3              & $1$ \\
{\it Lit.} &  $-2.37$        & $-2.37$        & $ 0.27$        & $ 0.32$        & $ 0.32$        & $0.01$         & $ 0.11$        & $0.63\pm0.03$\\
& & & & & & & & \\
\hline
\end{tabular}
\end{minipage}\\
\medskip
\raggedright {\bf Notes: } [Fe/H] and [X/Fe] values were calculated
{\it line by line} relative to the Solar values, as derived with the
EWs in Table \ref{table:LineList}.\\
{\bf References: } Literature abundances are from MB08, \citet{Pritzl2005},
\citet{Sneden1997}, \citet{Kraft1998}, \citet{Carretta2004},
\citet{Jasniewicz2004}, \citet{CohenMelendez2005} and the references
listed in Paper I.\\
$^{a}$ The Ba and Eu abundances vary between stars in
some of these GCs such that the integrated abundances may not match
the GC averages.\\
$^{b}$ These abundances were calculated via spectrum syntheses; see
Paper I. \\
$^{c}$ 47~Tuc's [\ion{Fe}{1}/H] abundances are slightly lower than
expected, which may be due to the treatment of damping in ILABUNDS
(A. McWilliam, private communication).\\
$^{d}$ M15's \ion{Ti}{1} lines were not sufficiently strong to
determine a robust [\ion{Ti}{1}/\ion{Fe}{1}] ratio.  To investigate
the effects on \ion{Ti}{1} at M15's metallicity, EWs were determined
to match the [\ion{Ti}{2}/\ion{Fe}{2}] abundance.  These values are
only used to calculate M15's systematic offsets in \ion{Ti}{1}.\\
\end{table*}

\subsection{Random Errors}\label{subsec:Errors}
The random abundance errors were calculated as in \citet{Shetrone2003}
and \citet{Sakari2011}. For each element, three different
uncertainties were calculated and compared.

\begin{enumerate}
\item {\it The line-to-line abundance scatter.}  For a single element
there is some standard deviation, $\sigma$, about the mean abundance.
The uncertainty in the mean abundance is therefore $\delta_{\rm{X}} =
\sigma/\sqrt{N}$, where $N$ is the number of spectral lines.
\item {\it The EW uncertainty.}  The error of an EW measurement in a
particular spectrum can be estimated with the \citet{Cayrel} formula;
note that an additional 10\%~$\cdot$~EW error is included (see
\citealt{Shetrone2003} and \citealt{Sakari2011}).  The abundances were
recalculated with larger and smaller EWs, and the offset in the mean
abundance, $\sigma_{\rm{EW}}$, was divided by $\sqrt{N}$ to give the
uncertainty in the mean abundance, $\delta_{\rm{EW}}$.
\item {\it The iron line-to-line scatter.}  Because there are many
iron lines, the iron line-to-line scatter provides an estimate of the
\textit{minimum} abundance uncertainty, $\delta_{\rm{Fe}}$.  For an
element with few detectable spectral lines the above error types may
underestimate the true abundance error.
\end{enumerate}

The largest of these three uncertainties ($\delta_{\rm{X}}$,
$\delta_{\rm{EW}}$, and $\delta_{\rm{Fe}}$) is adopted as the final
\textit{random} abundance error for that element.

\subsection{Comparisons with Literature Abundances}\label{subsec:LitAbunds}
Paper I demonstrated that the EW-based Fe abundances are in excellent
agreement with literature values, while the \ion{Eu}{2} abundances
fall within the literature ranges.  Table \ref{table:OriginalAbunds}
demonstrates that for the most part the integrated Ca, Ti, Ni, and Ba
abundances agree well with the literature abundances from individual
stars.  The 47~Tuc [\ion{Ti}{1}/\ion{Fe}{1}] and
[\ion{Ti}{2}/ion{Fe}{2}] ratios do not agree with the values from
MB08---however, these discrepancies seem to be due to 1) line choice
and 2) techniques for calculating differential [X/Fe] ratios.

\subsection{Systematic Offsets: A Description of Appendices
\ref{sec:CMD}, \ref{sec:HRD}, and
\ref{sec:Both}}\label{subsec:SystematicErrors}
The systematic errors are determined by changing the atmospheric
parameters by various amounts, and comparing the new abundances to the
original baseline abundances in Table \ref{table:OriginalAbunds}
(unless otherwise noted).  These differences are calculated as $\Delta
[\rm{X/Fe}]~=~[\rm{X/Fe}]~-~[\rm{X/Fe}]_{\rm{orig}}$. 

The specific details of the systematic errors calculations are
presented in Appendices \ref{sec:CMD}-\ref{sec:Both}.  As discussed in
Section \ref{sec:Intro}, the types of systematic uncertainties depend
on the analysis type.  Appendix \ref{sec:CMD} first investigates the
errors that only occur in a CMD-based analysis, while Appendix
\ref{sec:HRD} investigates the uncertainties in an HRD
analysis. Appendix \ref{sec:Both} then describes the errors that are
present in both types of analyses.

The largest systematic offsets are summarized in Table
\ref{table:Summary}.  Offsets $\ge 0.05$ dex are in bold.  The
specific magnitude of the errors can vary between clusters (due to
metallicity, HB morphology, etc.).  Occasionally the worst case
scenarios are considered, in which case the errors are likely to be
upper limits. It is unclear how to combine the individual errors; in
particular, it is unclear if all the errors are independent, and
should be combined in quadrature.

\section{Discussion}\label{sec:Discussion}
IL spectral analyses provide chemical abundances for individual GCs; 
{\it high resolution} IL spectroscopy is particularly well suited for
\textit{chemical tagging}.  Chemical tagging utilizes detailed
chemical abundances to identify chemically peculiar stars and GCs that
likely originated in dwarf galaxies (see
\citealt{Freeman,Cohen2004,Sakari2011}). In the MW, detailed
abundances and kinematic info can link together individual stars and
GCs that were accreted from the same dwarf galaxy.  Because of the
nature of IL analyses, it may not be possible to link GCs to specific
streams---however, it should be possible to separate dwarf-associated
GCs from those formed in a massive galaxy.  Chemically distinct GCs
can only be identified if the abundance ratios are sufficiently robust
to systematic uncertainties, ideally within 0.1 dex.\footnote{For
instance, Milky Way halo stars at $[\rm{Fe/H}]\la -1$ have
$[\alpha/{\rm{Fe}}]~=~+~0.3$, while dwarf galaxy stars at similar
metallicity have $[\alpha/{\rm{Fe}}] \sim 0$.  For a $3\sigma$
confirmation that a GC is chemically more like a dwarf galaxy, useful
ratios should have systematic errors $\la 0.1$ dex.}  This section
summarizes the accuracy of each abundance ratio and discusses
implications for future extragalactic studies.

\subsection{Summary of Results: Abundance Accuracy}\label{subsec:Summary}
Table \ref{table:Summary} provides a summary of the largest effects
on the chemical abundance ratios, based on the tests described in the
Appendices.  The values in the table are upper limits; the values only
apply to specific cases or worst case scenarios that will not apply to
all GCs.  Furthermore, many of the uncertainties vary between clusters
as a result of, e.g., metallicity or HB effects.  From the upper
limits in Table \ref{table:Summary} it is clear that abundance ratios
are more stable to uncertainties than others. The accuracy of the
individual abundance ratios are discussed in detail below.  Of course,
the results presented here are dependent upon the observed lines and
their properties, and may vary if different wavelength regions and/or
spectral lines are observed.

\begin{table*}
\centering
\begin{minipage}{175mm}
\caption{Summary of results.\label{table:Summary}}
  \begin{tabular}{@{}lccccccccc@{}}
  \hline
& $ \Big | \Delta[$\ion{Fe}{1}/H$] \Big | $ \hspace{-0.17in} & $ \Big | \Delta[$\ion{Fe}{2}/H$] \Big |  $ \hspace{-0.17in} & $\Big | \Delta[$\ion{Ca}{1}/\ion{Fe}{1}$] \Big | $ \hspace{-0.17in} & $ \Big | \Delta[$\ion{Ti}{1}/\ion{Fe}{1}$] \Big | $ \hspace{-0.17in} & $ \Big | \Delta[$\ion{Ti}{2}/\ion{Fe}{2}$] \Big | $ \hspace{-0.17in} & $ \Big | \Delta[$\ion{Ni}{1}/\ion{Fe}{1}$] \Big | $\hspace{-0.17in}  & $ \Big | \Delta[$\ion{Ba}{2}/\ion{Fe}{2}$] \Big | $ \hspace{-0.17in} & $ \Big | \Delta[$\ion{Eu}{2}/\ion{Fe}{2}$] \Big | $ \hspace{-0.17in} & $\Big | \Delta [$\ion{Ba}{2}/\ion{Eu}{2}$] \Big |$\\
\hline
{\bf CMD-based} & & & & & & & &  & \\
{\bf analyses} & & & & & & & &  & \\
Minimum errors$^{a}$  & {\bf $\le$0.12} & {\bf $\le$0.20} & {\bf $\le$0.06} & {\bf $\le$0.09} & {\bf $\le$0.14} & $\le$0.04 & {\bf $\le$0.22} & {\bf $\le$0.11} & {\bf $\le$0.17} \\
CTRs$^{a,b}$          & {\bf $\le$0.07} & $\le$0.01 & $\le$0.02 & $\le$0.03 & $\le$0.03 & $\le$0.04 & {\bf $\le$0.06} & $\le$0.04 & $\le$0.04 \\
Input photometry      & $\le$0.04 & {\bf $\le$0.07} & $\le$0.02 & $\le$0.04 & $\le$0.04 & $\le$0.01 & {\bf $\le$0.07} & $\le$0.02 & {\bf $\le$0.06} \\
Incompleteness        & {\bf $\le$0.07} & {\bf $\le$0.07} & {\bf $\le$0.05} & {\bf $\le$0.06} & {\bf $\le$0.07} & {\bf $\le$0.04} & {\bf $\le$0.07} & $\le$0.03 & $\le$0.04 \\
Sampling$^{c,d}$        & {\bf $\le$0.22} & {\bf $\le$0.10} & {\bf $\le$0.09} & $-$   & {\bf $\le$0.10} & $\le$0.03 & {\bf $\le$0.21} & {\bf $\le$0.09} & {\bf $\le$0.14} \\
& & & & & & & &  & \\
{\bf HRD-based } & & & & & & & &  & \\
{\bf analyses} & & & & & & & &  & \\
HRD vs. CMD$^{a}$     & {\bf $\le$0.11} & {\bf $\le$0.19} & {\bf $\le$0.05} & {\bf $\le$0.20} & {\bf $\le$0.10} & $\le$0.04 & {\bf $\le$0.12} & {\bf $\le$0.08} & {\bf $\le$0.10} \\
Age/[Fe/H] Errors$^a$&{\bf $\le$0.16} & {\bf $\le$0.16} & {\bf $\le$0.07} & {\bf $\le$0.12} & {\bf $\le$0.10} & $\le$0.04 & {\bf $\le$0.19} & {\bf $\le$0.08} & {\bf $\le$0.10}\\
Diff. Isochrones  & $\le$0.02 & $\le$0.04 & $\le$0.02 & $\le$0.02 & $\le$0.01 & $ $0.0 & $\le$0.03 & $\le$0.01 & $\le$ 0.02\\
IMF                   & $\le$0.04 & $\le$0.01 & $\le$0.02 & {\bf $\le$0.06} & {\bf $\le$0.05} & $\le$0.02 & {\bf $\le$0.07} & {\bf $\le$0.05} & $\le$0.02\\
Cluster $M_V^{e}$     & {\bf $\le$0.36} & {\bf $\le$0.10} & {\bf $\le$0.10} & {\bf $\le$0.41} & {\bf $\le$0.14} & {\bf $\le$0.10} & {\bf $\le$0.33} & {\bf $\le$0.23} & {\bf $\le$0.10}\\
HB morphology$^{a,d}$ & {\bf $\le$0.13} & {\bf $\le$0.28} & $\le$0.04 & {\bf $\le$0.17} & {\bf $\le$0.08} & {\bf $\le$0.07} & {\bf $\le$0.14} & {\bf $\le$0.11} & {\bf $\le$0.12}\\
AGB prescription      & {\bf $\le$0.19} & {\bf $\le$0.15} & {\bf $\le$0.05} & {\bf $\le$0.23} & {\bf $\le$0.09} & {\bf $\le$0.13} & {\bf $\le$0.19} & {\bf $\le$0.14} & {\bf $\le$0.07}\\
Blue stragglers       & {\bf $\le$0.07} & {\bf $\le$0.07} & $\le$0.02       & $\le$0.04       & $\le$0.04       & $\le$0.03       & {\bf $\le$0.05} & {\bf $\le$0.06} & $\le$0.02\\
Low mass cut-off$^{a}$ & {\bf $\le$0.13} & {\bf $\le$0.12} & $\le$0.04 & {\bf $\le$0.24} & {\bf $\le$0.07} & {\bf $\le$0.05} & {\bf $\le$0.11} & {\bf $\le$0.12} & {\bf $\le$0.05}\\
& & & & & & & &  & \\
{\bf All analyses} & & & & & & & &  & \\
CMD/HRD Boxes         & $\le$0.02 & $\le$0.01 & $\le$0.02 & $\le$0.03 & $\le$0.04 & $\le$0.03 & $\le$0.01 & {\bf $\le$0.07} & $\le$0.04 \\
Microturbulence  & {\bf $\le$0.11} & {\bf $\le$0.05} & $\le$0.03 & $\le$0.02 & {\bf $\le$0.08} & {\bf $\le$0.10} & {\bf $\le$0.16} & $\le$0.04 & {\bf $\le$0.16}\\
LPVs                  & $\le$0.01 & {\bf $\le$0.07} & $ $0.0    & $\le$0.03 & $\le$0.01 & $\le$0.03 & $ $0.0 & $$0.0 & $$0.0\\
CH stars$^{a,d}$      & $ $0.0    &  $ $0.0         & $\le$0.01 & $$0.0     & $$0.0     & $\le$0.01 & $ $0.0 & $\le$0.01 & $\le$0.01\\
Hot stars             & {\bf $\le$0.06} & $\le$0.04 & $\le$0.01 & $\le$0.01 & {\bf $\le$0.07} & $\le$0.03 & $\le$0.04 & $\le$0.04 & {\bf $\le$0.08}\\
Field stars$^{d}$     & {\bf $\le$0.10} & {\bf $\le$0.09} & {\bf $\le$0.09} & $\le$0.04 & {\bf $\le$0.07} & {\bf $\le$0.06} & {\bf $\le$0.10} & {\bf $\le$0.05} & {\bf $\le$0.09}\\
ODFNEW Atms    & {\bf $\le$0.05} & {\bf $\le$0.12} & $\le$0.03 & $\le$0.02 & $\le$0.03 & $\le$0.02 & $\le$0.03 & $\le$0.03 & $\le$0.03\\
CN-cycled Atms & {\bf $\le$0.05} & {\bf $\le$0.12} & $\le$0.03 & $\le$0.02 & $\le$0.03 & $\le$0.02 & $\le$0.03 & $\le$0.03 & $\le$0.03\\
& & & & & & & &  & \\
\hline
\end{tabular}
\end{minipage}\\
\medskip
\raggedright {\bf Notes: } The errors shown are upper limits from all
tests on all clusters.  The uncertainties for an individual GC will
depend on the properties of the GC.\\
\raggedright $^{a}$ Metallicity/cluster dependent result.\\
\raggedright $^{b}$ The $V$, $I$ errors with the
\citet{Alonso1996,Alonso1999} relations are not considered here; see
the text.\\
\raggedright $^{c}$ These error estimates are specific to M15's wedge
shaped pointing pattern, and are likely to be much higher than would
be expected for any extragalactic targets.\\
\raggedright $^{d}$ Recall that these error estimates consider the
worst case scenario.\\
$^{e}$ These large uncertainties arise in faint GCs due to stochastic
sampling of the brightest stars and should be mitigated by considering
fractional numbers of stars.\\
\end{table*}

\subsubsection{[Fe/H]}\label{subsubsec:FeH}
In CMD-based analyses, the largest systematic uncertainties in
[\ion{Fe}{1}/H] and [\ion{Fe}{2}/H] are $\sim 0.1$ and $\sim 0.2$
dex, respectively,\footnote{The clear offset in the $V$, $I$
\citealt{Alonso1996,Alonso1999} colour-temperature relations has been
neglected.} for all GCs.  The potential HRD-based offsets are much
larger, up to $\sim 0.1-0.4$ dex (depending on the GC) for
[\ion{Fe}{1}/H] and $\sim 0.2-0.3$ dex for [\ion{Fe}{2}/H].  The
HRD-based offsets are lowest for 47~Tuc, suggesting that red HB,
metal-rich GCs may have smaller systematic offsets in HRD-based [Fe/H]
ratios.

The [\ion{Fe}{1}/H] ratio is particularly sensitive to:
\begin{enumerate}
\item Sampling of input photometry
\item The usage of isochrones instead of resolved photometry
\item Uncertainties in isochrone parameters
\item Models of the AGB
\item Microturbulence variations in the brightest stars
\item The inclusion of bright field stars.
\end{enumerate}
With well-sampled IL spectra and photometry that extends at least to
the HB, the uncertainties in the parameters of the brightest stars are
reduced, and the individual systematic offsets should be $\sim
0.1$~dex.

The [\ion{Fe}{2}/H] ratio is strongly affected by bright RGB stars,
AGB and HB stars, hot stars, and model atmosphere chemistries.  The
offsets tend to be larger for [\ion{Fe}{2}/H] than [\ion{Fe}{1}/H],
and even with partially resolved photometry, the systematic errors in
[\ion{Fe}{2}/H] remain $\sim 0.2$ dex.  This further confirms the
suggestion by \citet{Colucci2009} that forcing the [\ion{Fe}{1}/H] and
[\ion{Fe}{2}/H] solutions to be equal will not lead to more accurate
isochrone solutions.

\subsubsection{[Ca/Fe] and [Ti/Fe]}\label{subsubsec:CaFe}
The \textit{largest} [\ion{Ca}{1}/\ion{Fe}{1}] offsets are $\sim 0.1$
dex, and are due to
\begin{enumerate}
\item Sampling of the brightest stars
\item Uncertainties in isochrone parameters
\item Treatment of the AGB
\item The inclusion of bright field stars.
\end{enumerate}
These results indicate that [\ion{Ca}{1}/\ion{Fe}{1}] is most affected
by the numbers and properties of the bright RGB stars.  For the
wavelength regions examined here the [\ion{Ca}{1}/\ion{Fe}{1}] ratio
is largely insensitive to the properties of hot stars.  With partially
resolved GCs and well-sampled IL spectra, the systematic errors in
[\ion{Ca}{1}/\ion{Fe}{1}] should be reduced to $\la 0.1$ dex,
depending on GC metallicity.

The [\ion{Ti}{1}/\ion{Fe}{1}] and [\ion{Ti}{2}/\ion{Fe}{2}] ratios, on
the other hand, are very sensitive to uncertainties in the
underlying stellar population, with offsets of as much as $\sim 0.2$
dex.  Like calcium, [\ion{Ti}{1}/\ion{Fe}{1}] is sensitive to the
numbers and properties of the brightest RGB stars.  \ion{Ti}{1} is
particularly affected in HRD-based analyses: in the initial
comparisons with CMD-based abundances, [\ion{Ti}{1}/\ion{Fe}{1}] is
persistently lower than individual stellar values by $\sim 0.1-0.2$
dex (see Appendix \ref{subsec:IsochroneOffsets}).  When uncertainties
in HB morphology and AGB prescription are considered,
[\ion{Ti}{1}/\ion{Fe}{1}] could therefore be uncertain by as much as
$0.2-0.4$ dex when isochrones are used.  In most tests, the
[\ion{Ti}{2}/\ion{Fe}{2}] uncertainties are typically constrained only
to within 0.15 dex.

\subsubsection{[Ni/Fe]}\label{subsubsec:NiFe}
Since Ni is an iron-peak element with an atomic structure similar to
Fe, it is not surprising that the [\ion{Ni}{1}/\ion{Fe}{1}] is
relatively stable to uncertainties in atmospheric parameters---for all
tests and GCs, the \textit{highest} systematic uncertainties in
[\ion{Ni}{1}/\ion{Fe}{1}] are only $\sim 0.1$ dex. Nickel appears to
be sensitive to both high and low mass stars, given that it is most
affected by:
\begin{enumerate}
\item Sampling when the total cluster magnitude is adjusted
\item AGB prescription
\item The HRD low mass cut-off
\item The presence of field stars.
\end{enumerate}
\noindent Despite these sensitivities, however, in general
[\ion{Ni}{1}/\ion{Fe}{1}] is quite robust in both CMD- and HRD-based
analyses.  With a well-modelled stellar population, the systematic
errors in [\ion{Ni}{1}/\ion{Fe}{1}] approach $\sim 0.05$ dex.

\subsubsection{[Ba/Fe] and [Eu/Fe]}\label{subsubsec:BaFe}
Both [\ion{Ba}{2}/\ion{Fe}{2}] and [\ion{Eu}{2}/\ion{Fe}{2}] are
particularly sensitive to uncertainties in the underlying population,
often in similar ways.  In CMD-based analyses,
[\ion{Ba}{2}/\ion{Fe}{2}] and [\ion{Eu}{2}/\ion{Fe}{2}] can be
constrained to $\sim 0.2$ and $\sim 0.1$ dex, respectively, for all
GCs.  The offsets are higher in HRD-based analyses (up to $\sim 0.3$
and $\sim 0.2$, respectively).

Both \ion{Ba}{2} and \ion{Eu}{2} are sensitive to uncertainties in the
brightest RGB stars and red/intermediate HB stars.  The strongest
effects are caused by:
\begin{enumerate}
\item Temperature and microturbulence uncertainties, including
various microturbulence relations
\item Sampling of the brightest stars, whether from uncertain input
photometry or from rounding errors in faint clusters
\item Isochrone age
\item AGB prescription.
\end{enumerate}
\noindent In all clusters, the \ion{Ba}{2} and \ion{Eu}{2} abundances
are insensitive to completeness of the lower main sequence, isochrone
offsets, atmospheric [$\alpha$/Fe], and properties of the blue HB
stars.

When Ba and Eu are affected in similar ways, the uncertainties in
[Ba/Eu] can be smaller than the individual uncertainties in
Ba and Eu.  This is true for, e.g., uncertainties in the AGB
prescription, the total cluster magnitude, and the lower mass cut-off.
Thus, in an HRD analysis [Ba/Eu] may have lower systematic errors than
the individual [\ion{Ba}{2}/\ion{Fe}{2}] and [\ion{Eu}{2}/\ion{Fe}{2}]
ratios.

\subsection{High Resolution vs. Lower Resolution Analyses}\label{subsec:HRvsLR}
High resolution ($R\ga 20,000$) IL spectral analyses provide two major
advantages over lower resolution studies:
\begin{enumerate}
\item More lines can be detected and resolved in a high resolution
spectrum. With more independent measurements, the random errors in
individual elemental abundances can be reduced.
\item Weaker features can be detected in high resolution IL spectra,
enabling abundances to be obtained for more elements.
\end{enumerate}
\noindent Despite these advantages, it requires more observing time to
obtain high resolution IL spectra of a sufficient S/N.  This paper has shown
that despite the increased precision offered by high resolution IL
spectroscopy, the low accuracy in integrated abundances may render
such sharp resolution unnecessary, depending on the science goals.

The cluster metallicity, [Fe/H], is an excellent example for when high
resolution may be unnecessary.  Although high resolution IL
spectroscopy can reduce random errors in [\ion{Fe}{1}/H] to
$\sim~0.02$ dex (depending on the S/N), the systematic errors can be
as large as $\sim 0.1-0.4$ dex depending on the analysis type,
cluster metallicity, etc.  Thus, for studies that focus only on [Fe/H]
(such as studies of population averages, bimodalities, or gradients,
e.g. \citealt{Caldwell2011}) the increased precision of high
resolution offers no benefit.  Similarly, for studies of large samples
of GCs where the abundances are averaged together, high resolution
provides no clear advantage (for example average values in certain
galaxy types or abundance correlations with GC properties, e.g.
\citealt{Puzia2008,Schiavon2013}).

The strength of high resolution IL spectroscopy is its ability to
provide accurate abundances for individual clusters.  High resolution
is therefore essential for examining the detailed chemical abundances
of GCs, e.g. for chemical tagging studies.

\subsection{Optimal Abundance Ratios for Chemical Tagging}\label{subsec:ChemicalTagging}
Based on the offsets presented in Table \ref{table:Summary} and the
discussion in Section \ref{subsec:Summary}, certain element ratios are
more useful for chemical tagging purposes.

\begin{description}
\item{\bf [Fe/H]: } Most chemical comparisons require knowledge of the
GC metallicity, [Fe/H].  Though the [\ion{Fe}{1}/H] ratio can
occasionally have large systematic errors, [\ion{Fe}{2}/H]
consistently also has large offsets, as well as larger statistical
errors (because there are fewer \ion{Fe}{2} lines).  Therefore, in
most cases, [\ion{Fe}{1}/H] will be the preferable choice to represent
the cluster metallicity.\\

\item{\bf [Ca/Fe]: } The [$\alpha$/Fe] ratio is particularly
useful for chemical tagging of dwarf galaxy stars and GCs (see, e.g.,
\citealt{Venn2004,Pritzl2005}), where Ca and Ti have both been used as
$\alpha$-indicators in individual stellar analyses.  (Though note that
the behaviour of Ca and Ti can be very different from other
$\alpha$-elements like O and Mg.) Given that [\ion{Ca}{1}/\ion{Fe}{1}]
is very stable to uncertainties in the underlying stellar population
for all GCs considered here, [\ion{Ca}{1}/\ion{Fe}{1}] is preferable
to [\ion{Ti}{1}/\ion{Fe}{1}] or [\ion{Ti}{2}/\ion{Fe}{2}] for probing
the [$\alpha$/Fe] ratios of extragalactic systems.\\

\item{\bf [Ni/Fe]: } The [Ni/Fe] ratio may be useful for identifying
chemically peculiar GCs.  In particular, Pal~12 and Ter~7, the two
metal-rich GCs that were accreted from the Sagittarius dwarf galaxy,
are underabundant in [Ni/Fe] like the Sgr field stars
\citep{Cohen2004,Sbordone2005,Sbordone2007}.  The integrated
[\ion{Ni}{1}/\ion{Fe}{1}] ratios are generally quite stable to
abundance uncertainties, and may therefore prove useful for
integrated chemical tagging.\\

\item{\bf [Ba/Fe], [Eu/Fe], [Ba/Eu]: }  Ba and Eu both form through
neutron captures onto iron-peak atoms.  In the Sun, 97\% of Eu forms
from rapid neutron captures (the r-process, i.e. where the neutron
flux is so high that the nucleus does not have time to decay between
captures) while 85\% of Ba forms from slow neutron captures (the
s-process; \citealt{Burris2000}).  The nucleosynthetic sites for the
two elements (and Fe) therefore differ, and the [Ba/Fe], [Eu/Fe], and
[Ba/Eu] ratios differ between stars in the Milky Way and those in
dwarf galaxies (see, e.g., \citealt{Venn2012}).  Though the systematic
uncertainties in [\ion{Ba}{2}/\ion{Fe}{2}], [\ion{Eu}{2}/\ion{Fe}{2}],
and [\ion{Ba}{2}/\ion{Eu}{2}] are quite large for all GCs, taken
together the three ratios could still prove useful for chemical
tagging since all three ratios are unlikely to have simultaneously
large offsets.
\end{description}

\subsection{CMD- vs. HRD-based Analyses}\label{subsec:CMDvsHRD}
Appendix \ref{subsec:IsochroneOffsets} indicates that systematic
offsets may occur between CMD-based and HRD-based analyses, since the
best-fitting HRD-based abundances are not always in agreement with
those from a CMD-based analysis (see Table \ref{table:Isochrones}),
with differences up to 0.2 dex.  These differences can be larger than
the uncertainties from identifying the best isochrone (i.e. the
offsets in Table \ref{table:Isochrones} are sometimes larger than the
uncertainties in Table \ref{table:IsochroneErrs}).  These offsets are
likely due to discrepancies between the input isochrone and the true
stellar populations---for example, changing the HB morphology can
bring [\ion{Ti}{1}/\ion{Fe}{1}] back into agreement with the CMD-based
ratio.  However, the necessary alterations to the input isochrones may
not be identifiable for unresolved GCs, particularly if the IL spectra
are noisy.\footnote{Noisy IL spectra will lead to a larger dispersion
in line-to-line \ion{Fe}{1} abundances.  A larger dispersion will then
complicate the process of minimizing trends with wavelength, REW, and
EP.}

Table \ref{table:Summary} indicates that if sampling problems are
reduced or eliminated then CMD-based chemical abundances are more
accurate than HRD-based abundances.  This result is driven by the
uncertainties in modelling the most evolved stars, notably the tip of
the RGB, HB, and AGB stars.  However, this approach is not currently
feasible for extragalactic targets, for which IL methods are
necessary.

Appendix \ref{subsec:PartiallyResolved} demonstrates that some of the
HRD-based offsets disappear when CMDs of the brightest stars are
combined with isochrones.  This is  important for IL analyses of
nearby extragalactic GC systems, e.g. GCs in M31
\citep{Mackey2007,Mackey2013}, particularly if those GCs have blue or
intermediate HBs.  Thus, if accurate and uncontaminated CMDs can be
obtained for the brightest stars in a GC, the systematic errors in
integrated abundances can be reduced.

\subsection{A Case Study: Partially Resolved Clusters in M31}\label{subsec:M31GCs}
To illustrate how the results of this paper can be applied to IL
studies, Table \ref{table:SummaryM31} summarizes the systematic errors
for the target clusters {\it if spectra of this quality were obtained
from GCs in M31} and if those GCs had partially resolved {\it HST}
photometry to constrain the age, [Fe/H], HB morphology, AGB
prescription, total observed magnitude, and the presence of severely
different interloping field stars.  The ideal science case would be to
perform a chemical tagging analysis on these clusters.  Table
\ref{table:SummaryM31} therefore only shows the systematic
uncertainties in the optimal abundance ratios for chemical tagging:
[\ion{Fe}{1}/H], [\ion{Ca}{1}/\ion{Fe}{1}], [\ion{Ni}{1}/\ion{Fe}{1}],
[\ion{Ba}{2}/\ion{Fe}{2}], [\ion{Eu}{2}/\ion{Fe}{2}], and
[\ion{Ba}{2}/\ion{Eu}{2}].  It is not clear how to combine these
errors in a meaningful way---however, if the errors are assumed to be
independent then they can be conservatively added together in
quadrature (though this may overestimate the errors).  These total
systematic errors are also shown in Table \ref{table:SummaryM31}.

\begin{table*}
\centering
\begin{minipage}{165mm}
\begin{center}
\caption{Summary of errors for partially resolved clusters at the
distance of M31.\label{table:SummaryM31}}
  \begin{tabular}{@{}lcccccc@{}}
  \hline
& $ \Big | \Delta[$\ion{Fe}{1}/H$] \Big | $ & $\Big | \Delta[$\ion{Ca}{1}/\ion{Fe}{1}$] \Big | $ & $ \Big | \Delta[$\ion{Ni}{1}/\ion{Fe}{1}$] \Big | $ & $ \Big | \Delta[$\ion{Ba}{2}/\ion{Fe}{2}$] \Big | $ & $ \Big | \Delta[$\ion{Eu}{2}/\ion{Fe}{2}$] \Big | $ &  $ \Big | \Delta[$\ion{Ba}{2}/\ion{Eu}{2}$] \Big |$\\
\hline
{\bf 47 Tuc} & & & & & & \\
Partially Resolved Errors     & 0.09 & 0.06 & 0.03 & 0.04 & 0.08 & 0.04 \\
Different Isochrones          & 0.03 & 0.01 & 0.01 & 0.03 & 0.02 & 0.02 \\
IMF                           & 0.04 & 0.02 & 0.02 & 0.07 & 0.05 & 0.02\\
Blue Stragglers               & 0.02 & 0.0  & 0.01 & 0.01 & 0.01 & 0.0\\
Low Mass Cut-Off               & 0.05 & 0.04 & 0.05 & 0.09 & 0.12 & 0.03\\
Microturbulence Relations     & 0.07 & 0.01 & 0.0  & 0.0  & 0.04 & 0.04\\
LPVs                          & 0.01 & 0.0  & 0.03 & 0.0  & 0.0  & 0.0 \\
Field Stars                   & 0.09 & 0.09 & 0.06 & 0.10 & 0.05 & 0.05\\
$\alpha$-enhanced atmospheres & 0.05 & 0.03 & 0.01 & 0.01 & 0.02 & 0.01\\
{\it Total}$^{a}$                   & 0.17 & 0.12 & 0.09 & 0.16 & 0.17 & 0.09\\
 & & & & & & \\
{\bf M3} & & & & & & \\
Partially Resolved Errors     & 0.04 & 0.04 & 0.06 & 0.03 & 0.01 & 0.02\\
Different Isochrones          & 0.05 & 0.02 & 0.03 & 0.06 & 0.03 & 0.03\\
IMF                           & 0.02 & 0.0  & 0.01 & 0.02 & 0.02 & 0.0 \\
Blue Stragglers               & 0.02 & 0.0  & 0.01 & 0.03 & 0.06 & 0.03\\
Low Mass Cut-Off               & 0.10 & 0.02 & 0.04 & 0.11 & 0.10 & 0.01\\
Microturbulence Relations     & 0.04 & 0.01 & 0.0  & 0.01 & 0.0  & 0.01\\
Field Stars                   & 0.04 & 0.0  & 0.01 & 0.03 & 0.04 & 0.01\\
$\alpha$-enhanced atmospheres & 0.0  & 0.01 & 0.02 & 0.02 & 0.01 & 0.01\\
{\it Total}$^{a}$                   & 0.13 & 0.05 & 0.08 & 0.14 & 0.13 & 0.05\\
 & & & & & & \\
{\bf M13} & & & & & & \\
Partially Resolved Errors     & 0.05 & 0.06 & 0.05 & 0.07 & 0.02 & 0.05\\
Different Isochrones          & 0.05 & 0.02 & 0.03 & 0.06 & 0.03 & 0.03\\
IMF                           & 0.02 & 0.0  & 0.01 & 0.02 & 0.02 & 0.0 \\
Blue Stragglers               & 0.03 & 0.02 & 0.02 & 0.03 & 0.03 & 0.0\\
Low Mass Cut-Off               & 0.10 & 0.01 & 0.01 & 0.07 & 0.08 & 0.01\\
Microturbulence Relations     & 0.04 & 0.01 & 0.0  & 0.01 & 0.0  & 0.01\\
Hot Stars                     & 0.06 & 0.01 & 0.03 & 0.04 & 0.04 & 0.08\\
Field Stars                   & 0.04 & 0.0  & 0.01 & 0.03 & 0.04 & 0.01\\
$\alpha$-enhanced atmospheres & 0.0  & 0.01 & 0.02 & 0.02 & 0.01 & 0.01\\
{\it Total}$^{a}$                   & 0.15 & 0.07 & 0.07 & 0.13 & 0.11 & 0.06\\
 & & & & & & \\
{\bf NGC~7006} & & & & & & \\
Partially Resolved Errors     & 0.04 & 0.03 & 0.08 & 0.04 & 0.07 & 0.11\\
Different Isochrones          & 0.05 & 0.02 & 0.03 & 0.06 & 0.03 & 0.03\\
IMF                           & 0.02 & 0.0  & 0.01 & 0.02 & 0.02 & 0.0 \\
Blue Stragglers               & 0.02 & 0.01 & 0.01 & 0.02 & 0.0  & 0.02\\
Low Mass Cut-Off               & 0.09 & 0.0  & 0.05 & 0.08 & 0.09 & 0.01\\
Microturbulence Relations     & 0.04 & 0.01 & 0.0  & 0.01 & 0.0  & 0.01\\
Field Stars                   & 0.04 & 0.0  & 0.01 & 0.03 & 0.04 & 0.01\\
$\alpha$-enhanced atmospheres & 0.0  & 0.0  & 0.01 & 0.03 & 0.03 & 0.0\\
{\it Total}$^{a}$                   & 0.13 & 0.04 & 0.10 & 0.12 & 0.13 & 0.12\\
 & & & & & & \\
{\bf M15} & & & & & & \\
Partially Resolved Errors     & 0.0  & 0.01 & 0.03 & 0.02 & 0.01 & 0.03\\
Different Isochrones          & 0.05 & 0.02 & 0.02 & 0.05 & 0.04 & 0.02\\
IMF                           & 0.03 & 0.01 & 0.02 & 0.02 & 0.02 & 0.02\\
Blue Stragglers               & 0.07 & 0.02 & 0.03 & 0.05 & 0.01 & 0.04\\
Low Mass Cut-Off               & 0.13 & 0.04 & 0.01 & 0.11 & 0.05 & 0.06\\
Microturbulence Relations     & 0.05 & 0.01 & 0.03 & 0.02 & 0.03 & 0.0 \\
Hot Stars                     & 0.06 & 0.01 & 0.03 & 0.04 & 0.04 & 0.08\\
Field Stars                   & 0.10 & 0.04 & 0.03 & 0.07 & 0.04 & 0.03\\
$\alpha$-enhanced atmospheres & 0.02 & 0.0  & 0.01 & 0.0  & 0.03 & 0.03\\
{\it Total}$^{a}$                   & 0.20 & 0.07 & 0.07 & 0.16 & 0.10 & 0.12\\
 & & & & & & \\
\hline
\end{tabular}
\end{center}
\end{minipage}\\
\medskip
\raggedright $^{a}$ Total errors are conservatively estimated by adding the other errors in quadrature.\\
\end{table*}

\begin{figure*}
\begin{center}
\centering
\subfigure{\includegraphics[scale=0.7]{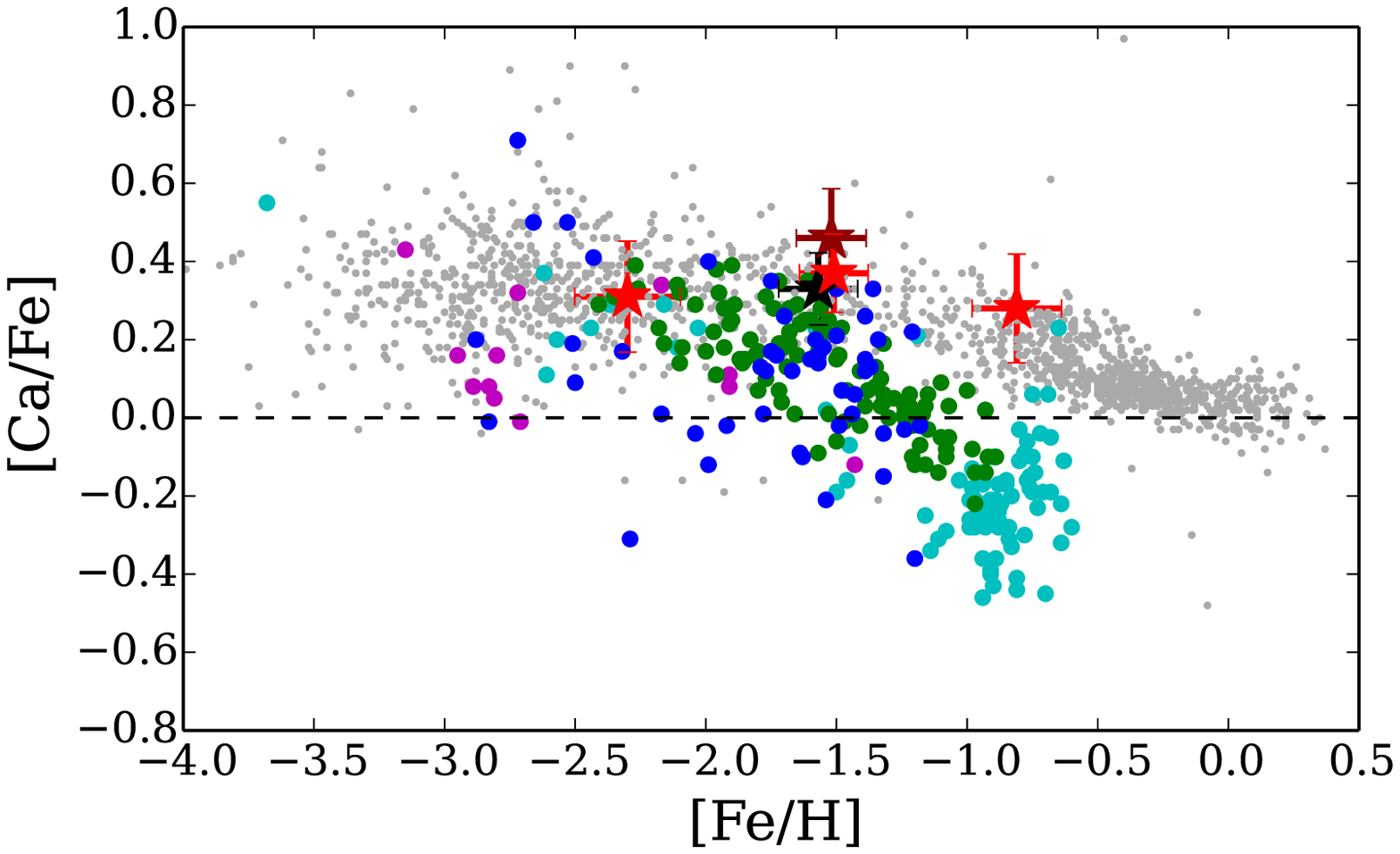}\label{fig:CaFe}}
\subfigure{\includegraphics[scale=0.7]{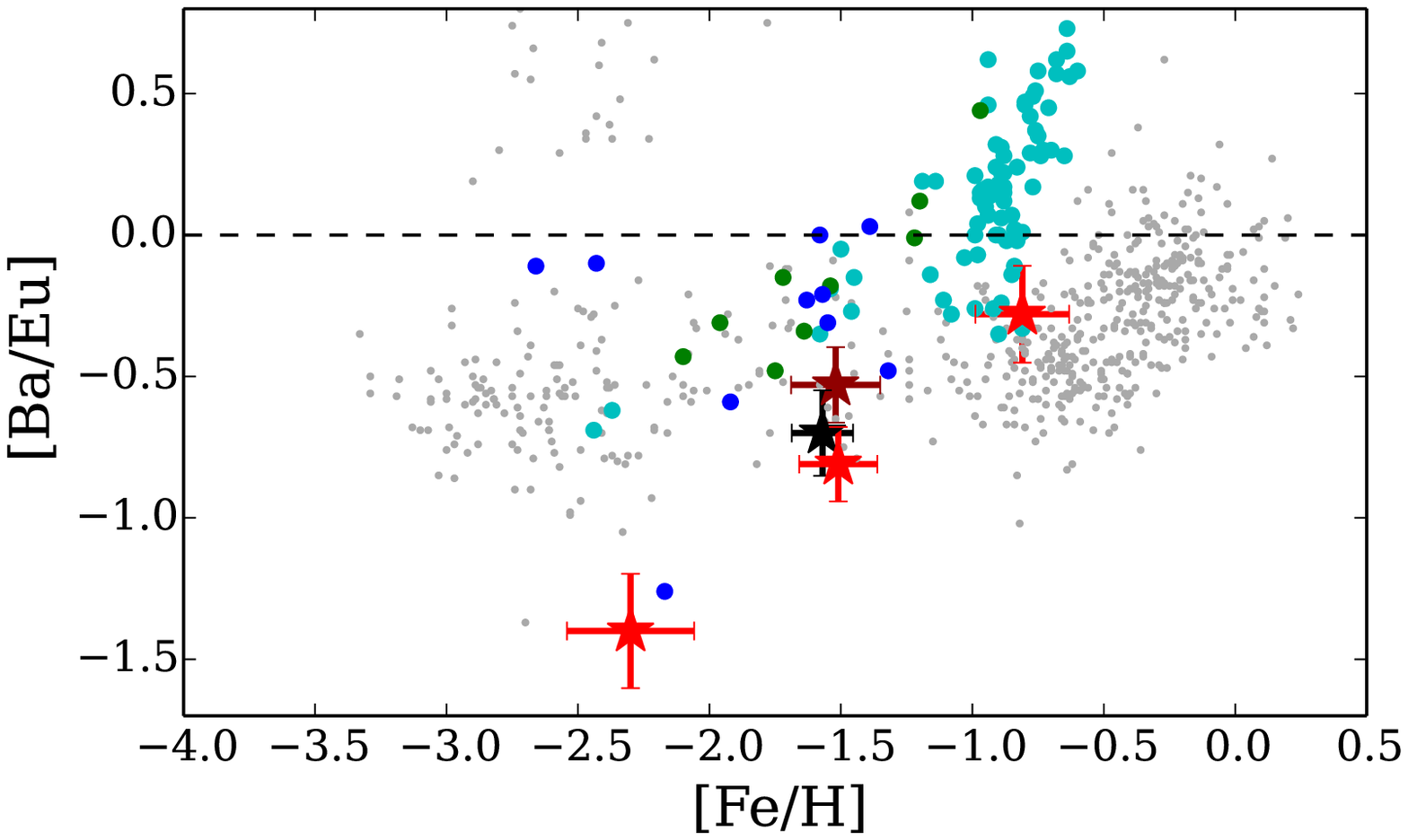}\label{fig:BaEuFe}}
\caption{Chemical comparisons between Galactic and dwarf galaxy
stars, illustrating that chemical tagging is still possible even when
systematic errors are taken into account.  The integrated abundances
from the target GCs are shown with red stars for 47~Tuc, M3, and M15,
with a black star for M13, and with a maroon star for NGC~7006.  The
error bars are the systematic {\it and} random errors (added in
quadrature), {\it assuming the GCs were partially resolved at the
distance of M31} (see Table \ref{table:SummaryM31}).  The grey points
are Milky Way stars.  The cyan, green, blue, and magenta points are
Fornax, Sculptor, Carina, and Sextans stars, respectively.  All points
are from the compilation assembled by \citet{Venn2012}.  This
comparison shows that, even including the systematic errors,
individual GCs can be chemically tagged based on their integrated
abundances, provided that their abundances are distinct from the Milky
Way stars.\label{fig:M31GCs}}
\end{center}
\end{figure*}

As an additional illustration of these errors, the cluster [Ca/Fe] and
[Ba/Eu] abundances from Table \ref{table:OriginalAbunds} are compared
to MW and dwarf galaxy abundances in Figure \ref{fig:M31GCs}, {\it
using the partially resolved systematic errors from Table
\ref{table:SummaryM31}}.  The error bars show the total random {\it
and} systematic errors, combined in quadrature.  Note that though
47~Tuc has a high [Ba/Eu], it is still consistent with the MW field
stars.  M15's low [Ba/Eu] ratio is likely due to the star-to-star
chemical variations within the massive cluster (see Paper I).  With
the systematic errors included these Galactic targets would appear
consistent with the Galactic field stars, even if they were located at
M31's distance.  Similarly, GCs associated with dwarf galaxies could
be distinguished, even with systematic errors considered, if they are
$\alpha$-deficient and/or [Ba/Eu]-enhanced.

\section{Conclusions}\label{sec:Conclusion}
This paper presents a detailed investigation of the systematic
uncertainties in high-resolution integrated abundance analyses that
occur when GC stellar populations are modelled.  High resolution HET
and LCO IL spectra (covering $\sim 5320-7340$ \AA) of the Galactic GCs
47~Tuc, M3, M13, NGC~7006, and M15 were combined with \textit{HST}
photometry and theoretical isochrones to investigate abundance
accuracies over a wide range in metallicity and HB morphology.  The
stability of Fe, Ca, Ti, Ni, Ba, and Eu abundances is determined
through IL analyses with various alterations to the underlying stellar
population.

The tests in this paper show that:
\begin{enumerate}
\item The accuracy in integrated abundances can approach that of
individual stellar analyses \textit{if the stellar population is
well-modelled.}  The minimum systematic errors in the abundance ratios
are $\la 0.05$ dex in [\ion{Ca}{1}/\ion{Fe}{1}] and
[\ion{Ni}{1}/\ion{Fe}{1}]; $\la 0.1$ dex in [\ion{Fe}{2}/H],
[\ion{Ti}{1}/\ion{Fe}{1}], [\ion{Ti}{2}/\ion{Fe}{2}],  and
[\ion{Eu}{2}/\ion{Fe}{2}]; and $\la 0.2$ dex in [\ion{Fe}{1}/H] and
[\ion{Ba}{2}/\ion{Fe}{2}].
\item CMD-based analyses are most sensitive to inaccuracies in the
input photometry, especially sampling of the brightest stars and
incompleteness in the low mass stars.  In the worst case scenario, the
accuracy in integrated CMD-based abundances is $\la 0.1$ dex in
[\ion{Fe}{2}/H], [\ion{Ca}{1}/\ion{Fe}{1}], [\ion{Ti}{1}/\ion{Fe}{1}],
[\ion{Ti}{2}/\ion{Fe}{2}], [\ion{Ni}{1}/\ion{Fe}{1}], and
[\ion{Eu}{2}/\ion{Fe}{2}], and $\la 0.2$ dex in [\ion{Fe}{1}/H] and
[\ion{Ba}{2}/\ion{Fe}{2}].  It is therefore crucial to select input
photometry that matches the regions scanned by IL spectra.
\item HRD-based analyses are highly sensitive to sampling of the
highest and lowest mass stars, AGB prescription, and HB morphology.
The uncertainties can be as high as $\la 0.1$ dex in
[\ion{Ca}{1}/\ion{Fe}{1}], [\ion{Ti}{2}/\ion{Fe}{2}], and
[\ion{Ni}{1}/\ion{Fe}{1}]; $\la 0.2$ dex in [\ion{Eu}{2}/\ion{Fe}{2}];
$\la 0.3$ dex in [\ion{Fe}{2}/H] and [\ion{Ba}{2}/\ion{Fe}{2}]; and
$\la 0.4$ dex in [\ion{Fe}{1}/H] and [\ion{Ti}{1}/\ion{Fe}{1}].
\end{enumerate}

These results have several important implications for IL analyses of
extragalactic GCs in distant systems, for both analysis methods.
\begin{enumerate}
\item Certain abundance ratios are less sensitive to systematic
uncertainties and are therefore more useful for chemical tagging
studies.
 	\begin{description}
	\item{$\bullet$} The [\ion{Fe}{1}/H] ratio should serve as the best
[Fe/H] indicator.
	\item{$\bullet$} The [\ion{Ca}{1}/\ion{Fe}{1}] ratio is an excellent
[$\alpha$/Fe] indicator.
	\item{$\bullet$} The [\ion{Ni}{1}/\ion{Fe}{1}] ratios are very
stable to uncertainties
 	\item{$\bullet$} Individually, [\ion{Ba}{2}/\ion{Fe}{2}],
[\ion{Eu}{2}/\ion{Fe}{2}], and [\ion{Ba}{2}/\ion{Eu}{2}] have large
systematic uncertainties.  Together, however, the three ratios may
prove useful for chemical tagging.
	\end{description}
\item HRD-based abundances may be systematically offset from CMD-based
abundances, making comparisons between studies/clusters difficult.
\item CMDs of only the brightest stars in a GC can be used
to constrain properties of evolved stars, providing more accurate
chemical abundance ratios in GCs with blue or intermediate HBs.
\item In an HRD-based analysis, high resolution does not provide an
advantage for certain abundance ratios, such as [Fe/H].  Lower
resolution ($R \la 6500$) IL spectroscopy appears to be sufficient for
[Fe/H] determinations, investigations of [Fe/H] distributions, and
studies with large sample sizes. 
\end{enumerate}

\section*{Acknowledgements}
The authors thank R. Bernstein and A. McWilliam for the use of their
47~Tuc spectrum and their interest in this work.  The authors also
thank the anonymous referee for suggestions that improved this
manuscript.  CMS acknowledges funding from the Natural Sciences \&
Engineering Research Council (NSERC), Canada, via the Vanier CGS
program.  KAV acknowledges funding through the NSERC Discovery Grants
program.  AD is supported by the Australian Research Council (grant
FL110100012).  The Hobby-Eberly Telescope (HET) is a joint project of
the University of Texas at Austin, the Pennsylvania State University,
Stanford University, Ludwig-Maximilians-Universit\"{a}t M\"{u}nchen,
and Georg-August-Universit\"{a}t G\"{o}ttingen. The HET is named in
honour of its principal benefactors, William P. Hobby and Robert
E. Eberly. The authors wish to thank the night operations staff of the
HET for their assistance and expertise with these unusual
observations.  This work has made use of BaSTI web tools.

\appendix

\section{Systematic Offsets that Occur in a CMD-based Analysis}\label{sec:CMD}
The stars in nearby clusters can be fully resolved (like the Galactic
GCs studied here) or partially resolved (such as the GCs surrounding
M31; see \citealt{Mackey2007}). If the colours and magnitudes of the
brightest stars are known, they can be directly converted to
temperatures and surface gravities, avoiding the problems associated
with attempting to model the underlying stellar population.

The main advantage of a CMD-based analysis is that the basic
properties (i.e. age and [Fe/H]) may be estimated from the CMD.
The distribution of stars in the CMD is also known, removing the need
to model difficult subpopulations (e.g. the HB or the AGB) or the
relative numbers of dwarfs and giants.  The main disadvantages of a
CMD-based analysis are that observable properties must be converted to
intrinsic, physical quantities and that the CMDs do not completely
represent the observed regions. Errors in observed quantities can lead
to cluster-to-cluster systematic errors, while differences in the
employed conversion techniques/relations can lead to systematic
offsets between studies. Discrepancies between photometric and
spectroscopic observations (e.g. sampled regions or incompleteness)
can also lead to systematic uncertainties, as can poor resolution in
cluster cores.

This appendix investigates the systematic errors in integrated
abundances that occur only when using a CMD.  Two sources of error are
considered:

\begin{enumerate}
\item Errors that occur when observable quantities are converted to
physical values (Appendices \ref{subsec:Intrinsic},
\ref{subsec:ColorTemp}, and \ref{subsec:Photometry})
\item Errors that occur when the input photometry does not exactly
match the population observed in the IL spectra (Appendices
\ref{subsec:Photometry}, \ref{subsec:Incompleteness},
\ref{subsec:MismatchingPhotometry}; also see Appendix \ref{sec:Both}).
\end{enumerate}

\subsection{Minimum Errors in Photometric Parameters}\label{subsec:Intrinsic}
Conversions to photometric stellar parameters require estimates of a
cluster's distance modulus, reddening, turnoff mass, etc., all of
which have associated uncertainties that lead to unavoidable minimum
uncertainties in the photometric effective temperature,
$T_{\rm{eff}}$, and surface gravity, $\log g$.  Detailed abundance
analyses with individual stars also show that the
spectroscopically-determined microturbulence, $\xi$, and metallicity,
[Fe/H], cannot be perfectly constrained.  These errors in the
atmospheric parameters are typically on the order of $\Delta
T_{\rm{eff}}~=~\pm 100$~K, $\Delta \log g~=~\pm 0.2$~dex, $\Delta
\xi~=~\pm 0.2$~km s$^{-1}$, and $\Delta [M/H]~=~\pm 0.1$~dex
\textit{regardless of the methods used to determine these parameters}
(see, e.g., \citealt{Sakari2011}).   These abundance differences are
therefore good estimates of the minimum systematic errors that would
occur in a CMD-based IL abundance analysis.

These minimum changes to the atmospheric parameters lead to the
abundances shown in Table \ref{table:Intrinsic}.  Significant errors
($\ge 0.05$ dex) are in bold.  Note that the surface gravity and
microturbulence were changed independently from each other, even
though the microturbulence is determined through an empirical
relationship with the surface gravity.\footnote{The effects of the
microturbulence relation are investigated in Appendix
\ref{subsec:MicroturbulenceLaw}.}  With the empirical relation, a
change in the surface gravity of $\Delta \log g = 0.2$ dex would only
lead to a $\Delta \xi = 0.04$ km/s.

The abundance differences in Table \ref{table:Intrinsic} indicate
that:
\begin{enumerate}
\item The largest differences in [Fe/H] and [X/Fe] are $\sim 0.1$
dex. 
\item The model atmosphere metallicity has a negligible effect on
all abundance ratios.
\item The differences in the [\ion{Fe}{1}/H] and [\ion{Fe}{2}/H]
ratios are generally $<0.1$ dex, except in 47~Tuc and M15, where
offsets are $\sim 0.1$ dex.
\item The [\ion{Fe}{1}/H] and [\ion{Fe}{2}/H] ratios can be
significantly (i.e. $|\Delta [\rm{Fe/H}]| > 0.05$) affected by the changes
in temperature, surface gravity, and microturbulence.
\item The relative [\ion{Ca}{1}/\ion{Fe}{1}] and
[\ion{Ni}{1}/\ion{Fe}{1}] ratios are largely unaffected by these
errors in the atmospheric parameters.
\item The [\ion{Ti}{1}/\ion{Fe}{1}] ratio is moderately affected by
temperature, while [\ion{Ti}{2}/\ion{Fe}{2}] is affected primarily by
microturbulence (though the [Ti/Fe] errors are all $<0.1$ dex); the
surface gravity effects are negligible.
\item The [\ion{Ba}{2}/\ion{Fe}{2}] and [\ion{Eu}{2}/\ion{Fe}{2}]
ratios are most affected by temperature and microturbulence, though
[\ion{Ba}{2}/\ion{Fe}{2}] is constrained to within 0.12 dex, while
[\ion{Eu}{2}/\ion{Fe}{2}] is within 0.07 dex.
\end{enumerate}
Thus, the systematic errors from the intrinsic uncertainties in
atmospheric parameters are $\la 0.1$ dex for [\ion{Fe}{1}/H],
[\ion{Fe}{2}/H], [\ion{Ti}{1}/\ion{Fe}{1}], [\ion{Ti}{2}/\ion{Fe}{2}],
[\ion{Ba}{2}/\ion{Fe}{2}], and [\ion{Eu}{2}/\ion{Fe}{2}], and are
$<0.05$ dex for [\ion{Ca}{1}/\ion{Fe}{1}] and
[\ion{Ni}{1}/\ion{Fe}{1}].

\begin{table*}
\centering
\begin{minipage}{165mm}
\caption{The offsets in the CMD-based abundances due to uncertainties
in the atmospheric parameters.\label{table:Intrinsic}}
  \begin{tabular}{@{}lccccccccc@{}}
  \hline
& & $\Delta[$\ion{Fe}{1}/H$]$ & $\Delta[$\ion{Fe}{2}/H$] $ & $\Delta[$\ion{Ca}{1}/\ion{Fe}{1}$]$ & $\Delta[$\ion{Ti}{1}/\ion{Fe}{1}$]$ & $\Delta[$\ion{Ti}{2}/\ion{Fe}{2}$]$ & $\Delta[$\ion{Ni}{1}/\ion{Fe}{1}$]$ & $\Delta[$\ion{Ba}{2}/\ion{Fe}{2}$]$ & $\Delta[$\ion{Eu}{2}/\ion{Fe}{2}$]$\\
\hline
47~Tuc & $+\Delta T$       & $+$0.04 & {\bf $-$0.10} & $+$0.04 & {\bf $+$0.07} & {\bf $+$0.07} & $ $ 0.0 & {\bf $+$0.10} & {\bf $+$0.07} \\
       & $-\Delta T$       & $-$0.03 & {\bf $+$0.11} & $-$0.04 & {\bf $-$0.08} & {\bf $-$0.07} & $ $ 0.0 & {\bf $-$0.11} & $-$0.06 \\
       &$+\Delta\log g$    & $+$0.01 & {\bf $+$0.10} & $-$0.02 & $ $0.0  & $-$0.02 & $+$0.03 & $-$0.03 & $-$0.02\\
       &$-\Delta\log g$    & $-$0.02 & {\bf $-$0.11} & $+$0.03 & $+$0.02 & $+$0.03 & $-$0.02 & $+$0.04 & $+$0.03\\
       &$+\Delta \xi$      & {\bf $-$0.09} & {\bf $-$0.05} & $-$0.01 & $+$0.02 & {\bf $-$0.07} & $+$0.01 & {\bf $-$0.11} & $+$0.02\\
       &$-\Delta \xi$      & {\bf $+$0.08} & {\bf $+$0.06} & $+$0.03 & $ $0.0  & {\bf $+$0.07} & $+$0.01 & {\bf $+$0.11} & $-$0.04\\
       &$+\Delta[\rm{M/H}]$& $ $0.0  & $+$0.03 & $ $0.0  & $ $0.0  & $-$0.01 & $+$0.01 & $ $0.0  & 0.0\\
       &$-\Delta[\rm{M/H}]$& $-$0.02 & $-$0.04 & $+$0.03 & $+$0.02 & $+$0.01 & $ $0.0  & $+$0.01 & 0.0\\
 & & & & & & & & \\
M3     & $+\Delta T$       & {\bf $+$0.08} & {\bf $-$0.05} & $ $0.0  & {\bf $+$0.06} & $+$0.04 & $ $0.0  & {\bf $+$0.07} & {\bf $+$0.05}\\
       & $-\Delta T$       & {\bf $-$0.06} & {\bf $+$0.07} & $-$0.02 & {\bf $-$0.09} & $-$0.04 & $ $0.0  & {\bf $-$0.07} & $-$0.03 \\
       &$+\Delta\log g$    & $-$0.01 & {\bf $+$0.09} & $-$0.02 & $ $0.0  & $-$0.02 & $+$0.02 & $-$0.04 & $+$0.01 \\
       &$-\Delta\log g$    & $+$0.01 & {\bf $-$0.08} & $+$0.01 & $-$0.01 & $+$0.02 & $-$0.03 & $+$0.03 & $+$0.02 \\
       &$+\Delta \xi$      & {\bf $-$0.05} & $-$0.02 & $-$0.02 & $+$0.01 & {\bf $-$0.05} & $ $0.0  & {\bf $-$0.12} & $+$0.02\\
       &$-\Delta \xi$      & {\bf $+$0.05} & $+$0.03 & $+$0.01 & $-$0.02 & {\bf $+$0.05} & $+$0.01 & {\bf $+$0.11} & $+$0.01\\
       &$+\Delta[\rm{M/H}]$& $ $0.0  & $+$0.04 & $-$0.01 & $-$0.02 & $-$0.01 & $+$0.01 & $-$0.01 & $+$0.02\\
       &$-\Delta[\rm{M/H}]$& $ $0.0  & $-$0.01 & $ $0.0  & $ $0.0  & $+$0.01 & $ $0.0  & $-$0.01 & $ $0.0 \\
& & & & & & & & \\
M13    & $+\Delta T$       & {\bf $+$0.09} & {\bf $-$0.05} & $-$0.01 & {\bf $+$0.05} & $+$0.03 & $-$0.01 & {\bf $+$0.07} & {\bf $+$0.05}\\
       & $-\Delta T$       & {\bf $-$0.07} & {\bf $+$0.07} & $ $0.0  & {\bf $-$0.07} & $-$0.04 & $ $0.0  & {\bf $-$0.08} & $-$0.02 \\
       &$+\Delta\log g$    & $ $0.0  & {\bf $+$0.09} & $-$0.02 & $-$0.01 & $-$0.02 & $+$0.02 & $-$0.04 & $+$0.02 \\
       &$-\Delta\log g$    & $+$0.01 & {\bf $-$0.08} & $+$0.02 & $ $0.0  & $+$0.02 & $-$0.02 & $+$0.03 & $+$0.02 \\
       &$+\Delta \xi$      & {\bf $-$0.05} & $-$0.03 & $-$0.01 & $+$0.02 & $-$0.04 & $ $0.0  & {\bf $-$0.10} & $+$0.03 \\
       &$-\Delta \xi$      & {\bf $+$0.06} & $+$0.03 & $ $0.0  & $-$0.03 & {\bf $+$0.05} & $-$0.01 & {\bf $+$0.10} & $+$0.02 \\
       &$+\Delta[\rm{M/H}]$& $ $0.0  & $+$0.03 & $ $0.0  & $-$0.01 & $-$0.01 & $ $0.0  & $ $0.0 & $+$0.03 \\
       &$-\Delta[\rm{M/H}]$& $ $0.0  & $-$0.02 & $+$0.01 & $+$0.01 & $+$0.01 & $ $0.0  & $ $0.0 & $+$0.02 \\
& & & & & & & & \\
NGC~7006&$+\Delta T$       & {\bf $+$0.08} & {\bf $-$0.06} & $+$0.01 & {\bf $+$0.05} & $+$0.04 & $ $0.0  & {\bf $+$0.09} & {\bf $+$0.05} \\
       & $-\Delta T$       & {\bf $-$0.07} & {\bf $+$0.06} & $-$0.02 & {\bf $-$0.07} & $-$0.04 & $ $0.0  & {\bf $-$0.07} & $-$0.03 \\
       &$+\Delta\log g$    & $-$0.01 & {\bf $+$0.08} & $-$0.02 & $ $0.0  & $-$0.01 & $+$0.03 & $-$0.03 & $+$0.01 \\
       &$-\Delta\log g$    & $+$0.01 & {\bf $-$0.08} & $+$0.02 & $ $0.0  & $+$0.01 & $-$0.02 & $+$0.04 & $+$0.01 \\
       &$+\Delta \xi$      & {\bf $-$0.06} & $-$0.03 & $+$0.02 & $+$0.03 & $-$0.03 & $+$0.01 & {\bf $-$0.11} & $+$0.02 \\
       &$-\Delta \xi$      & {\bf $+$0.06} & $+$0.02 & $+$0.01 & $-$0.03 & {\bf $+$0.05} & $ $0.0  & {\bf $+$0.12} & $+$0.01 \\
       &$+\Delta[\rm{M/H}]$& $ $0.0  & $+$0.03 & $-$0.01 & $-$0.02 & $ $0.0  & $+$0.01 & $+$0.01 & $+$0.01 \\
       &$-\Delta[\rm{M/H}]$& $ $0.0  & $-$0.02 & $ $0.0  & $ $0.0  & $+$0.01 & $ $0.0  & $ $0.0  & 0.0\\
& & & & & & & & \\
M15    & $+\Delta T$       & {\bf $+$0.10} & $-$0.03 & $-$0.04 & $+$0.02 & $+$0.03 & $+$0.02 & {\bf $+$0.06} & {\bf $+$0.07}\\
       & $-\Delta T$       & {\bf $-$0.11} & $+$0.02 & $+$0.03 & $-$0.02 & $-$0.02 & $-$0.02 & {\bf $-$0.06} & $-$0.02\\
       &$+\Delta\log g$    & $-$0.03       & {\bf $+$0.06} & $ $0.0  & $+$0.01 & $-$0.01 & $+$0.02 & $-$0.02 & $+$0.02\\
       &$-\Delta\log g$    & $+$0.02       & {\bf $-$0.07} & $ $0.0  & $ $0.0  & $+$0.01 & $-$0.01 & $+$0.01 & $+$0.02\\
       &$+\Delta \xi$      & {\bf $-$0.06} & $-$0.01       & $+$0.01 & {\bf $+$0.06} & $-$0.02 & $+$0.03 & {\bf $-$0.09} & $+$0.01\\
       &$-\Delta \xi$      & {\bf $+$0.06} & $ $0.0        & $-$0.01 & {\bf $-$0.05} & $+$0.04 & $-$0.03 & {\bf $+$0.11} & $+$0.03\\
       &$+\Delta[\rm{M/H}]$& $-$0.01       & $ $0.0        & $ $0.0  & $ $0.0  & $+$0.01 & $ $0.0  & $+$0.01 & $+$0.03\\
       &$-\Delta[\rm{M/H}]$& $+$0.01       & $-$0.01       & $-$0.01 & $ $0.0  & $ $0.0  & $ $0.0  & $ $0.0  & $+$0.01\\
& & & & & & & & \\
47~Tuc & $\Delta T,\log g$& $-$     & {\bf $<$0.20} & {\bf $<$0.06} & $-$     & $-$     & $<$0.04 & $-$     & $-$\\
       & $\Delta T, \xi$   & {\bf $<$0.12} & $-$     & $-$     & {\bf $<$0.06} & {\bf $<$0.14} & $-$     & {\bf $<$0.22} & {\bf $<$0.11}\\
\hline
\end{tabular}
\end{minipage}\\
\medskip
\raggedright {\bf Notes: } The uncertainties in the atmospheric
parameters are the typical values found in individual stellar
analyses: $\Delta T = \pm 100$ K, $\Delta \log g = \pm 0.2$ dex,
$\Delta \xi = \pm 0.2$ km/s, and $\Delta [\rm{M/H}] = \pm 0.1$ dex.
The last two rows tabulate the maximum abundance differences that
occur when two parameters are changed together.  Abundance differences
are calculated relative to the baseline abundances in Table
\ref{table:OriginalAbunds}, as described in Section
\ref{subsec:SystematicErrors}.  Significant offsets ($\ge 0.05$ dex)
are bolded.
\end{table*}

Of course, the atmospheric parameters are not independent. It is thus
instructive to see how the final abundances change as two parameters
are varied together---these tests were performed only on 47~Tuc (which
has the largest individual offsets).  For each element ratio, the two
parameters that individually showed the strongest changes in Table
\ref{table:Intrinsic} were varied \textit{together}.  The maximum
differences for all abundance ratios are shown at the bottom of Table
\ref{table:Intrinsic}.

These results show that within the $1\sigma$ boxes,
[\ion{Ni}{1}/\ion{Fe}{1}] is negligibly affected by the atmospheric
parameters, while [\ion{Ca}{1}/\ion{Fe}{1}] and
[\ion{Ti}{1}/\ion{Fe}{1}] are moderately affected ($\la 0.1$ dex).
The [\ion{Fe}{1}/H], [\ion{Fe}{2}/H], [\ion{Ti}{2}/\ion{Fe}{2}],
[\ion{Ba}{2}/\ion{Fe}{2}], and [\ion{Eu}{2}/\ion{Fe}{2}] ratios are
all significantly ($0.1 < \Delta [\rm{X/Fe}] < 0.22$) affected by the
changes in atmospheric parameters.

\subsection{Colour-Temperature Relations}\label{subsec:ColorTemp}
In a CMD-based analysis, the observed stellar colours are transformed
to effective temperatures via Colour-Temperature Relations (CTRs).
Several studies have calibrated these relations for different
photometric filters and different stellar types, over ranges in colour
and metallicity. This appendix investigates the effects on the
abundances caused by changing these CTRs.  To investigate metallicity
dependencies, 47~Tuc, M3, and M15 were used for these tests.  The
relations of \citet[for dwarfs and giants,
respectively]{Alonso1996,Alonso1999}, \citet[for dwarfs and
giants]{RamirezMelendez2005}, and \citet[for dwarfs only---the
Ram\'{i}rez \& Melendez relation was used for giants]{Casagrande2010}
are considered.  Recall that for the abundances presented in MB08,
Paper I, and Section \ref{sec:ILABUNDS} the $(B-V)$ relations of
\citet{Alonso1996,Alonso1999} were used for 47~Tuc, while the $(V-I)$
relations from \citet{RamirezMelendez2005} were used for the other
clusters.  The CTRs are only valid for the regions in which
they were calibrated; for stars whose colours fall outside the
calibrated regions, MB08 and Paper I utilized the Kurucz grid of
stellar models to determine effective temperatures.  The effects of
extrapolated relations and only values from the Kurucz grid are also
considered.

\begin{table*}
\centering
\begin{minipage}{165mm}
\caption{Differences in CMD-based abundance ratios with various
Colour-Temperature Relations.\label{table:ColorTemp}}
  \begin{tabular}{@{}lcccccccc@{}}
  \hline
& $\Delta[$\ion{Fe}{1}/H$]$ & $\Delta[$\ion{Fe}{2}/H$] $ &
$\Delta[$\ion{Ca}{1}/\ion{Fe}{1}$]$ & $\Delta[$\ion{Ti}{1}/\ion{Fe}{1}$]$ &
$\Delta[$\ion{Ti}{2}/\ion{Fe}{2}$]$ & $\Delta[$\ion{Ni}{1}/\ion{Fe}{1}$]$ & $\Delta[$\ion{Ba}{2}/\ion{Fe}{2}$]$ & $\Delta[$\ion{Eu}{2}/\ion{Fe}{2}$]$\\
\hline
{\bf 47~Tuc} & & & & & & & & \\
Extrapolated A96/99 &  $ $0.0  & $ $0.0  & $ $0.0  & $ $0.0  & $ $0.0  & $ $0.0  & $ $0.0  & $+$0.01\\
RM05                &  $-$0.01 & $+$0.01 & $-$0.01 & $-$0.03 & $-$0.01 & $ $0.0  & $-$0.02 & 0.0\\
C10+RM05            &  $+$0.01 & $ $0.0  & $+$0.01 & $ $0.0  & $+$0.01 & $ $0.0  & $+$0.01 & $+$0.01\\
Kurucz only         &  {\bf $+$0.08} & $-$0.04  & $+$0.03 & {\bf $+$0.05} & {\bf $+$0.07} & $+$0.01 & {\bf $+$0.10} & {\bf $+$0.07}\\
 & & & & & & & &\\
{\bf M3} & & & & & & & &\\
A96/99              &  {\bf $+$0.46} & {\bf $-$0.05} & {\bf $-$0.05} & {\bf $+$0.25} & {\bf $+$0.17} & {\bf $+$0.08} & {\bf $+$0.36} & {\bf $+$0.21}\\
Extrapolated RM05   &  $-$0.03 & $+$0.01 & $ $0.0  & $-$0.01 & $-$0.02 & $ $0.0  & {\bf $-$0.05} & $+$0.01\\
C10+RM05            &  $+$0.01 & $ $0.0  & $ $0.0  & $-$0.01 & $+$0.01 & $-$0.01 & $+$0.01 &  0.0\\
Kurucz only         &  $+$0.01 & {\bf $-$0.08} & {\bf $+$0.06} & {\bf $-$0.09} & {\bf $+$0.11} & {\bf $-$0.10} & {\bf $-$0.05} & {\bf $-$0.15}\\
 & & & & & & & & \\
{\bf M15} & & & & & & & & \\
A96/99              &  {\bf $+$0.44} & $+$0.02 & {\bf $-$0.17} & $ $-    & {\bf $+$0.15} & {\bf $+$0.12} & {\bf $+$0.34} & {\bf $+$0.25}\\
Extrapolated RM05   &  {\bf $-$0.07} & $ $0.0  & $+$0.02 & $ $-    & $-$0.03 & $+$0.04 & {\bf $-$0.06} & $-$0.03\\
C10+RM05            &  $ $0.0  & $ $0.0  & $+$0.01 & $ $-    & $+$0.01 & $ $0.0  & $ $0.0  & $-$0.01\\
Kurucz only         &  {\bf $+$0.14} & $ $0.0  & {\bf $-$0.05} & $+$0.02 & {\bf $+$0.05} & $+$0.04 & {\bf $+$0.10} & {\bf $-$0.06}\\
 & & & & & & & & \\
\hline
\end{tabular}
\end{minipage}\\
\medskip
\raggedright {\bf Notes: } Extrapolated relations carry the CTRs
outside the colour ranges in which they were calibrated.  Abundance
differences are calculated relative to the baseline abundances in
Table \ref{table:OriginalAbunds}, as described in Section
\ref{subsec:SystematicErrors}.\\
{\bf References: } A96 = \citet{Alonso1996}, A99 =
\citet{Alonso1999}, RM05 = \citet{RamirezMelendez2005}, C10 =
\citet{Casagrande2010}\\
\end{table*}

Table \ref{table:ColorTemp} shows the offsets that occur when
different CTRs are used.  With the exception of the Kurucz-only case,
the differences for 47~Tuc are all negligible ($\la 0.03$ dex).  The
M3 and M15 results are \textit{very} discrepant when the
\citet{Alonso1996,Alonso1999} relations are employed.  This is
consistent with the large offsets between the
\citet{Alonso1996,Alonso1999} CTRs versus the
\citet{RamirezMelendez2005} and \citet{Casagrande2010} CTRs. Of the
three relations, the \citet{RamirezMelendez2005} and
\citet{Casagrande2010} relations are likely to be more accurate, since
Alonso et al. had to rely on uncertain transformations between
photometric systems (see the discussion by \citealt{Casagrande2010}).
The Kurucz only relations are also quite discrepant, suggesting that
empirical relations (specifically the \citealt{RamirezMelendez2005}
and/or \citealt{Casagrande2010} CTRs) may be a better choice for
CMD-based studies.

Other than the large offsets from the $(V-I)$ Alonso et al. relations
and from the Kurucz only abundances, the differences from the other
relations (including the extrapolated relations) are insignificant,
except for [\ion{Ba}{2}/\ion{Fe}{2}], which is affected by $\sim 0.05$
when the \citet{RamirezMelendez2005} relation is extrapolated outside
the calibrated regions in M3 and M15.

\subsection{Different Photometric Data Sets}\label{subsec:Photometry}

\begin{figure*}
\begin{center}
\centering
\subfigure[$V$, $I$ CMD]{\includegraphics[scale=0.5]{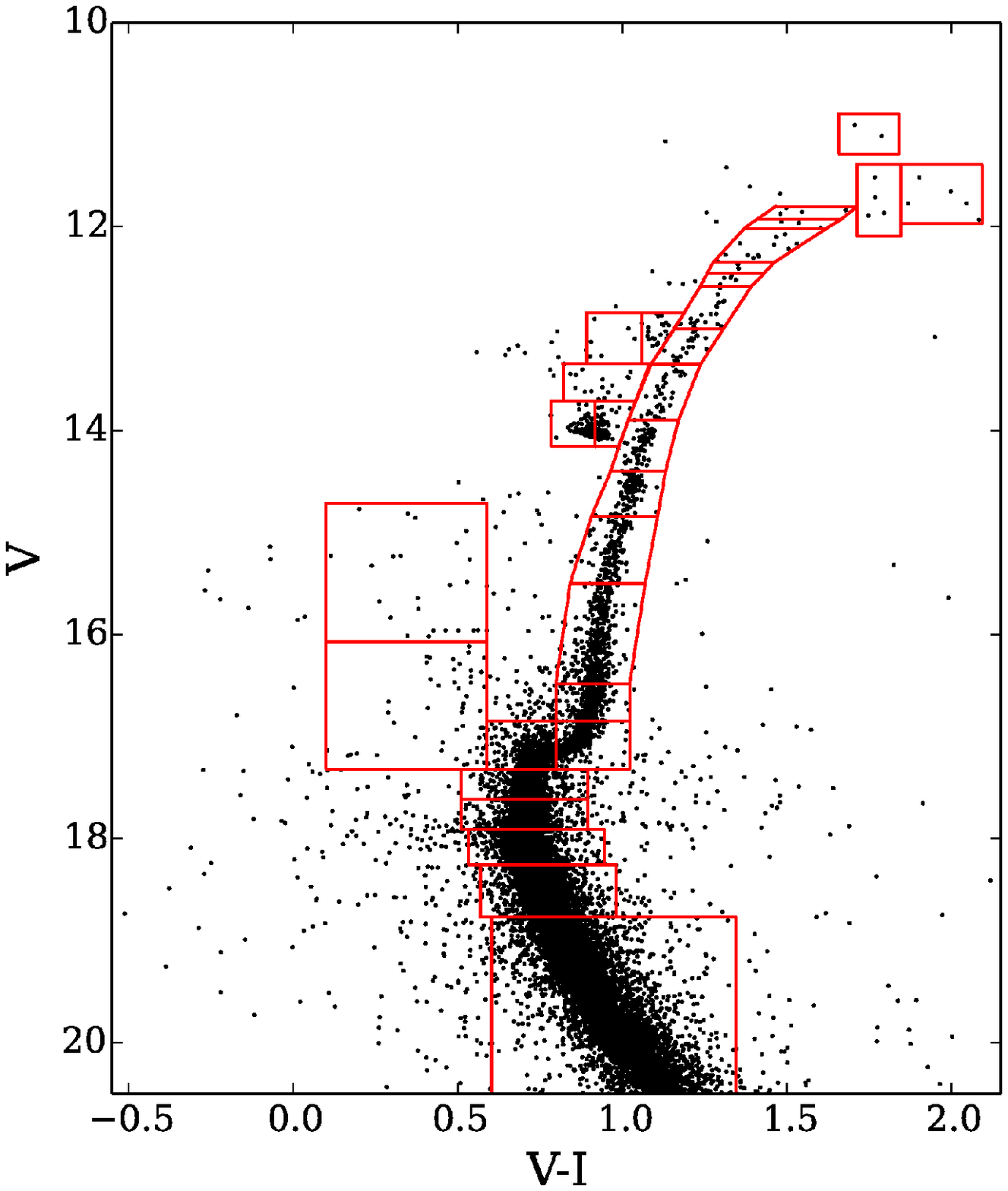}\label{fig:VIa}}
\subfigure[HRD for $B$, $V$ and $V$, $I$ photometry]{\includegraphics[scale=0.5]{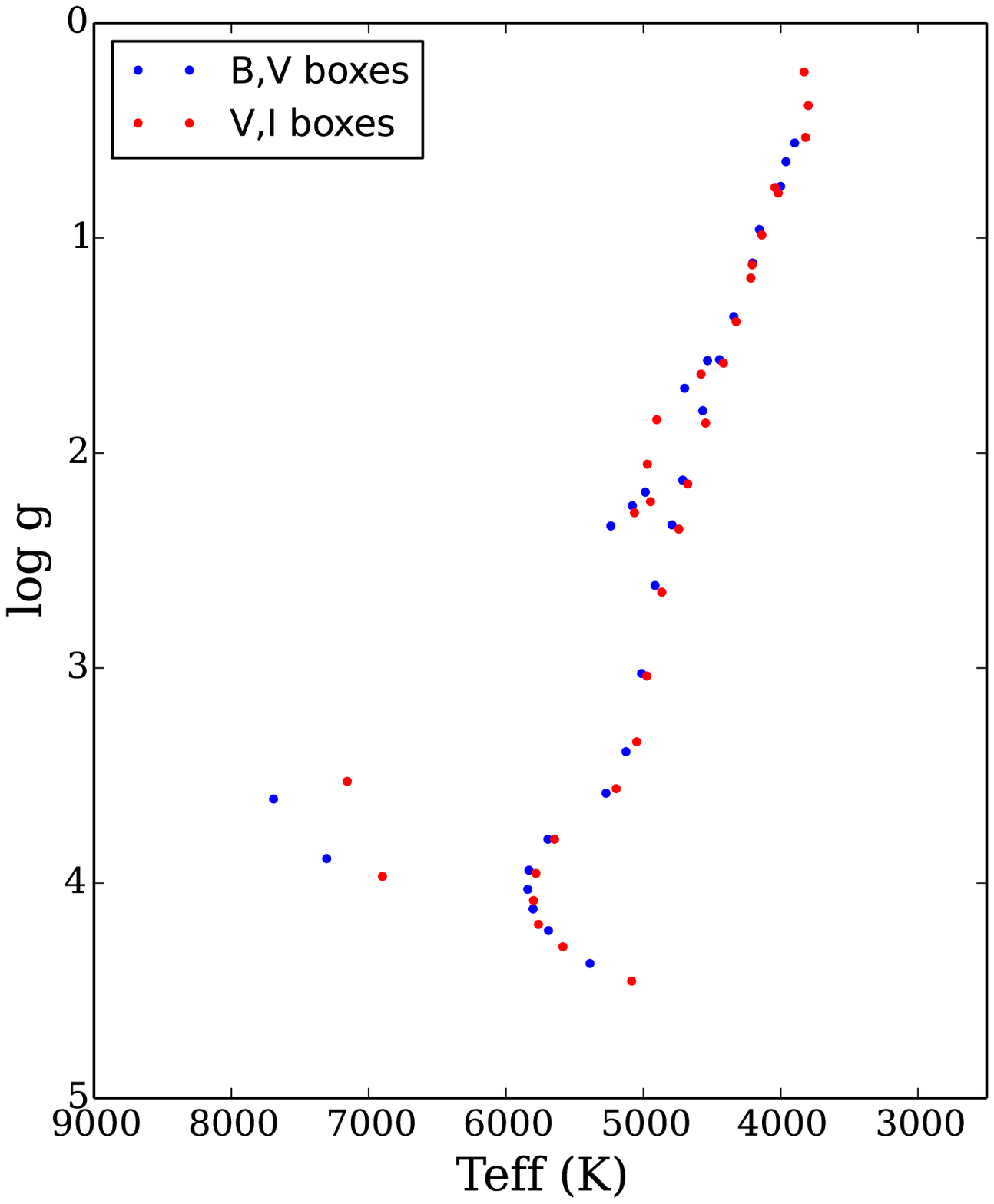}\label{fig:VIb}}
\caption{The $V$, $I$ photometry for 47~Tuc. \textit{Left: } The
thirty CMD boxes and the Johnson $V$, $I$ photometry from the ACS
Galactic Globular Cluster Treasury \citep{Sarajedini2007}.
\textit{Right: } An HRD of the $B$, $V$ (blue) and $V$, $I$ (red) CMD
boxes.\label{fig:VI}}
\end{center}
\end{figure*}

This appendix investigates the effects of different photometric
data sets (i.e. $V$, $I$ instead of $B$, $V$, taken with different
instruments at different times).  This test is only performed on
47~Tuc because $B$, $V$ CMDs of the cores are not available for the
other GCs.  Recall that the original 47~Tuc abundances were found with
the $B$, $V$ photometry from \citet{Guhathakurta1992}. Figure
\ref{fig:VIa} presents the boxes for the {\it HST} 47~Tuc $V$, $I$ CMD
from the ACS Galactic Globular Cluster Treasury
(e.g. \citealt{Sarajedini2007}). The CTRs of
\citet{RamirezMelendez2005} were used to determine atmospheric
parameters for the $V$, $I$ photometry; an HRD showing the box
averages for the two data sets is shown in Figure \ref{fig:VIb}. The
agreement between the parameters of each box is generally good, with
the exception of the brightest RGB and blue straggler boxes.

The bright RGB boxes in the $V$, $I$ CMD contain M giants.  Because of
TiO blanketing and the breakdown of the M giant $(B-V)$ CTR the M
giants appear mixed with the K giants in the $B$, $V$ CMD.  For this
reason, M giants need to be treated differently if a $B$, $V$ CMD is
employed. MB08 showed that, for the core of 47~Tuc, the TiO blanketing
in the M giant spectra significantly reduced their impact on the IL
spectrum, such that only small errors in the derived abundances would
result from the omission of the two M  giants.\footnote{However, the
mostly negligible abundance effects of the two M giants in the 47~Tuc
spectrum may not translate to similar effects in more metal-rich GCs,
which have a larger fraction of M giants.  Therefore the presence of M
giants needs to be considered carefully when analyzing the IL spectra
of GCs more metal-rich than 47~Tuc.}  The M giants were {\it not}
removed from the $V$, $I$ photometry, which accounts for the
differences in the brightest boxes.

The differences in blue stragglers are likely only due to sampling.
These boxes represent a small ($\la 1$\%) portion of the total light,
and therefore have an insignificant effect on the final abundances.
Small variations between $B$, $V$ and $V$, $I$ photometry may also be
due to the Bond-Neff effect \citep{BondNeff}, since 47~Tuc does have a
significant population of CN-strong stars.

The abundance differences are listed in Table
\ref{table:Photometry}. In general, these differences are not drastic,
with the exception of [\ion{Fe}{1}/H] and [\ion{Ba}{2}/\ion{Fe}{2}],
which differ by $\sim 0.07$ dex. The small differences in \ion{Fe}{1}
and \ion{Ba}{2} and the negligible differences in the other
abundances indicate that the M giants do not need to be removed from
the $V$, $I$ data, and that (as in MB08) the TiO molecular lines do
not need to be included for GCs at 47~Tuc's metallicity.  Only 47~Tuc
was considered for this test---however, variations between $B$, $V$
and $V$, $I$ may be metallicity dependent, or dependent upon the
populations in a given GC.

\begin{table*}
\centering
\begin{minipage}{165mm}
\caption{Differences in CMD-based abundance ratios as a result of
various alterations to the input photometry.\label{table:Photometry}}
  \begin{tabular}{@{}lccccccccc@{}}
  \hline
& & $\Delta[$\ion{Fe}{1}/H$]$ & $\Delta[$\ion{Fe}{2}/H$] $ &
$\Delta[$\ion{Ca}{1}/\ion{Fe}{1}$]$ & $\Delta[$\ion{Ti}{1}/\ion{Fe}{1}$]$ & $\Delta[$\ion{Ti}{2}/\ion{Fe}{2}$]$ & $\Delta[$\ion{Ni}{1}/\ion{Fe}{1}$]$ & $\Delta[$\ion{Ba}{2}/\ion{Fe}{2}$]$ & $\Delta[$\ion{Eu}{2}/\ion{Fe}{2}$]$ \\
\hline
$V,I$ data   & & & & & & & & & \\
47~Tuc       &  & $-$0.04 & {\bf $+$0.07} & $-$0.02 & $-$0.04 & $-$0.04 & $+$0.01 & {\bf $-$0.07} & $+$0.02\\
& & & & & & & & & \\
Completeness & & & & & & & & \\
47~Tuc     &  & {\bf $+$0.07} & {\bf $+$0.07} & $+$0.01       & {\bf $+$0.06} & {\bf $+$0.07} & $+$0.04 & {\bf $+$0.07} & $+$0.03\\
M3         &  & $ $0.0        & $+$0.01       & $+$0.01       & $ $0.0        & $ $0.0        & $+$0.01 & $-$0.01       & 0.0\\
M13        &  & $-$0.01       & $ $0.0        & $-$0.01       & $-$0.01       & $+$0.01       & $-$0.01 & $ $0.0        & $+$0.01\\
NGC~7006   &  & $+$0.03       & $+$0.01       & {\bf $-$0.05} & $+$0.01       & $+$0.02       & $-$0.02 & $+$0.01       & 0.0\\
M15        &  & $-$0.03       & $ $0.0        & $-$0.02       & $+$0.03       & $+$0.01       & $+$0.04 & $ $0.0        & $-$0.01 \\
& & & & & & & & & \\
Sampling & & & & & & & & & \\
M15        &  & {\bf $<$0.22} & {\bf $<$0.10} & {\bf $<$0.09} & {\bf $<$0.06} & {\bf $<$0.10} & $<$0.03 & {\bf $<$0.21} & {\bf $<$0.09}\\
& & & & & & & & & \\
\hline
\end{tabular}
\end{minipage}\\
\medskip
\raggedright {\bf Notes: } Abundance differences are calculated
relative to the baseline abundances in Table
\ref{table:OriginalAbunds}, as described in Section
\ref{subsec:SystematicErrors}.\\
\end{table*}

\subsection{Incompleteness}\label{subsec:Incompleteness}
Even the highest quality {\it HST} data suffer from incompleteness of
the faintest stars.  The effects of incompleteness were tested by
increasing the numbers of stars in the lower main sequence boxes in
order to match the theoretical luminosity functions (assuming no mass
segregation; this means that this test may add too many low mass
stars).  The abundance differences (tabulated in Table
\ref{table:Photometry}) are all $\la 0.1$ dex.  The only GC affected
by incompleteness is 47~Tuc; it is also the GC whose IL spectrum
covers the smallest portion of the cluster (see Table
\ref{table:Observations}), suggesting that mass segregation may be
more important for 47~Tuc than for the other clusters.  In 47~Tuc, the
[\ion{Fe}{1}/H], [\ion{Fe}{2}/H], [\ion{Ti}{1}/\ion{Fe}{1}],
[\ion{Ti}{2}/\ion{Fe}{2}], and [\ion{Ba}{2}/\ion{Fe}{2}] ratios are
affected by $< 0.1$ dex---the other abundances are largely
unaffected.

\subsection{Sampling the Input Photometry}\label{subsec:MismatchingPhotometry}
This appendix investigates the effects if the input photometry does not
perfectly match the population observed in the IL spectra.  This is
especially problematic in cases where the spectrograph fibres must be
scanned across the cluster.  The input photometry can be cleaned based
on distance from the cluster centre, but irregular coverage patterns
(see Paper I) will lead to differences between the input photometry
and the observed population.

As a test of this effect, it is assumed that there are no constraints
on the area that was scanned in M15's wedge-shaped coverage pattern.
Note that this is a somewhat unrealistic worst-case scenario; however,
it serves as a useful test of how sensitive the abundances are to
stochastic effects on the upper RGB.  To select the input photometry,
one hundred $80^{\circ}$ wedges were selected by assuming a
random\footnote{A random value was selected using NumPy's
\textit{random} routines.} starting angle between $0$ and
$360^{\circ}$.  ILABUNDS was then rerun  on each of the 100 wedges,
producing new abundances for each run.  The largest offsets from
the mean are listed in Table \ref{table:Photometry}.  The abundance
differences can be quite large, especially for [\ion{Fe}{1}/H] and
[\ion{Ba}{2}/\ion{Fe}{2}], where the maximum offsets are $\sim
0.2$ dex.  However, the [\ion{Fe}{2}/H], [\ion{Ca}{1}/\ion{Fe}{1}],
[\ion{Ti}{2}/\ion{Fe}{2}], 
[\ion{Ni}{1}/\ion{Fe}{1}], and [\ion{Eu}{2}/\ion{Fe}{2}] abundances
are less sensitive to this effect (with maximum differences $\la 0.1$
dex).  The primary differences between each run are the numbers and
properties of bright RGB stars.  Thus, these tests indicate that
[\ion{Fe}{1}/H] and [\ion{Ba}{2}/\ion{Fe}{2}] are particularly
sensitive to sampling of the upper RGB.

This test on M15's wedge-shaped pointing pattern illustrates the
importance of adequately selecting stars that truly match the observed
population.  The relative numbers of stars at various evolutionary
stages are important, as are slight differences in colours and
magnitudes.  Because each cluster is unique, this effect cannot be
removed through a differential analysis.  However, observations that
cover more of the cluster (e.g. extragalactic observations) or whose
pointing patterns are more regular (e.g. in the case of M3, M13,
and NGC~7006) will not suffer from the problem as severely as M15,
since the sampling differences between photometric data sets will be
less extreme.  Note, however, that the IL observations of M3, M13,
NGC~7006, and M15 utilized discrete pointings across the cluster (see
Paper I), albeit with short integration times.  This means that the IL
spectra are non-trivially weighted by the stars at those pointings.
This effect is extremely difficult to account for---however, as the
exposure times were short and the uncovered areas were small, this
effect should not be too large.

In a CMD-based analysis of resolved GCs sampling problems can be
alleviated by 1) symmetrically observing GCs and 2) using deep
photometry that has been accurately sampled to match the IL spectra.
Unfortunately, the second option is not  possible for unresolved
extragalactic targets.  The next appendix investigates systematic
offsets that occur when populations are unresolved, and have to be
modelled with theoretical isochrones.

\section{Systematic Offsets that Occur in an HRD-Based Analysis}\label{sec:HRD}
CMDs cannot be obtained for unresolved clusters, and theoretical
isochrones must be used to generate HRDs (i.e. temperatures and
surface gravities) for the underlying populations.  The main
advantage of an HRD-based analysis is that the stars are modelled in
the theoretical plane, and there is no need to convert observable
properties to physical quantities. The main disadvantage in a
HRD-based analysis of an unresolved target is that very little is
known about the GC \textit{a priori}, and diagnostics must be used to
revise the model of the underlying stellar population. This appendix
investigates systematic errors that occur when the stellar populations
are incorrectly modelled.  These errors include:
\begin{enumerate}
\item Uncertainties in identifying the best-fitting isochrones
(Appendix \ref{subsec:IsochroneOffsets})
\item Uncertainties that occur when the theoretical isochrones are
populated with stars (Appendix \ref{subsec:Sampling})
\item Uncertainties in modelling evolved stars (Appendices
\ref{subsec:HBMorphology} and \ref{subsec:AGBstars}) and main
sequence stars (Appendices \ref{subsec:BlueStragglers} and
\ref{subsec:LowMassCut-Off}).
\end{enumerate}
\noindent Discrepancies between the real population and
the modelled population may vary between clusters in the same study,
making it difficult to remove these effects though differential
analyses.  Finally, the case of a partially resolved GC is
investigated in Appendix \ref{subsec:PartiallyResolved}.

\subsection{HRD-based abundances}\label{subsec:IsochroneOffsets}
MB08 and \citet{Colucci2009,Colucci2011a,Colucci2012,Colucci2013} have
pioneered high resolution IL spectral analyses of unresolved GCs.  Their
algorithm for identifying the HRD that best represents an underlying
stellar population involves iterating upon isochrone parameters until
the following criteria are met:
\begin{enumerate}
\item The isochrone [Fe/H] matches the output integrated
[\ion{Fe}{1}/H] ratio
\item Any trends in \ion{Fe}{1} abundance with wavelength, reduced
EW (REW),\footnote{REW$= \rm{EW}/\lambda$} or excitation potential
(EP) are minimized (similar to individual stellar analyses; see
MB08)
\item The line-to-line abundance spreads from \ion{Fe}{1} and
\ion{Fe}{2} lines are minimized.
\end{enumerate}

\noindent Colucci et al. have demonstrated that a best-fitting HRD can
be identified based on these criteria.  Furthermore,
\citet{Cameron2009} argue that HRD-based abundances of Galactic GCs
are in good agreement with CMD-based ones (which are, in turn, in
agreement with literature abundances from individual stars).  This
paper focuses on the abundance uncertainties that arise as a result of
uncertainties in identifying the best input isochrones.

First, the best-fitting standard BaSTI isochrones are identified for
the targets GCs.  No modifications were made to the default BaSTI HBs
(see Appendix \ref{subsec:HBMorphology}).    As in Colucci et al,
isochrones with extended AGBs and mass loss parameters of
$\eta~=~0.2$ were initially adopted (see Appendix
\ref{subsec:AGBstars} for tests with other AGB prescriptions, though
note that MB08 utilize $\eta~=~0.4$ isochrones based on the tests by
\citealt{Maraston2005}).   In order to match 47~Tuc's observed
luminosity function MB08 manually enhanced the number of AGB stars; no
AGB enhancements were included in these tests. When the
[\ion{Ca}{1}/\ion{Fe}{1}] ratio indicated $\alpha$-enhancement,
$\alpha$-enhanced isochrones were used.  The ``best-fitting'' solution
was deemed to be the one for which all slopes are minimized---note
that this choice is subjective, since the slopes are rarely
simultaneously minimized.

Table \ref{table:Isochrones} presents the parameters for the
best-fitting HRDs and comparisons with the CMD-based abundances.
Note that none of the REW slopes are sufficiently flat for M15.  With
the exception of M13, all solutions are younger than isochrone fits
indicate (see, e.g., \citealt{Dotter2010,Vandenberg2013}).  The
best-fitting HRD abundances can be significantly offset from the
CMD-based abundances---in particular, the [\ion{Ti}{1}/\ion{Fe}{1}]
values are persistently lower than those from the CMD analyses and
those from individual stars.  The [\ion{Fe}{1}/H],
[\ion{Fe}{2}/H], [\ion{Ti}{2}/\ion{Fe}{2}], [\ion{Ba}{2}/\ion{Fe}{2}],
and [\ion{Eu}{2}/\ion{Fe}{2}] ratios are also significantly
affected. Only [\ion{Ca}{1}/\ion{Fe}{1}] and [\ion{Ni}{1}/\ion{Fe}{1}]
agree well with the CMD-based abundances.

\begin{table*}
\centering
\begin{minipage}{165mm}
\caption{Parameters of the ``best-fitting'' HRDs, and abundance comparisons with the CMD-based abundances.\label{table:Isochrones}}
  \begin{tabular}{@{}lccccccccccc@{}}
  \hline
& age & [Z/H] & $\Delta[$\ion{Fe}{1}/H$]$ & $\Delta[$\ion{Fe}{2}/H$] $ &
$\Delta[$\ion{Ca}{1}/\ion{Fe}{1}$]$ & $\Delta[$\ion{Ti}{1}/\ion{Fe}{1}$]$ & $\Delta[$\ion{Ti}{2}/\ion{Fe}{2}$]$ & $\Delta[$\ion{Ni}{1}/\ion{Fe}{1}$]$ & $\Delta[$\ion{Ba}{2}/\ion{Fe}{2}$]$ & $\Delta[$\ion{Eu}{2}/\ion{Fe}{2}$]$ \\
\hline
47~Tuc     & 10 & -0.35 & {\bf $+$0.07} & {\bf $+$0.19} & {\bf $-$0.05} & {\bf $-$0.10} & $-$0.01 & $+$0.02 & {\bf $-$0.05} & {\bf $-$0.08}\\
M3         &  9 & -1.27 & {\bf $-$0.05} & $-$0.04       & $-$0.01       & {\bf $-$0.20} & {\bf $-$0.05} & $-$0.03 & {\bf $-$0.10} & {\bf $-$0.06}\\
M13        & 12 & -1.27 & {\bf $-$0.11} & {\bf $-$0.13} & $+$0.03       & {\bf $-$0.19} & {\bf $-$0.10} & $-$0.04 & {\bf $-$0.12} & $-$0.02\\
NGC~7006   &  7 & -1.27 & {\bf $+$0.06} & $-$0.02       & $-$0.02       & {\bf $-$0.14} & {\bf $+$0.05} & $-$0.03 & $+$0.03 & {\bf $-$0.05}\\
M15        &  9 & -1.79 & $+$0.02       & {\bf $+$0.06} & $-$0.01       & $-$0.04       & $ $0.0  & $+$0.01 & $+$0.04 & $+$0.03 \\
& & & & & & & & & \\
\hline
\end{tabular}
\end{minipage}\\
\medskip
\raggedright {\bf Notes: } Abundance differences are calculated
relative to the CMD-based abundances in Table
\ref{table:OriginalAbunds}, as described in Section
\ref{subsec:SystematicErrors}.\\
\end{table*}

\subsubsection{Uncertainties in identifying the best HRDs}\label{subsubsec:CMDcomparison}
The best-fitting HRDs are those which \textit{best} meet the above
criteria; however, multiple solutions meet these criteria, and there
is a range of possible abundances.  In this appendix, the selection
criteria of Colucci et al. are broadened to assess the possible
abundance ranges.

The first two criteria for identifying the best HRD each have
associated uncertainties.  Not only does the integrated
[\ion{Fe}{1}/H] have its own uncertainty, there may be systematic
offsets between spectroscopically determined [Fe/H] values and between
those determined from isochrone fits---these [Fe/H] values could be
off by as much as 0.2 dex.\footnote{For example, from high resolution
spectroscopic analyses of M3, \citet{CohenMelendez2005} find an
average $[\rm{Fe/H}] \sim -1.4$ while \citet{Sneden2004} find
$[\rm{Fe/H}] \sim -1.6$.  Isochrone fits with the DSED isochrones also
indicate values of $[\rm{Fe/H}] \sim -1.6$ \citep{Dotter2010}.}
Additionally, [\ion{Fe}{1}/H] is not necessarily indicative of the
cluster [Fe/H], because of NLTE effects. Thus, it may not be ideal to
force the integrated [\ion{Fe}{1}/H] abundance to equal the isochrone
[Fe/H].

The least-squares fits to the \ion{Fe}{1} abundances versus
wavelength, REW, and EP also have their own uncertainties, such that
multiple solutions produce flat fits (i.e. with no significant trend
in the \ion{Fe}{1} abundances).  Furthermore the dispersion in
\ion{Fe}{1} abundances ensures that multiple solutions can produce
sufficiently flat slopes, even in individual stellar analyses.  This
means that it may not be reasonable to consider only the isochrones
that produce the flattest slopes.

Possible HRD solutions are identified in a similar way as Colucci et
al.:
\begin{enumerate}
\item BaSTI isochrones of all ages and metallicities were used to
generate synthetic stellar populations.
\item For each cluster, ILABUNDS was run on the synthesized
population.
\item Any isochrones whose output [\ion{Fe}{1}/H] ratios were within
0.2 dex of the input isochrone [Fe/H] were deemed to be possible
solutions.  Note that Colucci et al. find the best [Fe/H] solution for
each age; for the purposes of this errors analysis, \textit{all}
possible [Fe/H]/age combinations are retained if they meet this
criterion.
\item For each possible solution, the fits to the \ion{Fe}{1}
abundance vs. wavelength, REW, and EP were calculated.  All solutions
whose slopes were $|m|\le 0.04$ (within the uncertainty) were considered to
be alternate solutions.
\end{enumerate}

Table \ref{table:IsochroneErrs} presents the maximum offsets of the
alternate solutions from the best-fitting HRD solutions.  The spreads
around the best-fitting HRD abundances are quite large, with every
element except Ni having significant differences. The
[\ion{Ca}{1}/\ion{Fe}{1}] and [\ion{Eu}{2}/\ion{Fe}{2}] ratios are
still fairly robust, with differences of 0.07 and 0.08 dex,
respectively.  The other abundances ratios can be significantly
affected by the isochrone parameters, depending on the cluster.  It is
also important to note that the offsets from the CMD-based abundances
are sometimes larger than the uncertainties quoted in Table
\ref{table:IsochroneErrs}, suggesting that there is some systematic
offset between the two methods.

\begin{table*}
\centering
\begin{minipage}{165mm}
\caption{Abundance ranges when all acceptable HRD solutions are considered.\label{table:IsochroneErrs}}
  \begin{tabular}{@{}lccccccccc@{}}
  \hline
& $\Big|\Delta[$\ion{Fe}{1}/H$]\Big|$ & $\Big|\Delta[$\ion{Fe}{2}/H$]\Big|$ & $\Big|\Delta[$\ion{Ca}{1}/\ion{Fe}{1}$]\Big|$ & $\Big|\Delta[$\ion{Ti}{1}/\ion{Fe}{1}$]\Big|$ & $\Big|\Delta[$\ion{Ti}{2}/\ion{Fe}{2}$]\Big|$ & $\Big|\Delta[$\ion{Ni}{1}/\ion{Fe}{1}$]\Big|$ & $\Big|\Delta[$\ion{Ba}{2}/\ion{Fe}{2}$]\Big|$ & $\Big|\Delta[$\ion{Eu}{2}/\ion{Fe}{2}$]\Big|$ \\
\hline
47~Tuc     & {\bf $<$0.12} & {\bf $<$0.12} & {\bf $<$0.07} & {\bf $<$0.13} & {\bf $<$0.12} & $<$0.03 & {\bf $<$0.19} & {\bf $<$0.08}\\
M3         & {\bf $<$0.08} & {\bf $<$0.15} & $<$0.04       & {\bf $<$0.07} & {\bf $<$0.05} & $<$0.04 & {\bf $<$0.10} & {\bf $<$0.05}\\
M13        & {\bf $<$0.06} &      $<$0.02  & $<$0.01       & {\bf $<$0.06} & {\bf $<$0.05} & $<$0.01 & {\bf $<$0.06} & $<$0.02\\
NGC~7006   & {\bf $<$0.07} & {\bf $<$0.16} & {\bf $<$0.07} & {\bf $<$0.12} & {\bf $<$0.08} & $<$0.04 & {\bf $<$0.08} & $<$0.03\\
M15        & {\bf $<$0.16} & {\bf $<$0.05} & {\bf $<$0.05} & $<$0.02       & {\bf $<$0.05} & $<$0.03 & {\bf $<$0.10} & {\bf $<$0.07}\\
& & & & & & & & & \\
\hline
\end{tabular}
\end{minipage}\\
\medskip
\raggedright {\bf Notes: } Abundance differences are calculated
relative to the best-fitting HRD abundances in Table
\ref{table:Isochrones}.\\
\end{table*}

\subsubsection{Comparisons between different isochrones}\label{subsubsec:IsochroneOffsets}
Different sets of isochrones predict slightly disparate distributions
of stars in an HRD, even for a common age and metallicity, which could
lead to slight discrepancies in the integrated abundances.  Here the
DSED and Victoria-Regina isochrones (see Section
\ref{subsubsec:Isochrones}) are compared to the BaSTI isochrones.
Tests are run on 47~Tuc, M3, and M15 to investigate metallicity
effects. Because neither the DSED nor the Victoria-Regina models
include evolved HB or AGB stars in their models, the HB/AGB boxes from
the resolved photometry are used instead of HRD boxes, \textit{in all
cases} (even with the BaSTI isochrones).  The isochrones were sampled
such that the number of RGB stars agreed with the number of resolved
RGB stars---this was necessary to ensure that the relative number of
HB/AGB and RGB stars was approximately correct. The isochrones with
the best-fitting BaSTI parameters (from Appendix
\ref{subsec:IsochroneOffsets}) were used.

The offsets from the BaSTI abundances are shown in
Table~\ref{table:DiffIsochrones}.  The differences between the BaSTI
and Victoria-Regina isochrones are insignificant in all cases.  The
DSED isochrones have larger offsets at low [Fe/H], depending on the
input [$\alpha$/Fe] ratio.  The BaSTI and Victoria-Regina models use
[$\alpha$/Fe] $= +0.4$ and $+0.3$, respectively, while DSED isochrones
can have [$\alpha$/Fe]~$=~+0.2$ or $+0.4$.  The [$\alpha$/Fe] $= +0.4$
DSED isochrones are in much better agreement than the $+0.2$ ones,
although the [$\alpha$/Fe]~$=~+0.4$ isochrone offsets can be $\sim
0.05$ dex.  This suggests that the slight differences in the treatment
of the upper RGB, subgiant branch, and main sequence turnoff do not
have a strong effect on any of the final, integrated abundances,
though the input [$\alpha$/Fe] abundance may be important.

\begin{table*}
\centering
\begin{minipage}{165mm}
\caption{Abundance offsets with different isochrones.\label{table:DiffIsochrones}}
  \begin{tabular}{@{}lccccccccc@{}}
  \hline
& $\Delta[$\ion{Fe}{1}/H$]$ & $\Delta[$\ion{Fe}{2}/H$]$ & $\Delta[$\ion{Ca}{1}/\ion{Fe}{1}$]$ & $\Delta[$\ion{Ti}{1}/\ion{Fe}{1}$]$ & $\Delta[$\ion{Ti}{2}/\ion{Fe}{2}$]$ & $\Delta[$\ion{Ni}{1}/\ion{Fe}{1}$]$ & $\Delta[$\ion{Ba}{2}/\ion{Fe}{2}$]$ & $\Delta[$\ion{Eu}{2}/\ion{Fe}{2}$]$ \\
\hline
47 Tuc & & & & & & & & & \\
Victoria-Regina           & $+$0.02 & $+$0.03 & $-$0.01 & $+$0.01 & $ $0.0  & $ $0.0  & $ $0.0  & $ $0.0  \\
DSED [$\alpha$/Fe] $=0.2$ & $+$0.02 & $-$0.01 & $+$0.01 & $+$0.02 & $+$0.01 & $ $0.0  & $+$0.03 & $+$0.01 \\
DSED [$\alpha$/Fe] $=0.4$ & $+$0.03 & $+$0.02 & $-$0.01 & $+$0.02 & $ $0.0  & $+$0.01 & $+$0.02 & $+$0.02 \\
& & & & & & & & & \\
M3 & & & & & & & & & \\
Victoria-Regina           & $-$0.01       & $+$0.04       & $-$0.01 & $-$0.01       & $-$0.01       & $+$0.01 & $-$0.01       & $-$0.01 \\
DSED [$\alpha$/Fe] $=0.2$ & {\bf $-$0.15} & {\bf $-$0.06} & $+$0.02 & {\bf $-$0.09} & {\bf $-$0.05} & $-$0.04 & {\bf $-$0.11} & {\bf $-$0.07} \\
DSED [$\alpha$/Fe] $=0.4$ & {\bf $-$0.05} & $+$0.04       & $-$0.02 & $-$0.01       & $-$0.04       & $+$0.03 & {\bf $-$0.06} & $-$0.03 \\
& & & & & & & & & \\
M15 & & & & & & & & & \\
Victoria-Regina           & $-$0.01       & $+$0.04       & $ $0.0        & $-$0.04       & $ $0.0        & $-$0.02 & $ $0.0        & $-$0.02 \\
DSED [$\alpha$/Fe] $=0.2$ & {\bf $-$0.35} & {\bf $-$0.14} & {\bf $+$0.13} & {\bf $+$0.07} & {\bf $-$0.14} & $-$0.01 & {\bf $-$0.27} & {\bf $-$0.14} \\
DSED [$\alpha$/Fe] $=0.4$ & {\bf $+$0.05} & $+$0.04       & $-$0.02       & $-$0.03       & $+$0.01       & $ $0.0  & {\bf $+$0.05} & $+$0.04 \\
& & & & & & & & & \\
\hline
\end{tabular}
\end{minipage}\\
\medskip
\raggedright {\bf Notes: } Abundance differences are calculated
relative to the abundances derived with the BaSTI isochrones and
resolved boxes of the HB and AGB (see the text).\\
\end{table*}

\subsection{Populating an Isochrone}\label{subsec:Sampling}
An isochrone provides the temperature and surface gravity at certain
mass intervals for a cluster of a given age and chemical composition.
To determine the integrated abundances of a cluster, ILABUNDS must
also know the number of stars in each mass bin.  Stars are assigned to
each mass bin assuming that the stellar masses are distributed
according to an initial mass function (IMF; Appendix
\ref{subsubsec:IMF}), with the total number of stars determined from
a cluster's total absolute $V$ magnitude (Appendix \ref{subsubsec:Mv}).

\subsubsection{IMF}\label{subsubsec:IMF}
For their analyses of unresolved systems, MB08 and Colucci et
al. utilize a \citet{Kroupa2002} IMF.  However, other forms of the IMF
exist, for example the \citet{Salpeter1955} and \citet{Chabrier2003}
IMFs, which differ most from the Kroupa IMF at the high mass
end ($M \ga 0.5 {\rm M}_{\sun}$). These alternate IMFs are used to
assign stars to the best-fitting HRDs from Appendix
\ref{subsec:IsochroneOffsets}.  The abundance differences are shown in
Table \ref{table:IMFs}.  The different IMFs have no significant effect
on M3.  For 47~Tuc, the Salpeter IMF only significantly alters
the [\ion{Ba}{2}/\ion{Fe}{2}] abundance (by $0.05$ dex), while the
Chabrier IMF has a $0.05 \la \Delta [\rm{X/Fe}] < 0.1$ dex effect on
[\ion{Ti}{1}/\ion{Fe}{1}], [\ion{Ti}{2}/\ion{Fe}{2}],
[\ion{Ba}{2}/\ion{Fe}{2}], and [\ion{Eu}{2}/\ion{Fe}{2}].  The
[\ion{Fe}{1}/H] ratio is also slightly affected by the Chabrier
IMF. M15's [\ion{Ti}{1}/\ion{Fe}{1}] ratios are affected by both
IMFs.  These results suggest that \ion{Fe}{1}, Ti, Ba, and Eu are
sensitive to the sampling of the highest mass stars.

\begin{table*}
\centering
\begin{minipage}{165mm}
\caption{Abundance differences as a result of the input IMF.\label{table:IMFs}}
  \begin{tabular}{@{}lccccccccc@{}}
  \hline
& $\Delta[$\ion{Fe}{1}/H$]$ & $\Delta[$\ion{Fe}{2}/H$]$ & $\Delta[$\ion{Ca}{1}/\ion{Fe}{1}$]$ & $\Delta[$\ion{Ti}{1}/\ion{Fe}{1}$]$ & $\Delta[$\ion{Ti}{2}/\ion{Fe}{2}$]$ & $\Delta[$\ion{Ni}{1}/\ion{Fe}{1}$]$ & $\Delta[$\ion{Ba}{2}/\ion{Fe}{2}$]$ & $\Delta[$\ion{Eu}{2}/\ion{Fe}{2}$]$ \\
\hline
47 Tuc & & & & & & & & & \\
Salpeter & $-$0.04 & $ $0.0  & $-$0.01 &      $-$0.04  &      $-$0.04  & $ $0.0  & {\bf $-$0.05} & $-$0.03 \\
Chabrier & $-$0.04 & $-$0.01 & $-$0.02 & {\bf $-$0.06} & {\bf $-$0.05}& $-$0.02 & {\bf $-$0.07} & {\bf $-$0.05}\\
& & & & & & & & & \\
M3 & & & & & & & & & \\
Salpeter & $+$0.02 & $ $0.0  & $ $0.0  & $+$0.04       & $+$0.01       & $+$0.01 & $+$0.02       &  $+$0.02\\
Chabrier & $+$0.01 & $-$0.01 & $ $0.0  & $+$0.03       & $+$0.01       & $+$0.01 & $+$0.02       &  $+$0.02\\
& & & & & & & & & \\
M15 & & & & & & & & & \\
Salpeter & $+$0.03 & $ $0.0  & $-$0.01 & {\bf $-$0.08} &  $+$0.01      & $+$0.01 & $+$0.02       &  $+$0.02\\
Chabrier & $+$0.01 & $ $0.0  & $ $0.0  & {\bf $-$0.06} &  $ $0.0       & $+$0.02 & $ $0.0        &  $+$0.02\\
& & & & & & & & & \\
\hline
\end{tabular}
\end{minipage}\\
\medskip
\raggedright {\bf Notes: } Abundance differences are calculated
relative to the best-fitting HRD abundances in Table
\ref{table:Isochrones}.\\
\end{table*}

\subsubsection{Total Magnitude}\label{subsubsec:Mv}
The total magnitude of the observed portion of the GC, $M_{V,{\rm
obs}}$, determines the total number of stars in the populated HRD.
Fainter GCs will have fewer stars to populate the HRD; certain boxes
along the isochrone may then have no stars while others may be
rounded up to one star, and the relative flux contributions from the
boxes will be disrupted.  This is shown in Table \ref{table:Mv}, where
the abundance differences from the best-fitting HRD values are shown
when different values of $M_{V,\rm{obs}}$ are considered.

It is clear from Table \ref{table:Mv} that lowering the total
magnitude (i.e. making the cluster brighter) only leads to small
offsets ($\la 0.1$ dex) while making the cluster fainter can lead to
large offsets in the [\ion{Fe}{1}/H], [\ion{Ti}{1}/\ion{Fe}{1}],
[\ion{Ba}{2}/\ion{Fe}{2}], and [\ion{Eu}{2}/\ion{Fe}{2}] ratios ($0.1
< \Delta [\rm{X/Fe}] < 0.4$ dex, with the largest differences
occurring for M13 and M15).   The [\ion{Fe}{2}/H],
[\ion{Ca}{1}/\ion{Fe}{1}], [\ion{Ti}{2}/\ion{Fe}{2}], and
[\ion{Ni}{1}/\ion{Fe}{1}] ratios are somewhat affected ($\la 0.1$ dex)
when the GC is made fainter.  These abundance differences are driven
by how the isochrone is populated, such that fainter GCs cannot
adequately populate the upper RGB. 

This test indicates that fainter clusters will be more susceptible to
abundance offsets if the cluster $M_V$ is not well constrained.  These
problems can be reduced by using photometry of the bright RGB, AGB,
and HB stars, and/or by sampling as much of the GC as possible.
However, additional tests show that these errors can be dramatically
reduced if {\it fractional} stars are used to populate the HRDs,
instead of integer numbers of stars.  Although this choice is
distinctly non-physical it seems to work for IL spectra of bright,
well-sampled GCs.  Whether it will be applicable to real,
intrinsically poorly-sampled GCs is uncertain.

\begin{table*}
\centering
\begin{minipage}{165mm}
\caption{Abundance offsets when the GC total magnitude is adjusted.\label{table:Mv}}
  \begin{tabular}{@{}lccccccccc@{}}
  \hline
& $\Delta[$\ion{Fe}{1}/H$]$ & $\Delta[$\ion{Fe}{2}/H$]$ & $\Delta[$\ion{Ca}{1}/\ion{Fe}{1}$]$ & $\Delta[$\ion{Ti}{1}/\ion{Fe}{1}$]$ & $\Delta[$\ion{Ti}{2}/\ion{Fe}{2}$]$ & $\Delta[$\ion{Ni}{1}/\ion{Fe}{1}$]$ & $\Delta[$\ion{Ba}{2}/\ion{Fe}{2}$]$ & $\Delta[$\ion{Eu}{2}/\ion{Fe}{2}$]$ \\
\hline
47 Tuc & & & & & & & & & \\
$\Delta M_{V,{\rm obs}} = +1$   & $-$0.03 & $+$0.02 & $-$0.03 & {\bf $-$0.06} & {\bf $-$0.05} & $-$0.02 & {\bf $-$0.07} & {\bf $-$0.05} \\
$\Delta M_{V,{\rm obs}} = +0.5$ & $+$0.04 & $ $0.0  & $+$0.01 & {\bf $+$0.05} &      $+$0.03  & $+$0.01 & $+$0.04       & $+$0.03 \\
$\Delta M_{V,{\rm obs}} = -0.5$ & $+$0.02 & $ $0.0  & $ $0.0  &      $+$0.01  &      $+$0.02  & $ $0.0  & $+$0.02       & $+$0.01 \\
$\Delta M_{V,{\rm obs}} = -1$   & $+$0.01 & $ $0.0  & $-$0.01 &      $ $0.0   &      $+$0.01  & $ $0.0  & $ $0.0        & $ $0.0 \\
& & & & & & & & & \\
M3 & & & & & & & & & \\
$\Delta M_{V,{\rm obs}} = +1$   & {\bf $+$0.36} & {\bf $+$0.05} & $ $0.0  & {\bf $+$0.41} & {\bf $+$0.14} & {\bf $+$0.10} & {\bf $+$0.33} & {\bf $+$0.23} \\
$\Delta M_{V,{\rm obs}} = +0.5$ & {\bf $+$0.21} & $+$0.02       & $+$0.02 & {\bf $+$0.29} & {\bf $+$0.08} & {\bf $+$0.06} & {\bf $+$0.20} & {\bf $+$0.14} \\
$\Delta M_{V,{\rm obs}} = -0.5$ & {\bf $+$0.06} & $ $0.0        & $+$0.01 & {\bf $+$0.10} &      $+$0.03  & $+$0.02       & {\bf $+$0.06} & {\bf $+$0.05} \\
$\Delta M_{V,{\rm obs}} = -1$   & {\bf $+$0.06} & $+$0.02       & $ $0.0  & {\bf $+$0.05} &      $+$0.02  & $+$0.01       & $+$0.04       & $+$0.02 \\
& & & & & & & & & \\
M15 & & & & & & & & & \\
$\Delta M_{V,{\rm obs}} = +1$   & {\bf $+$0.25} & {\bf $+$0.10} & {\bf $-$0.10} & {\bf $+$0.05} & {\bf $+$0.07} & {\bf $+$0.07} & {\bf $+$0.20} & {\bf $+$0.15} \\
$\Delta M_{V,{\rm obs}} = +0.5$ & {\bf $+$0.12} & $+$0.04       & {\bf $-$0.05} & $+$0.03       &      $+$0.03  & $+$0.04       & {\bf $+$0.10} & {\bf $+$0.07} \\
$\Delta M_{V,{\rm obs}} = -0.5$ & {\bf $-$0.11} & $ $0.0        & $+$0.02       & {\bf $-$0.21} &      $-$0.02  & {\bf $-$0.07} & {\bf $-$0.10} & {\bf $-$0.08} \\
$\Delta M_{V,{\rm obs}} = -1$   & $-$0.03       & $+$0.02       & $ $0.0        & {\bf $-$0.14} &      $ $0.0   & $-$0.04       & $-$0.03       & $-$0.03 \\
& & & & & & & & & \\
\hline
\end{tabular}
\end{minipage}\\
\medskip
\raggedright {\bf Notes: } Abundance differences are calculated
relative to the best-fitting HRD abundances in Table
\ref{table:Isochrones}.  Note that most of the abundance offsets for
the low magnitude clusters are dramatically reduced if fractional
stars are used to populate the HRDs.\\
\end{table*}

\subsection{Horizontal Branch Morphology}\label{subsec:HBMorphology}
As discussed in Paper I, it is difficult to model the HBs of
unresolved GCs, given the uncertain effects of the ``second
parameter'' \citep{Dotter2008,Dotter2010}.  Synthetic HBs with a range
of morphologies can be generated (e.g. from the BaSTI data base), but
require inputs for the average HB mass and the spread in HB masses,
both of which are not known \textit{a priori} and may not exactly
match the true HB stars.  In particular, if blue HB stars are not
properly accounted for, spectroscopic ages will be skewed to younger
ages to compensate for the absence of the hot stars
(e.g. \citealt{Lee2000,Ocvirk2010}).  At high resolution, MB08 argued
that blue HB stars could also confuse trends in \ion{Fe}{1} abundances
with EP, leading to incorrect [Fe/H] and age determinations.

It is therefore possible that HB morphology could measurably affect
the derived chemical abundances. Lower resolution studies have
concluded that IL spectral features can help constrain HB morphology,
e.g. the Balmer line ratios \citep{Schiavon2004} or specific indices
from ionized atoms (e.g. the \ion{Mg}{2} doublet at 2800 \AA
\hspace{0.025in} or the \ion{Ca}{2} H and K index;
\citealt{PercivalSalaris2011}). However, the IL spectra presented here
do not extend blueward enough to access these features.

The purpose of the tests presented below is not to identify or test
the best way to constrain HB morphology, but to isolate and examine
the abundance effects from HB morphology.  With an M31 GC at
$[\rm{Fe/H}] = -2.2$, \citet{Colucci2009} tested the effects of HB
morphology by manually moving red HB stars to blue HB star boxes in
their best-fitting HRDs.  For that particular GC, they found that
individual \ion{Fe}{1} abundances changed by $<0.05$ dex and that the
effect on the best-fitting isochrone parameters was negligible.  Here
these results are tested on the Galactic GCs.

\subsubsection{The Direct Effects of HB Stars on Abundances}\label{subsubsec:HBDirect}

To test the direct effects of HB morphology on chemical abundances,
the IL spectra and resolved photometry of the second parameter triad
M3, M13, and NGC~7006 are used.  The HB boxes for the three GCs are
swapped, while maintaining the same total number of HB stars for each
cluster. Worst case scenarios of purely red and purely blue HBs were
also considered for M13 and NGC~7006, respectively.  Finally,
synthetic HBs from the BaSTI data base were assigned to M13 and
NGC~7006, using masses of 0.5 and 0.8 $\rm{M}_{\sun}$ and mass
dispersions of 0.02 $\rm{M}_{\sun}$.  These differences are shown in
Table \ref{table:HB}; they are first organized by GC, then by HB
morphology.

Table \ref{table:HB} shows that:
\begin{enumerate}
\item The slight differences between M3 and NGC~7006's HBs lead to
negligible abundance offsets.
\item HBs that are too red raise the integrated [\ion{Fe}{1}/H], while
HBs that are too blue lower the [\ion{Fe}{1}/H].  The largest
differences are $\sim 0.1$ dex.
\item The [\ion{Fe}{2}/H] ratios are most affected when red HB stars
are added (or when intermediate HB stars are removed).  The largest
offsets are $\sim 0.2$ dex.
\item The [\ion{Ca}{1}/\ion{Fe}{1}] and [\ion{Ni}{1}/\ion{Fe}{1}]
ratios are mostly unaffected by HB morphology.
\item The [Ti/Fe] ratios are significantly affected only when the HBs
are significantly different from reality (e.g. the pure and synthetic
red cases for M13).
\item   The total offsets in [\ion{Ba}{2}/\ion{Fe}{2}] are $\la 0.1$
dex.  HBs that are too blue lower the output
[\ion{Ba}{2}/\ion{Fe}{2}].  However, HBs that are too red do not
always raise [\ion{Ba}{2}/\ion{Fe}{2}], because of the varying effects
on \ion{Fe}{2}.  When intermediate HB stars are added to M13, the
[\ion{Ba}{2}/\ion{Fe}{2}] ratio is increased; when they are removed or
altered in M3 and NGC~7006 [\ion{Ba}{2}/\ion{Fe}{2}] is decreased.  It
therefore appears that [\ion{Ba}{2}/\ion{Fe}{2}] is most affected by
the presence or absence of intermediate HB stars.  This is not only
driven by the [\ion{Fe}{2}/H] abundance.
\item HBs that are too blue lower [\ion{Eu}{2}/\ion{Fe}{2}], while
redder HBs raise [\ion{Eu}{2}/\ion{Fe}{2}].  Again, these effects are
not driven by [\ion{Fe}{2}/H] differences.
\end{enumerate}

The alternate HBs also affect the trends of \ion{Fe}{1} abundances
with wavelength, REW, and EP, such that the slopes are generally made
steeper when the HB is improperly modelled.  These slope changes imply
that different HB models will lead to alternate best-fitting
isochrones.

\begin{table*}
\centering
\begin{minipage}{165mm}
\caption{Abundance differences as a result of HB
morphology.\label{table:HB}}
  \begin{tabular}{@{}lccccccccc@{}}
  \hline
& & $\Delta[$\ion{Fe}{1}/H$]$ & $\Delta[$\ion{Fe}{2}/H$] $ & $\Delta[$\ion{Ca}{1}/\ion{Fe}{1}$]$ & $\Delta[$\ion{Ti}{1}/\ion{Fe}{1}$]$ & $\Delta[$\ion{Ti}{2}/\ion{Fe}{2}$]$ & $\Delta[$\ion{Ni}{1}/\ion{Fe}{1}$]$ & $\Delta[$\ion{Ba}{2}/\ion{Fe}{2}$]$ & $\Delta[$\ion{Eu}{2}/\ion{Fe}{2}$]$\\
\hline
M3   & & & & & & & & & \\
M13's HB      &  & {\bf $-$0.05} & $ $0.0        & $ $0.0  & $ $0.0        & $-$0.02       & $-$0.01 & {\bf $-$0.06} & $-$0.02 \\
NGC~7006's HB &  & $ $0.0        & $ $0.0        & $ $0.0  & $+$0.01       & $-$0.01       & $+$0.01 & $-$0.01       & $+$0.01 \\
& & & & & & & & & \\
M13   & & & & & & & & & \\
Purely blue HB&  & $-$0.04       & $-$0.02       & $ $0.0  & $+$0.01       & $-$0.03       & $+$0.01       & $-$0.04       & $+$0.01 \\
Synthetic Blue&  & {\bf $-$0.06} & {\bf $-$0.09} & $+$0.02 & $-$0.01       & $+$0.01       & $-$0.04       & $-$0.03       & $-$0.02 \\
M3's HB       &  & {\bf $+$0.06} & $-$0.01       & $-$0.01 & $+$0.01       & $+$0.02       & $ $0.0        & {\bf $+$0.05} & {\bf $+$0.05} \\
NGC 7006's HB &  & {\bf $+$0.07} & $-$0.01       & $-$0.01 & $+$0.01       & $ $0.0        & $ $0.0        & {\bf $+$0.05} & {\bf $+$0.06} \\
Synthetic Red &  & $+$0.02       & {\bf $-$0.17} & $-$0.01 & {\bf $+$0.05} & {\bf $-$0.06} & $+$0.02       & $+$0.01       & {\bf $+$0.11} \\
Purely red HB &  & $+$0.02       & {\bf $-$0.08} & $-$0.02 & {\bf $+$0.08} & {\bf $-$0.07} & $+$0.04       & $-$0.01       & {\bf $+$0.10} \\
& & & & & & & & & \\
NGC 7006   & & & & & & & & & \\
Purely blue HB&  & {\bf $-$0.11} & $-$0.02       & $-$0.01 & $+$0.03       & {\bf $-$0.07} & $+$0.02 & {\bf $-$0.11} & {\bf $-$0.04} \\
Synthetic Blue&  & {\bf $-$0.08} & $-$0.02       & $ $0.0  & $+$0.02       & $-$0.03       & $+$0.01 & {\bf $-$0.07} & $-$0.03 \\
M13's HB      &  & {\bf $-$0.07} & $ $0.0        & $ $0.0  & $+$0.01       & $-$0.03       & $ $0.0  & {\bf $-$0.07} & {\bf $-$0.05} \\
M3's HB       &  & $ $0.0        & $ $0.0        & $ $0.0  & $+$0.01       & $ $0.0        & $ $0.0  & $-$0.01       & $-$0.02 \\
Synthetic Red &  & $-$0.03       & {\bf $-$0.05} & $ $0.0  & $+$0.04       & $-$0.03       & $+$0.03 & $-$0.03       & $+$0.01 \\
Purely red HB &  & $-$0.04       & $-$0.04       & $-$0.01 & {\bf $+$0.06} & {\bf $-$0.05} & $+$0.03 & {\bf $-$0.05} & $+$0.01 \\
& & & & & & & & & \\
\hline
\end{tabular}
\end{minipage}\\
\medskip
\raggedright {\bf Notes: } Abundance differences are calculated
relative to the baseline abundances in Table
\ref{table:OriginalAbunds}, as described in Section
\ref{subsec:SystematicErrors}.  Tests are organized by cluster, then
by HB morphology, with the bluest HBs listed first.\\
\end{table*}

\subsubsection{The Indirect Effects of HB Stars on Isochrone Age and [Fe/H]}\label{subsubsec:HBIndirect}
To test the indirect effects of HB stars on isochrone age and [Fe/H],
the default HBs were replaced with extremely blue and extremely red
synthetic HBs (from the BaSTI synthetic HB generator), and ILABUNDS was
rerun on the new populations.  The parameters of the new isochrones
and the subsequent abundance offsets are shown in Table
\ref{table:HRIndirect}.

These results show that for the GCs with intermediate HB morphologies
(M3 and NGC~7006), HBs that are too red lead to underpredictions of
the GC age (most likely to compensate for the lack of hot, blue stars
in the models) while HBs that are too blue lead to overpredictions of
the GC age (likely for the opposite reason).  These findings agree
well with the findings of \citet{Lee2000,Ocvirk2010}, i.e. that when
blue HB stars are not properly accounted for, IL analyses will
converge on ages that are too young.  For M13 and M15 (the clusters
with blue HBs) the extreme blue and extreme red cases both converge on
old ages.  To understand this effect, the default HB morphologies of
the original best-fitting isochrones must be investigated.  For M13
and M15 the original HBs are significantly redder than the real HBs;
M15's default HB also extends slightly blueward of the synthetic red
HB tested here. The fact that the synthetic pure red and blue HBs both
push the isochrones to old ages suggests that the presence (or
absence) of intermediate HB stars have a more significant effect than
the bluest HB stars.  This agrees with the findings of
\citet{Colucci2009}, who tested these effects on a GC with both blue
and red HB stars and found that the bluest HB stars had a negligible
effect. Thus, convergence on a correct age (within $\sim 5$ Gyr)
requires modelling the intermediate age HB stars (at least
approximately) correctly.

However, regardless of how the HBs are modelled, all isochrones
converge on reasonable isochrone metallicities.  Furthermore, certain
abundance ratios are relatively insensitive to the adopted isochrone
age.  While [\ion{Fe}{1}/H], [\ion{Ti}{1}/\ion{Fe}{1}],
[\ion{Ba}{2}/\ion{Fe}{2}], and [\ion{Eu}{2}/\ion{Fe}{2}] are very
sensitive to changes in HB morphology (with offsets $\ga 0.1$ dex),
[\ion{Ca}{1}/\ion{Fe}{1}], [\ion{Ti}{2}/\ion{Fe}{2}], and
[\ion{Ni}{1}/\ion{Fe}{1}] are much less sensitive, with offsets $<0.1$
dex.

\begin{table*}
\centering
\begin{minipage}{165mm}
\caption{Abundance differences and parameters of the best-fitting HRDs
when synthetic HBs are used.\label{table:HRIndirect}}
  \begin{tabular}{@{}lccccccccccc@{}}
  \hline
& age & [Z/H] & $\Delta[$\ion{Fe}{1}/H$]$ & $\Delta[$\ion{Fe}{2}/H$] $ &
$\Delta[$\ion{Ca}{1}/\ion{Fe}{1}$]$ & $\Delta[$\ion{Ti}{1}/\ion{Fe}{1}$]$ & $\Delta[$\ion{Ti}{2}/\ion{Fe}{2}$]$ & $\Delta[$\ion{Ni}{1}/\ion{Fe}{1}$]$ & $\Delta[$\ion{Ba}{2}/\ion{Fe}{2}$]$ & $\Delta[$\ion{Eu}{2}/\ion{Fe}{2}$]$ \\
\hline
M3 & & & & & & & & & \\
Red HB   &  8 & -0.96 & {\bf $+$0.11} & {\bf $+$0.05} & $-$0.02 & {\bf $+$0.15} & $ $0.0        & {\bf $+$0.07} & {\bf $+$0.11} & {\bf $+$0.12}\\
Blue HB  & 14 & -1.27 & {\bf $+$0.08} & $+$0.01       & $+$0.02 & {\bf $+$0.16} & {\bf $+$0.07} & $+$0.03       & {\bf $+$0.12} & {\bf $+$0.06}\\
 & & & & & & & & & \\
M13 & & & & & & & & & \\
Red HB   & 12 & -1.27 & {\bf $+$0.18} & $-$0.02 & $-$0.02 & {\bf $+$0.20} & {\bf $+$0.05} & {\bf $+$0.05} & {\bf $+$0.16} & {\bf $+$0.13}\\
Blue HB  & 14 & -1.27 & {\bf $+$0.10} & $+$0.04 & $ $0.0  & {\bf $+$0.10} & {\bf $+$0.08} & $ $0.0        & {\bf $+$0.10} & $+$0.02\\
 & & & & & & & & & \\
NGC~7006 & & & & & & & & & \\
Red HB   &  5 & -0.96 & {\bf $+$0.13} & {\bf $+$0.06} & $ $0.0  & {\bf $+$0.10} & $+$0.03 & {\bf $+$0.05} & {\bf $+$0.14} & {\bf $+$0.09}\\
Blue HB  & 14 & -0.96 & $+$0.02       & {\bf $+$0.28} & $-$0.03 & {\bf $-$0.12} & $+$0.01 & $+$0.01       & $-$0.04       & {\bf $+$0.08}\\
 & & & & & & & & & \\
M15 & & & & & & & & & \\
Red HB   & 14 & -1.79 & {\bf $+$0.06} & $-$0.02 & $-$0.03 & {\bf $+$0.06} & $+$0.01 & {\bf $+$0.07} & $+$0.03 & {\bf $+$0.08}\\
Blue HB  & 14 & -1.79 & $+$0.01       & $+$0.03 & $ $0.0  & $+$0.02       & $+$0.02 & $ $0.0        & $ $0.0  & $-$0.01\\
& & & & & & & & & \\
\hline
\end{tabular}
\end{minipage}\\
\medskip
\raggedright {\bf Notes: } Abundance differences are calculated
relative to the best-fitting HRD-based abundances in Table
\ref{table:Isochrones}.\\ 
\end{table*}

\subsection{Asymptotic Giant Branch Stars}\label{subsec:AGBstars}
With the BaSTI isochrones the AGB can be modelled in various ways.
First, different mass loss parameters of $\eta = 0.2$ or $\eta =
0.4$ can be selected.  Second, the AGB can be extended through all
thermal pulse phases or can be terminated after the first few pulses
(where the former is denoted as the ``Extended'' case and the latter
as the ``Normal'' case; see the BaSTI website).  Given their tests
with Galactic GCs \citep{Cameron2009}, Colucci et al. utilize Extended
AGB isochrones with $\eta = 0.2$.  This appendix investigates the
abundance offsets that arise when the other AGB prescriptions are
used.  Note that MB08 required an enhancement in the number of AGB
stars in order to match the observed luminosity function and
abundances of 47~Tuc.  That enhancement is not included here.

\subsubsection{The Direct Effects of AGB Stars on Abundances}\label{subsubsec:AGBIndirect}
The AGB prescriptions were first altered while maintaining the
best-fitting isochrone parameters from Appendix
\ref{subsubsec:CMDcomparison}.  These offsets are shown in Table
\ref{table:AGB}.  The AGB prescription has a small effect on 47~Tuc's
abundances, and a much larger effect on M3 and M15's abundances.  The
ratios that are most affected by the AGB models are [\ion{Fe}{1}/H],
[\ion{Ti}{1}/\ion{Fe}{1}], [\ion{Ba}{2}/\ion{Fe}{2}], and
[\ion{Eu}{2}/\ion{Fe}{2}] (with offsets $\sim 0.1-0.2$ dex, depending
on the cluster), while [\ion{Fe}{2}/H], [\ion{Ti}{2}/\ion{Fe}{2}], and
[\ion{Ni}{1}/\ion{Fe}{1}] are occasionally affected (0.1~-~0.15 dex).
For all GCs, the [\ion{Ca}{1}/\ion{Fe}{1}] ratio is largely
insensitive ($\la 0.05$ dex) to the AGB prescription.

The abundance offsets are not the same for a given AGB prescription.
For 47~Tuc, only the normal, $\eta = 0.4$ case significantly alters
the abundances.  For M3 both of the normal AGB cases ($\eta = 0.2$
and $0.4$) lead to large offsets, while for M15 both $\eta = 0.4$
cases create significant offsets.  In some cases the various AGB 
prescriptions bring the HRD-based abundances into better agreement
with the CMD-based abundances; for example, normal AGBs raise M3's
[\ion{Ti}{1}/\ion{Fe}{1}] ratio; however, other abundance ratios are
then sometimes brought out of agreement.  Thus, the systematic
uncertainties from a given AGB prescription are not the same for
all clusters, and adopting a uniform treatment of the AGB will not
remove intra-cluster systematic offsets.  Without resolved photometry
of the brightest AGB stars, it would be difficult to determine which
AGB prescription is most representative of a given cluster.

\begin{table*}
\centering
\begin{minipage}{165mm}
\caption{Abundance differences from modelling the AGB.\label{table:AGB}}
  \begin{tabular}{@{}llccccccccc@{}}
  \hline
 & \multicolumn{2}{c}{Isochrone} & & & & & & & & \\
AGB$^{a}$ & Age$^{b}$ & [Fe/H] & $\Delta[$\ion{Fe}{1}/H$]$ & $\Delta[$\ion{Fe}{2}/H$] $ & $\Delta[$\ion{Ca}{1}/\ion{Fe}{1}$]$ & $\Delta[$\ion{Ti}{1}/\ion{Fe}{1}$]$ & $\Delta[$\ion{Ti}{2}/\ion{Fe}{2}$]$ & $\Delta[$\ion{Ni}{1}/\ion{Fe}{1}$]$ & $\Delta[$\ion{Ba}{2}/\ion{Fe}{2}$]$ & $\Delta[$\ion{Eu}{2}/\ion{Fe}{2}$]$\\
 & & & & & & & & & & \\
\hline
47 Tuc   & & & & & & & & & & \\
E-0.4 & 10$^{c}$  & -0.70 &  $+$0.01      & $+$0.02       & $-$0.03       & $-$0.04       & $-$0.01       & $-$0.01 & $-$0.03       & $-$0.03 \\
N-0.2 & 10$^{c}$  & -0.70 &  $+$0.03      & $+$0.02       & $-$0.01       & $+$0.01       & $+$0.01       & $+$0.01 & $+$0.02       & $+$0.01 \\
N-0.4 & 10$^{c}$  & -0.70 & {\bf $+$0.07} & $ $0.0        & $+$0.01       & {\bf $+$0.06} & {\bf $+$0.06} & $+$0.01 & {\bf $+$0.08} & {\bf $+$0.05} \\
E-0.4 & 11        & -0.70 & $+$0.03       & $ $0.0        & $-$0.01       & $+$0.01       & $+$0.01       & $+$0.01 & $+$0.02       & $+$0.02 \\
N-0.2 & 12        & -0.70 & $+$0.01       & $+$0.03       & $-$0.02       & $-$0.01       & $-$0.01       & $ $0.0  & $-$0.03       & $-$0.02 \\
N-0.4 & 11        & -0.60 & $+$0.03       & {\bf $+$0.09} & {\bf $-$0.05} & $-$0.04       & $-$0.02       & $+$0.01 & $-$0.04       & $-$0.03 \\
& & & & & & & & & & \\
M3       & & & & & & & & & & \\
E-0.4 & 9$^{c}$   & -1.62 & {\bf $+$0.06} & $+$0.03       & $ $0.0        & $ $0.0        & $+$0.04       & $+$0.02       & {\bf $+$0.05} & $+$0.01\\
N-0.2 & 9$^{c}$   & -1.62 & {\bf $+$0.14} & $+$0.01       & $+$0.02       & {\bf $+$0.23} & {\bf $+$0.06} & {\bf $+$0.05} & {\bf $+$0.15} & {\bf$+$0.10}\\
N-0.4 & 9$^{c}$   & -1.62 & {\bf $+$0.18} & $+$0.02       & $+$0.01       & {\bf $+$0.22} & {\bf $+$0.09} & {\bf $+$0.05} & {\bf $+$0.19} & {\bf$+$0.11}\\
E-0.4 & 10        & -1.62 & $+$0.04       & $+$0.02       & $ $0.0        & $ $0.0        & $+$0.03       & $+$0.01       & $+$0.03       & $ $0.0     \\
N-0.2 & 8         & -1.31 & {\bf $+$0.08} & {\bf $+$0.15} & $+$0.04       & {\bf $+$0.08} & $-$0.01       & {\bf $+$0.06} & {\bf $+$0.05} & {\bf$+$0.07}\\
N-0.4 & 10        & -1.31 & {\bf $+$0.10} & {\bf $+$0.15} & $-$0.03       & {\bf $+$0.09} & $ $0.0        & {\bf $+$0.07} & {\bf $+$0.08} & {\bf$+$0.08}\\
& & & & & & & & & & \\
M15      & & & & & & & & & & \\
E-0.4 & 9$^{c}$   & -2.14 & {\bf $-$0.16} & $ $0.0        & $+$0.03       & {\bf $-$0.38} & $-$0.01       & {\bf $-$0.13} & {\bf $-$0.14} & {\bf$-$0.13}\\
N-0.2 & 9$^{c}$   & -2.14 & $+$0.03       & $ $0.0        & $-$0.01       & $+$0.01       & $+$0.02       & $+$0.02       & $+$0.02       & $+$0.02\\
N-0.4 & 9$^{c}$   & -2.14 & {\bf $+$0.13} & $+$0.03       & $-$0.03       & {\bf $-$0.06} & {\bf $+$0.08} & $-$0.03       & {\bf $+$0.11} & $+$0.04\\
E-0.4 & 10        & -2.62 & {\bf $-$0.19} & $+$0.01       & $+$0.03       & {\bf $-$0.06} & $-$0.04       & {\bf $-$0.11} & {\bf $-$0.18} & {\bf$-$0.14}\\
N-0.2 & 8         & -2.14 & $+$0.01       & $ $0.0        & $ $0.0        & $+$0.02       & $+$0.01       & $+$0.01       & $ $0.0        & $+$0.01\\
N-0.4 & 10        & -2.62 & {\bf $+$0.10} & $+$0.02       & $-$0.03       & {\bf $-$0.09} & {\bf $+$0.05} & $ $0.0        & $+$0.07       & $+$0.04\\
& & & & & & & & & & \\
\hline
\end{tabular}
\end{minipage}\\
\medskip
\raggedright {\bf Notes: } Abundance differences are calculated
relative to the best-fitting HRD abundances in Table
\ref{table:Isochrones}, which were determined with Extended, $\eta =
0.2$ isochrones.  Tests are organized by cluster, then by AGB
prescription, with the original isochrone age and [Fe/H] listed first,
followed by the new, best-fitting HRD.\\
\raggedright $^{a}$ The AGB prescription indicates which BaSTI
isochrones were utilized.  ``E-0.4'' denotes extended, $\eta = 0.4$
isochrones, ``N-0.2'' denotes normal, $\eta = 0.2$ isochrones, and
``N-0.4'' denotes normal, $\eta = 0.4$ isochrones.\\
\raggedright $^{b}$ Ages are in Gyr.\\
\raggedright $^{c}$ These tests utilized the original, best-fitting
isochrone parameters from Appendix \ref{subsubsec:CMDcomparison}.\\
\end{table*}

\subsubsection{The Indirect Effects of AGB Stars on Isochrone Age and [Fe/H]}\label{subsubsec:AGBIndirect}
To test how the AGB models affect the parameters for the best-fitting
HRDs, the isochrone parameters were allowed to vary.  These new
best-fitting parameters for each AGB prescription and the resulting
abundance differences are also shown in Table \ref{table:AGB}.  In all
cases new ages and/or metallicities are favoured, though they are not
significantly different from the original values.  This indicates that
the AGB prescription is not responsible for the young isochrone ages
for M3 and M15.

When the new isochrone parameters are selected for a given AGB
treatment, the abundances are generally brought into \textit{slightly}
better agreement with the original HRD-based abundances, particularly
for 47~Tuc.  For example, the large offsets in [\ion{Fe}{1}/H],
[\ion{Ti}{1}/\ion{Fe}{1}], [\ion{Ba}{2}/\ion{Fe}{2}], and
[\ion{Eu}{2}/\ion{Fe}{2}] from assuming the original age and [Fe/H]
are generally (though not always) reduced when new best-fitting HRDs
are adopted. However, in some cases the offsets are still quite large
(e.g. with M3's ``Normal'' AGBs), illustrating that the treatment of
the AGB could be problematic for high-resolution optical IL spectral
studies of unresolved GCs.

\subsection{Blue Stragglers}\label{subsec:BlueStragglers}
Isochrones do not contain models for blue stragglers (the stars that
appear to lie on the main sequence, blueward of the turnoff).  Though
there are few of these stars, they are brighter and hotter than main
sequence stars, and thus may have a non-negligible effect on the IL
spectral lines.  To test these effects the resolved blue straggler
boxes were included with the best-fitting isochrones.  The results are
shown in Table \ref{table:BS}, and are generally quite small, except
for a few cases where [\ion{Fe}{1}/H], [\ion{Fe}{2}/H],
[\ion{Ba}{2}/\ion{Fe}{2}], and [\ion{Eu}{2}/\ion{Fe}{2}] are affected
by up to 0.07 dex.  This suggests that the inclusion of blue
stragglers is not essential for the majority of elements, though the
singly ionized elements are mildly sensitive to them.  Furthermore,
the blue stragglers have only a slight effect on the \ion{Fe}{1}
trends with wavelength, REW, and EP, and therefore do not have a
significant effect on the isochrone age.

\begin{table*}
\centering
\begin{minipage}{165mm}
\caption{The effects of blue stragglers.\label{table:BS}}
  \begin{tabular}{@{}lccccccccc@{}}
  \hline
& $\Delta[$\ion{Fe}{1}/H$]$ & $\Delta[$\ion{Fe}{2}/H$] $ &
$\Delta[$\ion{Ca}{1}/\ion{Fe}{1}$]$ & $\Delta[$\ion{Ti}{1}/\ion{Fe}{1}$]$ & $\Delta[$\ion{Ti}{2}/\ion{Fe}{2}$]$ & $\Delta[$\ion{Ni}{1}/\ion{Fe}{1}$]$ & $\Delta[$\ion{Ba}{2}/\ion{Fe}{2}$]$ & $\Delta[$\ion{Eu}{2}/\ion{Fe}{2}$]$ \\
\hline
47~Tuc     &      $+$0.02  & $ $0.0        & $ $0.0  & $ $0.0  & $+$0.02 & $-$0.01 & $+$0.01       & $+$0.01\\
M3         &      $+$0.02  & {\bf $+$0.07} & $ $0.0  & $ $0.0  & $-$0.03 & $+$0.01 & $-$0.03       & {\bf $-$0.06}\\
M13        &      $+$0.03  & $+$0.04       & $-$0.02 & $-$0.02 & $-$0.02 & $-$0.02 & $-$0.03       & $-$0.03\\
NGC~7006   &      $+$0.02  & $+$0.01       & $+$0.01 & $-$0.01 & $+$0.02 & $-$0.01 & $+$0.02       & $ $0.0 \\
M15        & {\bf $+$0.07} & $+$0.02       & $-$0.02 & $-$0.04 & $-$0.04 & $-$0.03 & {\bf $+$0.05} & $+$0.01\\
& & & & & & & & & \\
\hline
\end{tabular}
\end{minipage}\\
\medskip
\raggedright {\bf Notes: } Abundance differences are calculated
relative to the best-fitting HRD abundances in Table
\ref{table:Isochrones}.\\
\end{table*}

\subsection{Lower mass cut-off}\label{subsec:LowMassCut-Off}
In their IL analysis of 47~Tuc, MB08 found that a lower mass cut-off
was necessary to reproduce the observed luminosity function
(ostensibly because the IL spectrum only covers the cluster core, and
mass segregation must be taken into account).  This appendix
investigates the effects of applying a lower mass cut-off such that all
stars fainter than $M_V = +4.7$ are removed from the synthetic
HRD---this was the cut-off adopted by MB08 to match 47~Tuc's observed
luminosity function. (Note that this test is essentially the opposite
of the incompleteness test in Appendix \ref{subsec:Incompleteness},
except that now new isochrones are identified.)  This cut-off was
applied to all the GCs, even though some of the IL spectra cover out
to further radii where there may still be fainter stars.  New
best-fitting isochrones were then identified.

The new isochrone parameters and the abundance offsets
from the original best-fitting HRDs are shown in Table
\ref{table:LowerMass}.  With the lower mass cut-off, the same
isochrones are identified for 47~Tuc, M3, and M15; for M13 a slightly
younger isochrone is preferred, while for NGC~7006 a more metal-rich,
younger isochrone is preferred.  Note that the slopes are never
sufficiently flat for M15, as with the original best-fitting HRD
(Appendix \ref{subsubsec:CMDcomparison}).  The [\ion{Fe}{1}/H],
[\ion{Ti}{1}/\ion{Fe}{1}], [\ion{Ba}{2}/\ion{Fe}{2}], and
[\ion{Eu}{2}/\ion{Fe}{2}] ratios are particularly affected (up to
$\sim 0.1-0.2$ dex) by the absence of the lowest mass stars.  However,
this may be because more high mass stars are needed to maintain the
same total cluster magnitude.

\begin{table*}
\centering
\begin{minipage}{165mm}
\caption{The effects of a lower mass cut-off.\label{table:LowerMass}}
  \begin{tabular}{@{}lccccccccccc@{}}
  \hline
& age & [Z/H] & $\Delta[$\ion{Fe}{1}/H$]$ & $\Delta[$\ion{Fe}{2}/H$] $ &
$\Delta[$\ion{Ca}{1}/\ion{Fe}{1}$]$ & $\Delta[$\ion{Ti}{1}/\ion{Fe}{1}$]$ & $\Delta[$\ion{Ti}{2}/\ion{Fe}{2}$]$ & $\Delta[$\ion{Ni}{1}/\ion{Fe}{1}$]$ & $\Delta[$\ion{Ba}{2}/\ion{Fe}{2}$]$ & $\Delta[$\ion{Eu}{2}/\ion{Fe}{2}$]$ \\
\hline
47~Tuc     & 10 & -0.35 & {\bf $-$0.05} & $ $0.0        & $-$0.04 & {\bf $-$0.07} & {\bf $-$0.07} & $-$0.03      & {\bf $-$0.09} & {\bf $-$0.12}\\
M3         &  9 & -1.27 & {\bf $+$0.10} & $-$0.03       & $+$0.02 & {\bf $+$0.24} & $+$0.03       & $+$0.04      & {\bf $+$0.11} & {\bf $+$0.10}\\
M13        & 11 & -1.27 & {\bf $+$0.10} & $-$0.02       & $-$0.01 & {\bf $+$0.16} & $+$0.02       & $+$0.01      & {\bf $+$0.07} & {\bf $+$0.08}\\
NGC~7006   &  5 & -0.96 & {\bf $+$0.09} & {\bf $+$0.12} & $ $0.0  & {\bf $+$0.10} & $ $0.0        &{\bf $+$0.05} & {\bf $+$0.08} & {\bf $+$0.09}\\
M15        &  9 & -1.79 & {\bf $-$0.13} & {\bf $-$0.05} & $+$0.04 & $-$0.01       & $-$0.04       & $-$0.01      & {\bf $-$0.11} & {\bf $-$0.05} \\
& & & & & & & & & \\
\hline
\end{tabular}
\end{minipage}\\
\medskip
\raggedright {\bf Notes: } Abundance differences are calculated
relative to the best-fitting HRD abundances in Table
\ref{table:Isochrones}.\\
\end{table*}

\subsection{Partially Resolved Clusters}\label{subsec:PartiallyResolved}

So far, the tests on HRD abundances have shown that the uncertainties
in HB, AGB, RGB, and lower main sequence stars can be prohibitively
large, with uncertainties as high as 0.4 dex in [\ion{Fe}{1}/H] and
[\ion{Ti}{1}/\ion{Fe}{1}], 0.3 dex in [\ion{Ba}{2}/\ion{Fe}{2}], 0.2
dex in [\ion{Fe}{2}/H], and 0.1 dex in [\ion{Ca}{1}/\ion{Fe}{1}],
[\ion{Ti}{2}/\ion{Fe}{2}], and [\ion{Ni}{1}/\ion{Fe}{1}] (depending on
the cluster). Observations of clusters outside of the Milky Way and
its dwarf satellite systems can provide photometry of the brightest
stars in a cluster; the HB morphology, AGB prescription, etc. can then
be characterized, eliminating or reducing many of these
uncertainties. However, even with partial photometry, stars fainter
than the HB still contribute a significant amount of light to the IL
spectra (see Tables 4-8 in Paper I).  Furthermore, stars in cluster
{\it cores} may not be resolvable.  Given the large errors associated
with sampling uncertainties, it may be preferable to model the stellar
populations with stellar isochrones that can be refined based on the
resolved photometry. This appendix investigates the effects of
combining observations of the upper CMD with models of the lower HRD.

The [Fe/H] of a partially resolved cluster can be estimated through
comparisons with Galactic GC fiducials (e.g. \citealt{Mackey2013}).
The [Fe/H] of the input isochrone can then be refined based on the
output from ILABUNDS, as for completely unresolved clusters.
Furthermore, a partially resolved GC's age can be somewhat constrained
from the upper CMD.  Here the ``best-fitting'' HRDs are found for all
target GCs, adopting the criterion that the isochrone must fit the
``observed portion'' of the CMD (taken to be the portion down to the
bottom of the HB).

\begin{figure*}
\begin{center}
\centering
\subfigure[47 Tuc]{\includegraphics[scale=0.35]{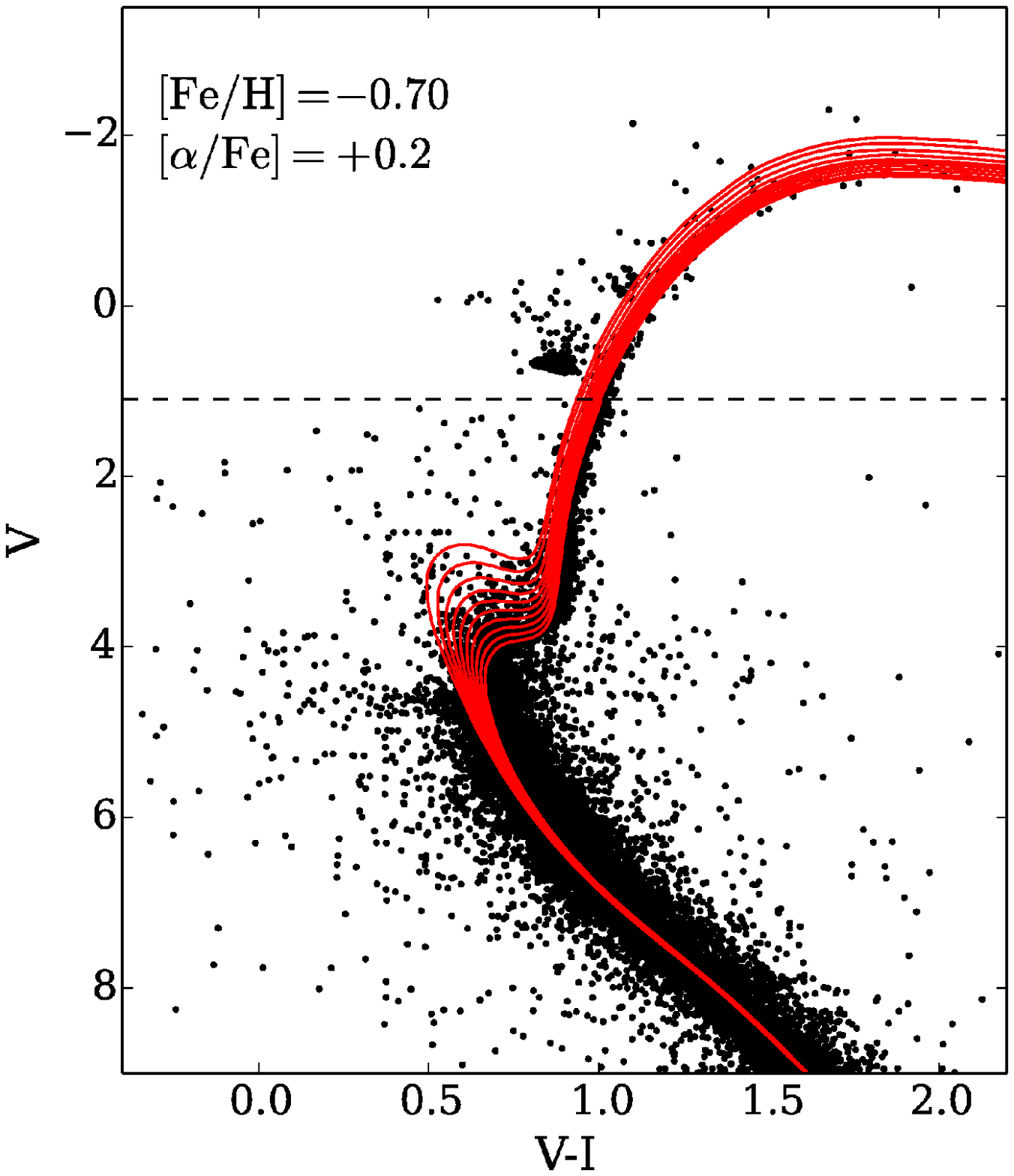}\label{fig:Partial47Tuc}}
\subfigure[M3]{\includegraphics[scale=0.35]{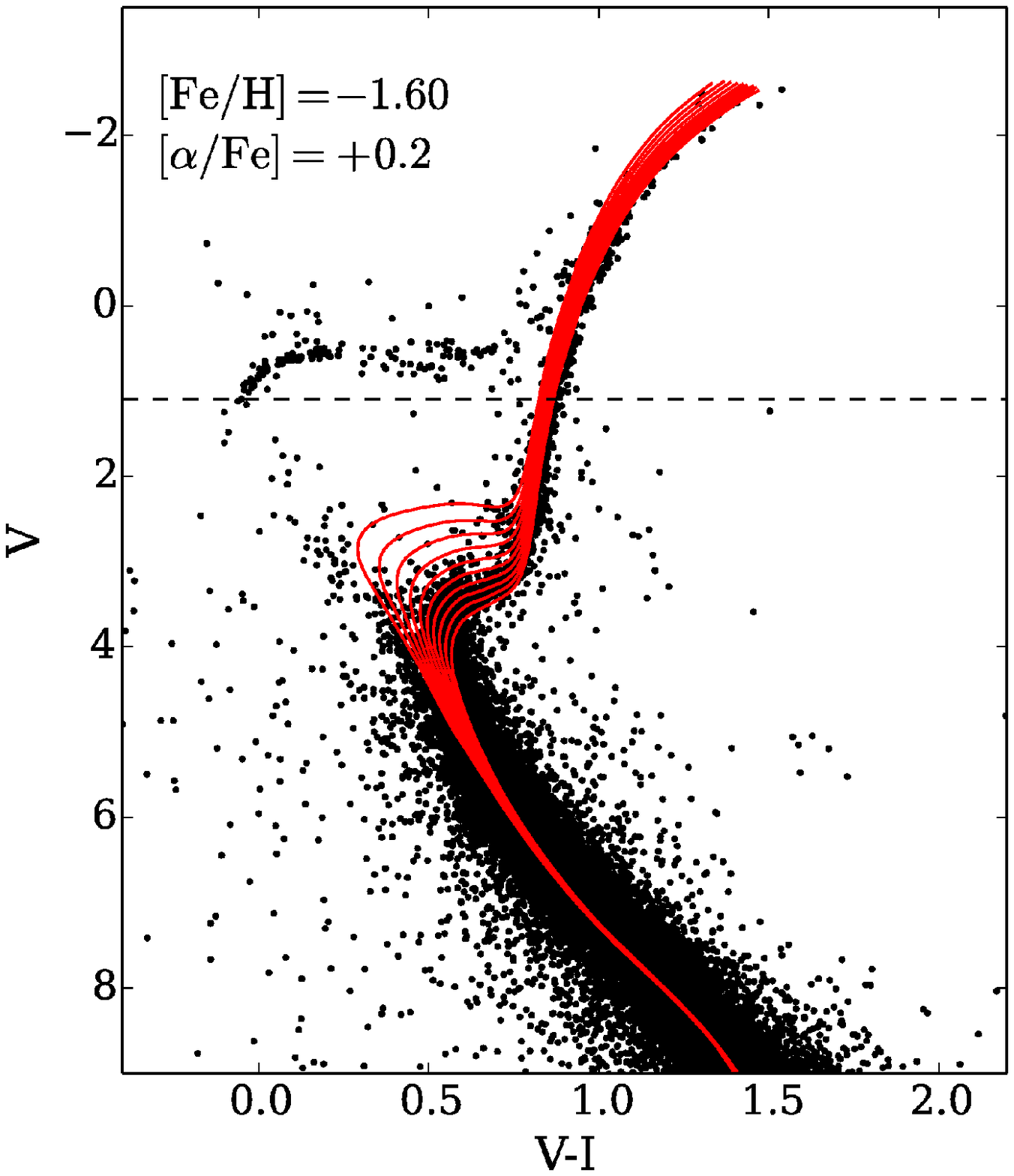}\label{fig:PartialM3}}
\subfigure[M15]{\includegraphics[scale=0.35]{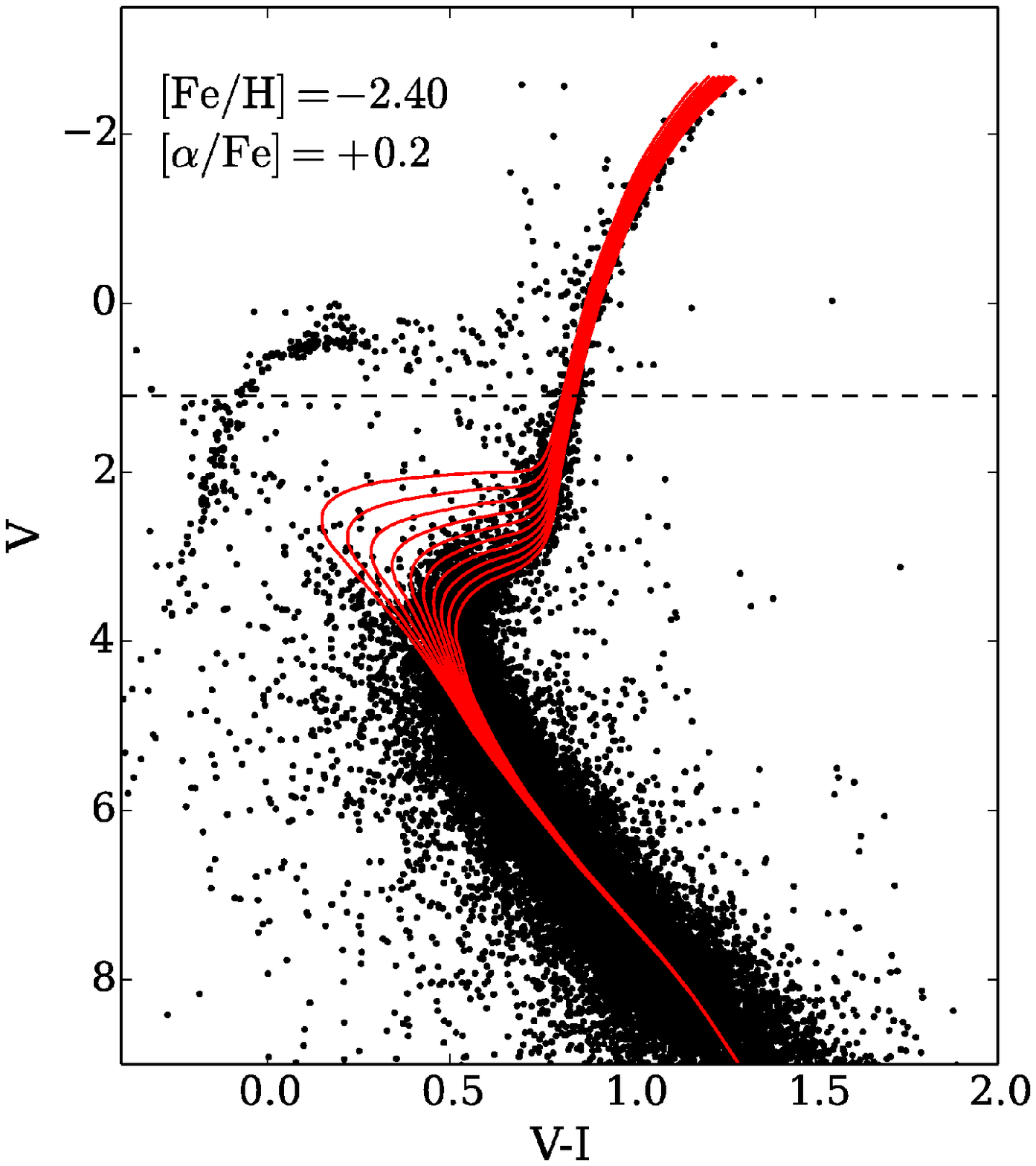}\label{fig:PartialM15}}
\caption{Examples of isochrones that might be used in an analysis of a
partially resolved cluster.  Here it is assumed that the GCs can only 
be observed to just below the HB, i.e. to the dashed line.   The
isochrones are from the DSED \citep{DSED} and have ages of 5, 6, 7, 8,
9, 10, 11, 12, 13, and 14 Gyr.\label{fig:PartiallyResolved}}
\end{center}
\end{figure*}

Initially, the best-fitting [Fe/H] values from the BaSTI isochrones
(see Appendix \ref{subsec:IsochroneOffsets}) were chosen, since they
fit the upper RGBs well (see Figure \ref{fig:PartiallyResolved}).
Synthetic HBs were selected to best match the observed HB.  The
isochrones were populated and the default HBs were replaced with the
synthetic ones.  ILABUNDS was then rerun on the new stellar
populations.

The differences from the CMD-based abundances are shown in Table
\ref{table:PartiallyResolved}.  All GCs converge on isochrone ages
that agree slightly better with results from resolved photometry.  For
M3, M13, NGC~7006, and M15 (whose HB's were not modelled accurately
with the default isochrones), the addition of synthesized HBs has
brought many of the abundances into better agreement with the
CMD-based ones. For 47~Tuc, however, the synthetic HBs introduce
larger discrepancies with the CMD-based values, suggesting that for
red HB GCs the default BaSTI HBs are likely to be sufficient.
NGC~7006's [\ion{Ti}{1}/\ion{Fe}{1}] ratio remains discrepant,
suggesting that the population is still not perfectly modelled.

Note that for nearby extragalactic clusters the {\it faint} detection
limit will be just below the HB, and the photometric uncertainties
will be much larger than in Figure \ref{fig:PartiallyResolved}.  This
means it will not be as easy to constrain the best-fitting
metallicities from the CMDs.  However, even if incorrect isochrone
metallicities are chosen for these GCs, the abundances converge back
on reasonable metallicities for the Galactic GCs.

\begin{table*}
\centering
\begin{minipage}{165mm}
\caption{Abundance differences with a partially resolved
cluster.\label{table:PartiallyResolved}}
  \begin{tabular}{@{}llccccccccc@{}}
  \hline
Age (Gyr) & [Z/H] & & $\Delta[$\ion{Fe}{1}/H$]$ & $\Delta[$\ion{Fe}{2}/H$] $ &
$\Delta[$\ion{Ca}{1}/\ion{Fe}{1}$]$ & $\Delta[$\ion{Ti}{1}/\ion{Fe}{1}$]$ &
$\Delta[$\ion{Ti}{2}/\ion{Fe}{2}$]$ &
$\Delta[$\ion{Ni}{1}/\ion{Fe}{1}$]$ &
$\Delta[$\ion{Ba}{2}/\ion{Fe}{2}$]$ & $\Delta[$\ion{Eu}{2}/\ion{Fe}{2}$]$\\
\hline
47~Tuc & & & & & & & & & \\
11 & -0.35 & & {\bf $+$0.09} & {\bf $+$0.17} & {\bf $-$0.06} & {\bf $-$0.12} & $ $0.0  & $+$0.03 & $-$0.04       & {\bf $-$0.08} \\
M3 & & & & & & & & & \\
13 & -1.27 & & $-$0.04      & {\bf $-$0.15} & $+$0.04 & $-$0.03 & $+$0.02 & {\bf $-$0.06} & $-$0.03       & $-$0.01 \\
M13 & & & & & & & & & \\
13 & -1.27 & & {\bf $-$0.05}& {\bf $-$0.12} & {\bf $+$0.06} & $-$0.04 & $-$0.03 & {\bf $-$0.05} & {\bf $-$0.07} & $-$0.04 \\
NGC7006 & & & & & & & & & \\
10 & -1.27 & & $+$0.04 & {\bf$+$0.08} & $+$0.03 & {\bf $-$0.11} & {\bf $+$0.09} & {\bf $-$0.08} & $+$0.04  & {\bf $-$0.07} \\
M15 & & & & & & & & & \\
10 & -1.79 & & $ $0.0       & {\bf $+$0.06} & $+$0.01 & $-$0.03 & $+$0.01 & $-$0.03 & $+$0.02 & $-$0.01 \\
& & & & & & & & & & \\
\hline
\end{tabular}
\end{minipage}\\
\medskip
\raggedright {\bf Notes: } All isochrones have $[\alpha/\rm{Fe}] =
+0.2$.  Abundance differences are calculated relative to the CMD-based
abundances in Table \ref{table:OriginalAbunds}, as described in
Section \ref{subsec:SystematicErrors}.\\
\end{table*}

\section{Systematic Offsets that Occur in all IL Analyses}\label{sec:Both}
Regardless of how the stellar population is modelled, some simplifying
assumptions must be made.  These include:
\begin{enumerate}
\item The methods used to generate the stellar parameters (Appendices
\ref{subsec:Boxes} and \ref{subsec:MicroturbulenceLaw})
\item The models of stellar subpopulations (Appendices 
\ref{subsec:Oddballs} and \ref{subsec:Hot})
\item The influence of foreground stars (Appendix \ref{subsec:FSs}) and
chemical variations in the model atmospheres (Appendix
\ref{subsec:AlphaFe}).
\end{enumerate}
Again, the validity of these assumptions can differ between GCs in a
given study.  This appendix tests specific assumptions that will affect
both CMD- and HRD-based analyses.

\subsection{CMD/HRD Boxes}\label{subsec:Boxes}

In both CMD and HRD-based methods the stars are binned together to
reduce computation time.  The effects from the coarseness and
definition of the boxes are investigated here.  First, an abundance
analysis is performed on 47~Tuc with \textit{no} CMD boxes (i.e. EWs
are computed for each star).  The abundance differences (tabulated in
Table \ref{table:boxes}) are completely negligible, suggesting that
boxing the CMD is an appropriate choice to speed up computations.
This is essential, since using the default number of 27 boxes speeds
up computations by a factor of 200 compared to the no box case.

\begin{figure*}
\begin{center}
\centering
\subfigure[47 Tuc boxes]{\includegraphics[scale=0.4]{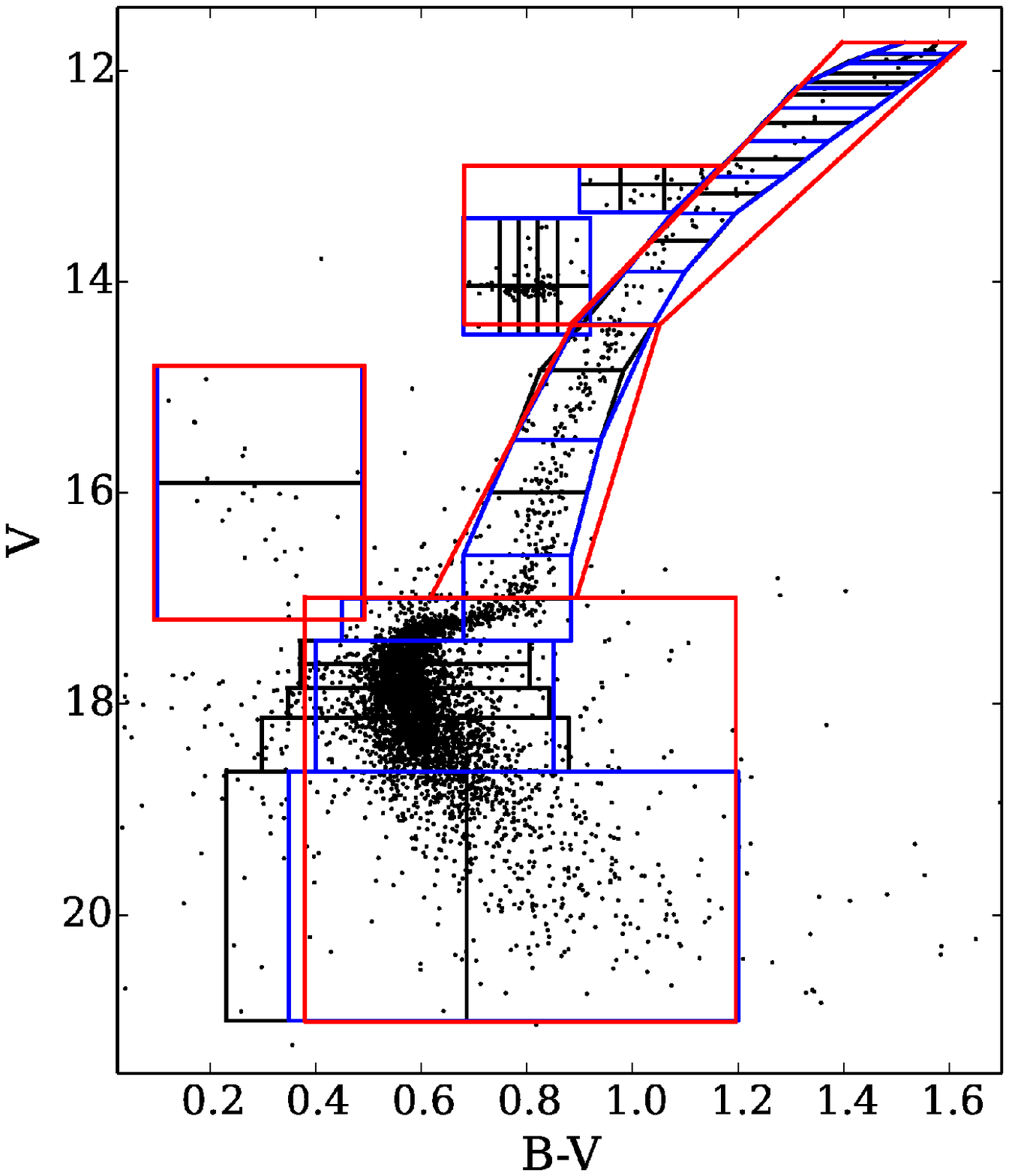}\label{fig:Finer}}
\subfigure[M13 boxes]{\includegraphics[scale=0.4]{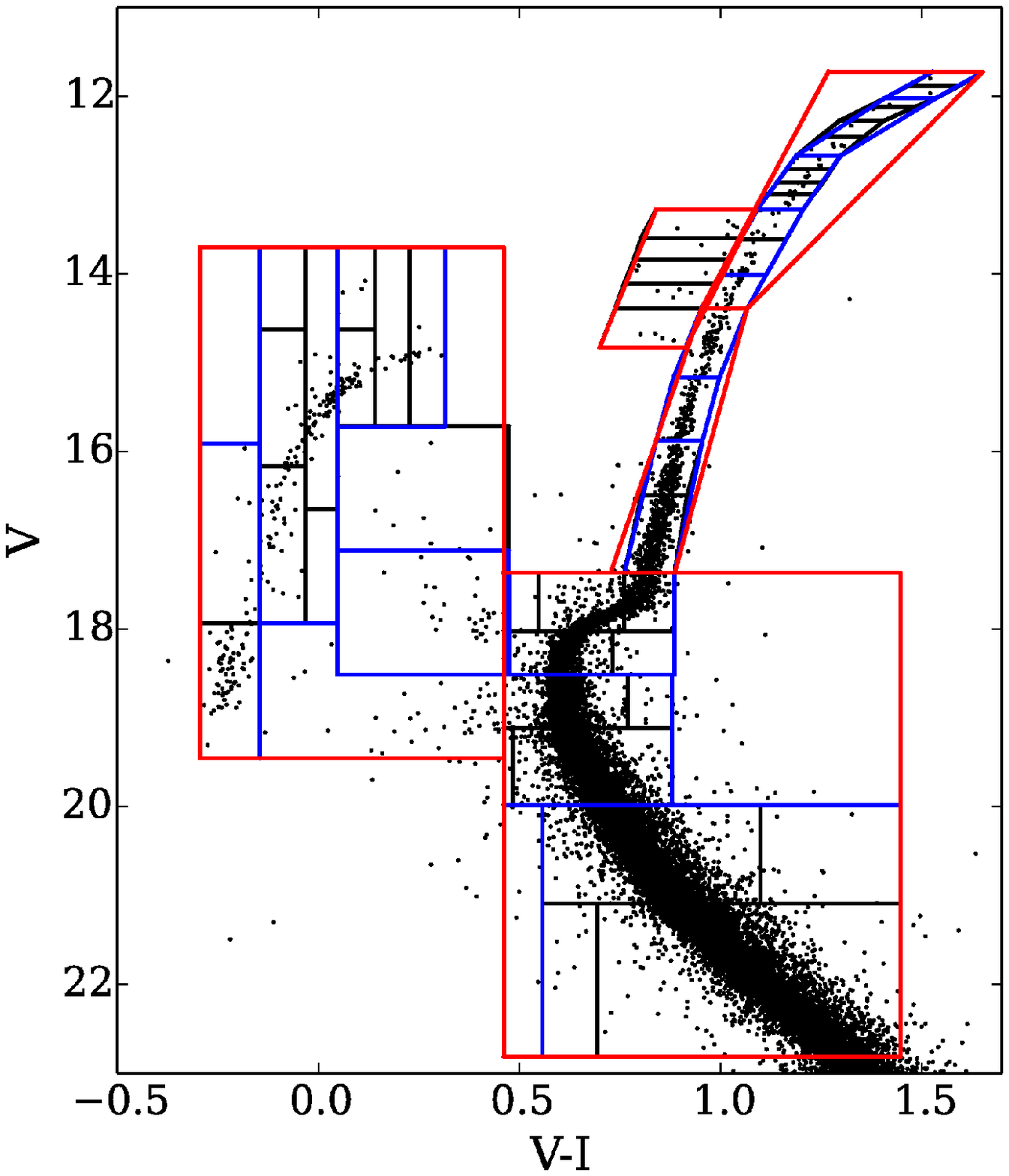}\label{fig:Coarser}}
\caption{Comparisons of box definitions for 47~Tuc (left) and M13
(right).  Finer boxes (in black) have increased resolution on the RGB,
HB, and AGB.  Coarser boxes are shown in blue and red.\label{fig:Boxes}}
\end{center}
\end{figure*}

Box definition was then investigated with 47~Tuc and M13, to compare
the effects of metallicity and HB morphology.  Finer and coarser boxes
are shown in Figure \ref{fig:Boxes} with the old 47~Tuc and M13 boxes
(in black).  The finer boxes were reshaped to provide finer coverage
of the upper red giant branch (RGB), horizontal branch (HB), and
asymptotic giant branch (AGB), and to include more stars in the main
sequence boxes.  The coarser boxes still maintain finer resolution of
the brightest stars.  These abundance differences are also shown in
Table \ref{table:boxes}.  As expected, the differences are negligible
for the cases with finer boxes.  For the coarser boxes, the
differences are significant for 47~Tuc when $5-17$ boxes are
considered while M13 is sensitive to the coarse five box case.  A
moderate number of boxes ($\sim 25-40$) therefore provides a
compromise between faster computing time and precision.  In fact,
using the default number of boxes ($\sim 30$) only slows computations
down by a factor of 2 over the coarsest box cases.

These tests were then performed on the synthetic HRDs.  The original
HRD-based abundances in Table \ref{table:Isochrones} were produced
using isochrones that were binned into boxes that each contained 3.5\%
of the total luminosity.  Table \ref{table:boxes} also shows the
effects if these HRD boxes are redefined.  For 47~Tuc, boxes of 2-20\%
lead to insignificant differences.  Surprisingly, the 1\% boxes have
large offsets---this seems to be a result of rounding errors when
individual boxes are assigned fractions of stars instead of round
numbers (as discussed in Appendix \ref{subsubsec:Mv}).  M13 is much
more sensitive to HRD box definitions, though the 2\% case seems to
still be due to rounding errors.  Thus, these results indicate that
the HRD-based abundances are also largely insensitive (with offsets
$\la0.05$ dex) to the precise box definitions.

\begin{table*}
\centering
\begin{minipage}{165mm}
\caption{Differences in 47~Tuc abundance ratios as a result of
different boxing methods.\label{table:boxes}}
  \begin{tabular}{@{}lccccccccc@{}}
  \hline
& & $\Delta[$\ion{Fe}{1}/H$]$ & $\Delta[$\ion{Fe}{2}/H$] $ &
$\Delta[$\ion{Ca}{1}/\ion{Fe}{1}$]$ & $\Delta[$\ion{Ti}{1}/\ion{Fe}{1}$]$ &
$\Delta[$\ion{Ti}{2}/\ion{Fe}{2}$]$ & $\Delta[$\ion{Ni}{1}/\ion{Fe}{1}$]$ & $\Delta[$\ion{Ba}{2}/\ion{Fe}{2}$]$ & $\Delta[$\ion{Eu}{2}/\ion{Fe}{2}$]$ \\
\hline
47 Tuc: CMD & & & & & & & & \\
No boxes         &  & $ $0.0  & $+$0.0  & $+$0.01 & $+$0.01 & $+$0.01 & $+$0.01 & $+$0.01 & $+$0.02 \\
Finer boxes (49) &  & $-$0.01 & $ $0.0  & $+$0.01 & $+$0.01 & $ $0.0  & $+$0.01 & $ $0.0  & $+$0.04 \\
Coarse  boxes (17)& & $-$0.01 & $+$0.01 & $+$0.01 & $+$0.01 & $ $0.0  & $ $0.0  & $ $0.0  & {\bf $+$0.05} \\
Coarser boxes (5)&  & $-$0.02 & $ $0.0  & $+$0.02 & $+$0.03 & $-$0.01 & $+$0.01 & $ $0.0  & {\bf $+$0.07} \\
 & & & & & & & & \\
M3: CMD & & & & & & & & \\
Finer boxes (40) &  & $-$0.03 & $+$0.01       & $ $0.0        & $ $0.0        & $-$0.01       & $ $0.0        & $-$0.03 & $ $0.0 \\
Coarse  boxes (16)& & $-$0.01 & $+$0.02       & $+$0.01       & $+$0.01       & $ $0.0        & $ $0.0        & $-$0.01 & $+$0.01 \\
Coarser boxes (5)&  & {\bf $+$0.09} & {\bf $+$0.08} & $+$0.02       & $+$0.04       & $+$0.04       & $ $0.0        & {\bf $+$0.05}  & $+$0.03 \\
 & & & & & & & & \\
47 Tuc: HRD & & & & & & & & \\
1\%         &  & {\bf $-$0.07} & $+$0.04       & {\bf $-$0.06} & {\bf $-$0.14} & {\bf $-$0.10} & $-$0.04       & {\bf $-$0.17} & {\bf $-$0.12} \\
2\%         &  & $-$0.01       & $+$0.01       & $-$0.02       & $-$0.03       & $-$0.01       & $-$0.01       & $-$0.04       & $-$0.03 \\
5\%         &  & $+$0.01       & $-$0.01       & $ $0.0        & $+$0.02       & $+$0.01       & $ $0.0        & $+$0.01       & $+$0.02 \\
10\%        &  & $-$0.01       & $ $0.0        & $ $0.0        & $ $0.0        & $-$0.01       & $ $0.0        & $+$0.02       & $ $0.0  \\
20\%        &  & $-$0.01       & $-$0.02       & $+$0.01       & $+$0.03       & $ $0.0        & $+$0.01       & $+$0.01       & $+$0.03 \\
 & & & & & & & & \\
M13: HRD & & & & & & & & \\
1\%         &  & {\bf $+$0.11} & $+$0.01       & $-$0.01       & {\bf $+$0.16} & $+$0.03 & $+$0.03 & {\bf $+$0.09} & {\bf $+$0.06} \\
2\%         &  & {\bf $+$0.05} & $ $0.0        & $ $0.0        & {\bf $-$0.06} & $-$0.03 & $-$0.01 & {\bf $-$0.05} & $-$0.04 \\
5\%         &  & {\bf $-$0.09} & $-$0.02       & $+$0.02       & $-$0.04       & $-$0.04 & $-$0.01 & {\bf $-$0.07} & {\bf $-$0.05} \\
10\%        &  & $-$0.03       & $-$0.03       & $ $0.0        & $ $0.0        & $-$0.03 & $ $0.0  & $-$0.03       & $ $0.0  \\
20\%        &  & $+$0.04       & $ $0.0        & $-$0.01       & $+$0.04       & $-$0.01 & $+$0.02 & $+$0.02       & $+$0.02 \\
 & & & & & & & & \\
\hline
\end{tabular}
\end{minipage}\\
\medskip
\raggedright {\bf Notes: } CMD-based abundance differences are
calculated relative to the baseline abundances in Table
\ref{table:OriginalAbunds}, as described in Section
\ref{subsec:SystematicErrors}, and use 27 boxes for 47~Tuc and 33
boxes for M13.  HRD-based abundance differences are calculated
relative to the best-fitting HRD-based values in Table
\ref{table:Isochrones}, and use box sizes of 3.5\%.\\
\end{table*}

\subsection{The Microturbulence Relation}\label{subsec:MicroturbulenceLaw}
Each box's microturbulent velocity is determined through an empirical
relation with the surface gravity; this relationship is based on a fit
to Arcturus and the Sun (see MB08 for details). ILABUNDS was rerun
with alternate empirical microturbulence relations from \citet[K09,
calibrated to GC and dwarf galaxy stars]{Kirby2009} and
\citet[G96]{Gratton1996}.  Note that the MB08 and K09 relations are
only dependent on $\log g$, though the G96 relation is dependent on
$\log g$ and $T_{\rm{eff}}$.  The differences in these relations will
lead to slight variations in the subpopulations.

The abundance offsets are shown in Table \ref{table:MicroturbLaws} for
47~Tuc, M3, and M15 (to investigate [Fe/H] effects). With the
exception of \ion{Ba}{2}, the largest abundance differences are all
$\la 0.1$ dex.  The differences between abundances with the MB08 and
K09 relations are mostly insignificant, supporting that the small
offset is negligible.  The G96 relation has a significant effect on
all abundances, depending on the cluster, where the offsets are
largest for M15.  It is not clear if it is valid to extend this
relationship to the hottest stars in the blue HB clusters.

\begin{table*}
\centering
\begin{minipage}{165mm}
\caption{Differences in abundance ratios as a result of
different microturbulence relations.\label{table:MicroturbLaws}}
  \begin{tabular}{@{}lcccccccc@{}}
  \hline
& $\Delta[$\ion{Fe}{1}/H$]$ & $\Delta[$\ion{Fe}{2}/H$] $ & $\Delta[$\ion{Ca}{1}/\ion{Fe}{1}$]$ & $\Delta[$\ion{Ti}{1}/\ion{Fe}{1}$]$ & $\Delta[$\ion{Ti}{2}/\ion{Fe}{2}$]$ & $\Delta[$\ion{Ni}{1}/\ion{Fe}{1}$]$ & $\Delta[$\ion{Ba}{2}/\ion{Fe}{2}$]$ & $\Delta[$\ion{Eu}{2}/\ion{Fe}{2}$]$\\
\hline
47~Tuc & & & & & & & & \\ 
MB08 with dispersion& {\bf $<$0.08} & {\bf $<$0.05} & $<$0.02 & $<$0.02 & {\bf $<$0.06} & $<$0.01 & {\bf $<$0.09} & $<$0.03\\
\citet{Kirby2009}   & {\bf $-$0.07} & $-$0.04       & $-$0.01 & $+$0.01 & {\bf $-$0.05} & $ $0.0  & $ $0.0        & $+$0.04 \\
\citet{Gratton1996} & {\bf $-$0.13} & {\bf $-$0.08} & $-$0.02 & $-$0.03 & {\bf $-$0.10} & $+$0.01 & {\bf $ $0.16} & {\bf $+$0.06} \\
   & & & & & & & & \\
M3 & & & & & & & & \\ 
\citet{Kirby2009}   & $-$0.04       & $-$0.01       & $-$0.01 & $+$0.01 & $-$0.04       & $ $0.0  & $-$0.01       & $ $0.0 \\
\citet{Gratton1996} & {\bf $-$0.12} & {\bf $-$0.05} & $-$0.03 & $+$0.04 & {\bf $-$0.11} & $ $0.0  & {\bf $ $0.27} & $+$0.01 \\
 & & & & & & & & \\
M15 & & & & & & & & \\ 
\citet{Kirby2009}   & {\bf $-$0.05} & $ $0.0  & $+$0.01       & $+$0.04       & $-$0.03       & $+$0.03        & $-$0.02       & $-$0.03 \\
\citet{Gratton1996} & {\bf $-$0.21} & $-$0.01 & {\bf $+$0.05} & {\bf $+$0.18} & {\bf $-$0.10} & {\bf $+$0.11}  & {\bf $-$0.30} & {\bf $+$0.06} \\
 & & & & & & & & \\
\hline
\end{tabular}
\end{minipage}\\
\medskip
\raggedright {\bf Notes: } Abundance differences are calculated
relative to the baseline abundances in Table
\ref{table:OriginalAbunds}, as described in Section
\ref{subsec:SystematicErrors}.\\
\end{table*}

The ``real'' microturbulent velocities are dispersed about these
relations.  Furthermore, each box contains stars with a dispersion of
microturbulent velocities.  To test these effects, each star in a
given box was assigned the same microturbulence value, which was
randomly selected from a Gaussian distribution with a standard
deviation of $0.2$ dex, centred on the MB08 relation.  These
microturbulence values were reselected 100 times.  The maximal
abundance offsets (also shown in Table \ref{table:MicroturbLaws}) are
$\la 0.1$ dex, with [\ion{Ba}{2}/\ion{Fe}{2}] having the greatest
difference.

\subsection{Anomalous Stars}\label{subsec:Oddballs}
Some cluster stars are distinctly different from the other cluster
stars.  This appendix investigates the effects of two different types
of oddball stars: long period variables (Appendix \ref{subsubsec:LPVs})
and carbon-enhanced stars (Appendix \ref{subsubsec:CHstars}).

\subsubsection{Long Period Variables}\label{subsubsec:LPVs}
As discussed in MB08 and Appendix \ref{subsec:Photometry}, the core
region of 47~Tuc contains two bright, cool M giants.  These stars are
long period variables (LPVs), stars which exhibit large brightness
variations over fairly long periods (days to years).  These LPVs are
only likely to exist in clusters at 47~Tuc's metallicity and above.
MB08 showed that these M giants are troublesome in the $B$, $V$
photometry because line blanketing reduces the $B$ and $V$ magnitudes
such that the stars appear to lie further down the RGB; including
those stars in boxes with incorrect atmospheric parameters led to
large abundance offsets. This problem does not occur in the $V$, $I$
photometry (see Appendix \ref{subsec:Photometry})---however, since the
M giants are LPVs, their atmospheric parameters change over time, such
that the properties of the M giants in the photometry/isochrone may
not match the conditions that were present when the IL spectra was
obtained.

The original $V$, $I$ abundances were calculated with the two bright M
giants at the tip of the RGB.  To test the worst case effects of long
period variability, these two stars were moved to boxes that were 1
mag fainter.  The abundance offsets (with respect to the abundances
from the $V$, $I$ photometry in Appendix \ref{subsec:Photometry}) are
shown in Table \ref{table:Both}.  The [\ion{Fe}{1}/H],
[\ion{Ti}{1}/\ion{Fe}{1}], [\ion{Ti}{2}/\ion{Fe}{2}], and
[\ion{Ba}{2}/\ion{Fe}{2}] ratios are all significantly affected,
though the differences are $<0.1$ dex.  The other ratios are largely
unaffected by the changes in the LPVs.

\begin{table*}
\centering
\begin{minipage}{165mm}
\caption{Differences in abundance ratios as a result of
various assumptions about the underlying stellar population.\label{table:Both}}
  \begin{tabular}{@{}lccccccccc@{}}
  \hline
& $\Delta[$\ion{Fe}{1}/H$]$ & $\Delta[$\ion{Fe}{2}/H$] $ & $\Delta[$\ion{Ca}{1}/\ion{Fe}{1}$]$ & $\Delta[$\ion{Ti}{1}/\ion{Fe}{1}$]$ & $\Delta[$\ion{Ti}{2}/\ion{Fe}{2}$]$ & $\Delta[$\ion{Ni}{1}/\ion{Fe}{1}$]$ & $\Delta[$\ion{Ba}{2}/\ion{Fe}{2}$]$ & $\Delta[$\ion{Eu}{2}/\ion{Fe}{2}$]$\\
\hline
LPVs$^{a}$ & & & & & & & & \\
47~Tuc        & $+$0.01 & {\bf$+$0.07} & $ $0.0  & $+$0.03 & $+$0.01 & $+$0.03 & $$0.0 & 0.0\\
& & & & & & & & \\
CH stars$^{a}$ & & & & & & & & \\
M15          & $ $0.0  & $ $0.0 & $+$0.01 & $ $0.0  & $ $0.0  & $+$0.01 & $ $0.0 & $+$0.01\\
& & & & & & & & \\
Hot Stars & & & & & & & & \\
M13: Abundances  & {\bf $+$0.06} & $+$0.04 & $-$0.01 & $ $0.0  & $-$0.01 & $-$0.03 & $+$0.04 & $-$0.04\\
M13: Rotation$^{a}$    & $+$0.02 & $+$0.02 & $+$0.01 & $-$0.01 & {\bf $+$0.07} & $+$0.01 & $ $0.0  & $-$0.02\\
& & & & & & & & \\
Field stars$^{a}$ & & & & & & & & \\
47~Tuc      &  {\bf $<$0.09} & {\bf $<$0.08} & \bf{$<$0.09} & $<$0.04 & {\bf $<$0.07} & {\bf $<$0.06} & {\bf $<$0.10} & {\bf $<$0.05}\\
NGC~7006    &  $<$0.04 & $<$0.01 & $<$0.0  & $<$0.03 & $<$0.02 & $<$0.01 & $<$0.03 & $<$0.04\\
M15         &  {\bf $<$0.10} & {\bf $<$0.09} & $<$0.04 & $-^{b}$ & $<$0.02 & $<$0.03 & {\bf $<$0.07} & $<$0.04\\
& & & & & & & \\
ODFNEW Atms & & & & & & & \\
47~Tuc     & {\bf $-$0.05} & {\bf $-$0.12} & $+$0.03 & $+$0.02 & $+$0.02 & $-$0.01 & $-$0.01 & $+$0.02\\
M3         & $ $0.0  & {\bf $-$0.07} & $+$0.01 & $+$0.01 & $+$0.02 & $-$0.02 & $-$0.02 & $-$0.01\\
M13        & $ $0.0  & {\bf $-$0.07} & $+$0.01 & $+$0.01 & $+$0.03 & $-$0.02 & $-$0.02 & $-$0.01\\
NGC~7006   & $ $0.0  & {\bf $-$0.07} & $ $0.0  & $+$0.02 & $+$0.02 & $-$0.01 & $-$0.03 & $-$0.03\\
M15        & $+$0.02 & $-$0.02       & $ $0.0  & $+$0.01 & $+$0.01 & $+$0.01 & $ $0.0  & $-$0.03\\
& & & & & & & \\
CN-cycled Atms & & & & & & & \\
47~Tuc     & {\bf $-$0.05} & {\bf $-$0.07} & $ $0.0 & $-$0.01 & $ $0.0 & $-$0.01 & $-$0.02 & $+$0.01\\
& & & & & & & \\
\hline
\end{tabular}
\end{minipage}\\
\medskip
\raggedright {\bf Notes: } Abundance differences are calculated
relative to the baseline abundances in Table
\ref{table:OriginalAbunds}, unless otherwise noted.\\
\raggedright $^{a}$ Baseline abundances were calculated separately
(see text).\\
\raggedright $^{b}$ Lines are too weak to measure in the synthesized
spectra.\\
\end{table*}

\subsubsection{Carbon Enhanced CH Stars}\label{subsubsec:CHstars}
Certain clusters (e.g. M15; \citealt{Shetrone1999}) have been observed
to have anomalous bright stars with strong CH bands (which have been
referred to as CH stars).  To test the effects of these stars, the
brightest star in M15 was made a CH star.  Note that this is not the
real CH star in M15; instead, this provides an indication of a worst
case scenario.  During the EW analysis, the brightest star was
assigned the [C/Fe], [N/Fe], and [O/Fe] abundances from
\citet{Shetrone1999} and the standard cluster abundances for the lines
of interest.  These C, N, and O abundances are then included in the
calculations for the continuous fluxes.  Note that the effects of
molecular lines would have to be investigated via spectrum syntheses
(see Paper I).

The abundance offsets are shown in Table \ref{table:Both}, and are
insignificant for all elements.

\subsection{Hot stars}\label{subsec:Hot}
The hottest stars in a cluster ($T_{\rm{eff}} \ga 8000$~K) can have
different properties from the other stars in the cluster.  The effects
of radiative levitation can drastically increase the surface
abundances of hot stars, possibly increasing the metal-poor surface
abundances of some elements to Solar composition
(e.g. \citealt{Behr2000,Behr2003,Lovisi2012}).  The hottest stars can
also have high rotation (up to $\sim 60$ km s$^{-1}$;
\citealt{Behr2003}), which broadens the line profiles and could affect
the shape of an IL spectral line.  In old GCs, the hottest stars are
often blue HB stars, which do not contribute much to the IL---Paper I
showed that these changes had a minimal effect on the synthesized
\ion{Mg}{1}, \ion{Na}{1}, and \ion{Eu}{2} lines.  This appendix
investigates the effects on the EWs of the Fe, Ca, Ti, Na, and Ba
lines.  Only M13 is considered for these tests, since 47~Tuc, M3, and
NGC~7006 do not have hot stars.

\subsubsection{Surface Composition}\label{subsubsec:HotAbunds}
For this test, all stars hotter than 8000~K were given Solar
composition, while all stars cooler than 8000~K were assigned the
standard cluster chemistry.  EWs were calculated for each box and
were combined as in the standard method---however, the initial
abundances were preserved, and \textit{no} iterations were done to
match the observed EWs.  ILABUNDS was then rerun on the new EWs.  The
differences from the original abundances provide indications of the
effects of the hottest HB stars.  These differences are listed in
Table \ref{table:Both}. With the exception of [\ion{Fe}{1}/H], all
abundance ratios are stable to within 0.04 dex.

\subsubsection{Rotation}\label{subsubsec:Rotation}
For stellar rotation, the same approach was employed as in Appendix 
\ref{subsubsec:HotAbunds}, except that stars hotter than 8000~K were
assigned rotational velocities of 60~km~s$^{-1}$ \textit{and} Solar
abundances.\footnote{Note that only considering rotation without
enhanced abundances leads to no differences in spectral features.}
Since rotation affects the shape of the line profiles, lines were
synthesized (in 10~\AA \hspace{0.025in} regions around the line of
interest).  Again, the boxes were combined and a new synthetic IL
spectra was produced.  To automate this process, EWs of the lines in
the new IL spectra were measured in DAOSPEC, and the new EWs were fed
to ILABUNDS.  Because the success of spectrum syntheses is highly
dependent on the input line list, the same procedure was applied
\textit{without the rotation enhancement in the hot stars}---these
abundances were used as the original abundances in the calculation of
the abundance differences, which are shown in Table \ref{table:Both}.
Table \ref{table:Both} shows that, with the exception of
[\ion{Ti}{2}/\ion{Fe}{2}], all abundances are stable to within 0.02
dex.  

\subsection{Field stars}\label{subsec:FSs}
There is always the possibility that an interloping field star could
contaminate the IL spectra from the cluster.  For Galactic clusters these
field stars would be in the Milky Way---for extragalactic GCs these
interloping field stars could also be in the host galaxy.  To test the
possible effects of field stars, the worst case scenario is considered,
i.e. that one of the brightest cluster stars is actually a field
star.  Three factors are varied:
\begin{enumerate}
\item \textit{Colour: } The field star is taken to be either the
brightest star on the RGB, or the brightest blue star (which may not
be included in any of the CMD boxes).
\item \textit{Composition: } The field star is considered to be either
Solar metallicity or a metal-poor star (with $[\rm{Fe/H}] = -2.5$).
In the latter case the field star is assumed to be $\alpha$-enhanced.
\item \textit{Luminosity Class: } The field star is taken to be either a
dwarf or a giant.  Physical parameters are then assigned to the field
star based on isochrone fits with the DSED isochrones.
\end{enumerate}

To test metallicity effects, 47~Tuc, NGC~7006, and M15 were all
considered for these tests.  Besancon models of the
Galaxy\footnote{\url{http://model.obs-besancon.fr/}}
\citep{BesanconRef} were used to find the average radial velocity of a
star at the same Galactic latitude and longitude as the target
GC---the artificial field stars were then assigned these radial
velocities.  For each spectral line, synthetic spectra were generated
for each CMD box (with the field star in its own box), EWs were
remeasured in the combined synthetic spectrum, and ILABUNDS was rerun
on the new EWs (this procedure is similar to that in Appendix
\ref{subsubsec:Rotation}).  Because the input line lists are
uncalibrated, the same procedure was performed on the original CMD
boxes; those abundances serve as the baseline values for the
comparisons.

The offsets are listed in Table \ref{table:Both}, and are generally
$\la 0.1$ dex.  For these resolved GCs, the abundance differences are
likely to be upper limits, since the worst case scenarios were
considered.  For unresolved GCs a brighter field star of a vastly
different colour could be included.  Targets should therefore be
inspected carefully for stellar contamination.  Extragalactic GCs will
have smaller Galactic field star contamination, but may also suffer
from contamination from its host galaxy. 

\subsection{Model Atmosphere Chemistry}\label{subsec:AlphaFe}

\subsubsection{$\alpha$-enhancement}\label{subsubsec:AlphaFe}
Spectroscopic analyses typically adopt $\alpha$-enhanced model
atmospheres for metal-poor stars, since the [$\alpha$/Fe] ratios in
Milky Way stars and clusters are enhanced (e.g. see
\citealt{Venn2004,Pritzl2005}).  To reflect this $\alpha$-enhancement,
the AODFNEW model atmospheres from the Kurucz data base have all
$\alpha$-elements (Ne, Mg, Si, S, Ar, Ca, and Ti) enhanced by 0.4
dex over the scaled-solar abundances.\footnote{Note that the high
Solar O abundance means that O is actually enhanced by $+0.54$.}
These $\alpha$-enhanced model atmospheres have therefore been used for
the baseline abundances of the target Galactic GCs, which are known to
be $\alpha$-enhanced.  The $\alpha$-enhanced atmospheres have also
been used for extragalactic targets whose $\alpha$-abundances indicate
enhancement (e.g. \citealt{Colucci2009}).

However, the IL abundance analyses have shown that some
$\alpha$-elements are \textit{not} enhanced in IL, such as Mg
(e.g. \citealt{Colucci2009}, Paper I).  This has been interpreted as a
chemical signature of the multiple populations in GCs, where there
is a second population that is enriched in products from, e.g., AGB
nucleosynthesis.  Thus, the abundances of, e.g., O and Mg, are
expected to be lower in the second generation stars, as has been
observed (e.g. \citealt{Carretta2009}).  This has the effect of
lowering the IL abundances if there are bright second generation
stars.  Since some of those elements are included in the model
atmosphere $\alpha$-enhancement, it may not be proper to use AODFNEW
atmospheres for all stars.  This effect is tested by using
solar-scaled ODFNEW atmospheres instead of AODFNEW ones.  The
abundance differences are tabulated in Table \ref{table:Both}.  For
the vast majority of elements the differences are insignificant.  Only
for \ion{Fe}{2} does the $\alpha$-enhancement make a difference, with
offsets up to $\sim 0.1$ dex.  This is likely because for the
brightest RGB stars, \ion{Fe}{2} is the dominant ionization stage, and
will be more affected by the presence or absence of free electrons. 

\subsubsection{Heavily CN-cycled atmospheres}\label{subsubsec:AlphaFe}
Stellar abundances (in particular, the C and N abundances) change as a
star evolves up the RGB and proceeds through the HB and AGB phases.
To test the worse case effects of C and N variations on the
atmospheric opacities, the heavily CN-cycled MARCS atmospheres
\citep{MARCS} were adopted for all boxes with $\log g < 3.5$ dex
(i.e. for all boxes that contained giants).  The results are shown in
Table \ref{table:Both}, and are only significant for [\ion{Fe}{1}/H]
and [\ion{Fe}{2}/H], though both are $\la 0.07$ dex.

\end{document}